\documentclass[preprint2]{aastex62}

\usepackage{hyperref,ifthen,psfig,graphicx}
\usepackage{epstopdf}
\usepackage{soul,color,amsmath}
\usepackage{ifthen,psfig,graphicx}
\usepackage{epstopdf,aas_macros}
\usepackage{soul,color,amsmath}
\usepackage{overpic}
\newcommand*{\mcc}[1]{\multicolumn{2}{c}{#1}}

\def\mathnew{\mathsurround=0pt}
\def\simov#1#2{\lower 2.5pt\vbox{\baselineskip0pt \lineskip-.5pt
\ialign{$\mathnew#1\hfil##\hfil$\crcr#2\crcr\sim\crcr}}}
\def\simless{\mathrel{\mathpalette\simov <}}
\def\simgreat{\mathrel{\mathpalette\simov >}}
\newcommand{\MeV}{Me\kern-0.11em V}
\newcommand{\keV}{ke\kern-0.11em V}

\newcommand{\cha}{{\it Chandra\/}}

\newcommand{\hst}{{\it HST\/}}

\newcommand{\ecmss}{erg\,cm$^{-2}$\,s$^{-1}$}

\newcommand{\es}{erg\,s$^{-1}$}
\newcommand{\kms}{\ensuremath{\rm km\;s}^{-1}}

\newcommand{\omegam}{$\Omega_{m,0}$\/}
\newcommand{\omegal}{$\Omega_{\Lambda,0}$\/}

\newcommand{\ciao}{{\it CIAO\/}}

\newcommand{\vpm}[3]{\ensuremath{{#1}^{+#2}_{-#3}}}
\newcommand{\vpmc}[3]{\ensuremath{{#1}}&\ensuremath{^{+#2}_{-#3}}}
\newcommand{\vmp}[3]{\ensuremath{{#1}^{+#3}_{-#2}}}

\makeatletter
\newcommand\iont[2]{{#1$\;${\small\expandafter\@slowromancap\romannumeral #2@\relax}}}
\makeatother
\makeatletter
\newcommand{\raisemath}[1]{\mathpalette{\raisem@th{#1}}}
\newcommand{\raisem@th}[3]{\raisebox{#1}{$#2#3$}}
\makeatother

\shorttitle{CHEERS Results from NGC 3393, III}
\shortauthors{Maksym et al.}

\begin{document}


\title{CHEERS Results from NGC 3393, III:\\ {\it Chandra} X-ray Spectroscopy of the Narrow Line Region}


\author[0000-0002-2203-7889]{W. Peter Maksym}
\affiliation{Harvard-Smithsonian Center for Astrophysics, 60 Garden S., Cambridge, MA 02138, USA}
\email{walter.maksym@cfa.harvard.edu; @StellarBones}

\author{Giuseppina Fabbiano}
\affiliation{Harvard-Smithsonian Center for Astrophysics, 60 Garden S., Cambridge, MA 02138, USA}

\author{Martin Elvis}
\affiliation{Harvard-Smithsonian Center for Astrophysics, 60 Garden S., Cambridge, MA 02138, USA}

\author{Margarita Karovska}
\affiliation{Harvard-Smithsonian Center for Astrophysics, 60 Garden S., Cambridge, MA 02138, USA}

\author{Alessandro Paggi}
\affiliation{Harvard-Smithsonian Center for Astrophysics, 60 Garden S., Cambridge, MA 02138, USA}
\affiliation{Dipartimento di Fisica, Universitˆ degli Studi di Torino, via Pietro Giuria 1, I-10125 Torino, Italy}
\affiliation{Istituto Nazionale di Fisica Nucleare, Sezione di Torino, via Pietro Giuria 1, I-10125 Torino, Italy}
\affiliation{INAF-Osservatorio Astrofisico di Torino, via Osservatorio 20, I-10025 Pino Torinese, Italy}

\author{John Raymond}
\affiliation{Harvard-Smithsonian Center for Astrophysics, 60 Garden S., Cambridge, MA 02138, USA}

\author{Junfeng Wang}
\affiliation{Department of Astronomy, Physics Building, Xiamen University Xiamen, Fujian, 361005, China}

\author{Thaisa Storchi-Bergmann}
\affiliation{Departamento de Astronomia, Universidade Federal do Rio Grande do Sul, IF, CP 15051, 91501-970 Porto Alegre, RS, Brazil}

\author{Guido Risaliti}
\affiliation{INAF - Arcetri Astrophysical Observatory, Largo E. Fermi 5, I-50125 Firenze, Italy}



\begin{abstract}


We present spatially resolved {\it Chandra} narrow-band imaging and imaging spectroscopy of NGC 3393.  This galaxy hosts a Compton-thick Seyfert 2 AGN with sub-kpc bipolar outflows that are strongly interacting with the circumnuclear gas.  We identify narrow-band excess emission associated with the \ion{Ne}{9} 0.905\,keV transition (with likely contributions due to intermediate-state iron emission) that points to strong shocks driven by AGN feedback.  Imaging spectroscopy resolves outflow-ISM interaction sites and the surrounding ISM at $\sim100$\,pc scales, and suggests the presence of a hot AGN wind above the plane at radii beyond the shock sites.  The cross-cone shows evidence for reprocessing of photoionization which has passed through gaps in the torus, and also for collisionally excited plasma which may be powered by a shock-confined equatorial outflow.  Deep X-ray observations at sub-arcsecond resolution (such as may be performed very efficiently by {\it Lynx}, which would also energetically resolve the complex line emission) are necessary to eliminate model degeneracies and reduce uncertainties in local feedback properties.

\end{abstract}




\section{Introduction}

The ubiquity of supermassive black holes (SMBHs) in galactic nuclei does not appear to arise from coincidence.  Rather, SMBHs and their host galaxies appear to form and co-evolve, with the properties and behavior of each informing the other \citep{HB14}.  In particular, rapidly accreting active galactic nuclei (AGN) produce photoionizing radiation, jets, and winds which can modulate star formation in the host galaxy and accretion by the SMBH \citep{Fabian12, GS17} via feedback processes which must be considered in addtion to those seen in rapid starbursts \citep{DS86,Silk97}.  

Although AGN feedback processes are easily observed on large scales in clusters, groups and massive galaxies \citep{Voit15}, statistical studies of SMBH obscuration show that radiative AGN feedback shapes galactic nuclei down to parsec scales \cite{Ricci17}.  The low density gas of the AGN extended narrow line region (ENLR) is sensitive to both kinematic outflows and radiative excitation associated with feedback, making studies of the ENLR particularly valuable.

Via the {\it CHEERS} ({\it CHandra Extended Emission Line Region Survey}) program, we are investigating the X-ray signatures of AGN feedback in nearby Seyfert Galaxies.  By spatially resolving X-ray emission within the ENLR, we can directly measure radiation from gas which has been collisionally ionized by AGN outflows or which arises from re-processed photoionizing emission  \citep{Paggi17,Fabbiano17,Fabbiano18a,Fabbiano18b}.  The velocities ($v\sim1000\,\kms$) of AGN ENLR outflows lead to characteristic energies of AGN feedback ($kT\sim1\,$keV), making X-rays the appropriate band for such measurements; in addition, measurement of multiple high ionization species in emission can be used to break degeneracies between models dominated by photoionized reflection or collisionally ionized plasma.  

\begin{sloppypar}
X-ray studies of systems with outflows smaller than $\sim$few kpc (\citealt{Wang11b,Wang11a,Wang11c,Paggi12,Bogdan17}; \citealt{Fabbiano18a}) have shown that the ENLR in these systems is not a single photoionized plasma but rather a multiphase gas where photoionized emission may co-exist simultaneously with contributions from collisionally ionized plasma stimulated by interactions between the outflows and the ENLR.  

This is the third paper in our series of {\it CHEERS} papers on NGC 3393, \citep[we refer to][as ``Paper I" and ``Paper II", respectively]{Maksym16,Maksym17}.  NGC 3393 is a nearby ($D=53\,$Mpc) barred spiral galaxy hosting a Seyfert 2 active nucleus\citep{DPW88,deV91}.  The AGN is Compton-thick (\citealt{Maiolino98, Guainazzi05, Burlon11,Koss15}), so extended soft X-ray emission emitted by the nuclear ISM can be spatially resolved without being dominated by the AGN accretion disk or corona.  Short {\it Chandra} observations \citep{Bianchi06,Levenson06} showed extended X-ray emission associated with biconical optical [\ion{O}{3}] emission.  \cite{Fabbiano11} claimed that deeper {\it CHEERS} revealed double hard point sources indicating a binary AGN with subarcsecond separation, though this result was challenged by \cite{Koss15}.  \cite{Koss15} did confirm the association between the NLR and soft X-rays using the {\it CHEERS} observations and zeroth-order {\it Chandra} imaging.  
\end{sloppypar}

Continuum-subtracted narrow-line observations using {\it HST} show that while the ENLR of the nuclear bicone is Seyfert-like, the surrounding medium forms a Low Ionization Nuclear Emission Region (LINER) cocoon (Paper I).  Subsequent analysis in Paper II shows that while the LINER could be caused by photoionization filtered by the bicone gas or by slow shocks from lateral expansion of a biconical outflow, fast \citep[$FWHM\simgreat1000\,\kms$][]{Fischer13} photoionizing shocks (as proposed by \citealt{Cooke00}) cannot be excluded as a contributor to excitation within the bicone.  This picture is supported by [\ion{O}{3}] and H$\alpha$ kinematics at sites of outflow-ISM interaction with enhanced X-ray emission.   Fast shocks are an alternative to the suggestion by \cite{Koss15} that emission at 1.8 keV (consistent with e.g. \ion{Mg}{12} and \ion{Si}{13}) indicates emission photoionized by the AGN.

We have investigated the spatially resolved spectroscopic properties of the X-ray emission associated with these processes.  {\it Chandra}'s sub-arcsecond angular resolution is critical for this analysis due to the complex and compact morphology of the system.  We find evidence for simultaneous contributions from reprocessing of photoionizing radiation and collisional feedback-driven plasma throughout the nucleus (including in the cross-cone, pointing to gaps in the torus, e.g. \citealt{Paggi12,Fabbiano18a}).  In particular, we find evidence for shocks as significant contributors to the extended X-ray emission.  The physical properties of that plasma indicate that the interactions between the AGN and gas in the inner galactic plane may be sufficient to unbind gas from the host galaxy, but it is not clear that this is the case (as suggested by evidence for trapped hot plasma which may be collisionally excited by a shock-confined equatorial outflow).  

Throughout this paper, we adopt concordant cosmological parameters\footnote{\raggedright{Distances are calculated according to http://www.astro.ucla.edu/~wright/CosmoCalc.html}} of
$H_0=70\ $km$^{-1}$ sec$^{-1}$ Mpc$^{-1}$, \omegam=0.3 and \omegal=0.7. All coordinates are J2000.  In all figures, celestial North is up and color scales are logarithmic (unless otherwise noted).  For distance evaluation we use the \cite{Theureau98} determination of redshift $z=0.0125$ from observations of the 21-cm neutral hydrogen emission line, such that NGC 3393 is at distance $D=53\;$Mpc with linear scale $257\;\rm{pc\,arcsec}^{-1}$.  The redshift corresponds to $7\,$eV for a photon emitted at $600\,eV$, which is negligible compared to the $\sim100\,$eV FWHM of ACIS-S, so redshift corrections are negligible.

\begin{table}[t]
\centering
\caption{Chandra ACIS-S Imaging Observations of NGC 3393}
\label{table:cxo-obstable}
\vspace{0.1in}
\begin{tabular}{ccc}
\tableline
Obsid	&	Date			& ks		\\
\tableline
04868	&	2004 Feb 28	& 29.33		\\ 
12290	&	2011 Mar 12	& 69.16		\\ 
\tableline
Total	&	... 			& 98.49		\\
\tableline 


\tableline

\end{tabular}
\end{table}

\section{Observations and Data Reduction}\label{sec-obs}

We have obtained two observations totaling $\sim100\,$ks of ACIS-S exposure from the {\it Chandra} Data Archive\footnote{http://cda.harvard.edu/chaser} (Table \ref{table:cxo-obstable}).  Since the {\it Chandra} gratings may introduce uncalibrated distortions to the PSF on sub-arcsecond scales\footnote{For discussion of the \cha\ zeroth order, see http://asc.harvard.edu/proposer/POG/html/HETG.html}, we do not use zeroth order grating images.  As in Paper II, we reprocessed the data with {\it Chandra} Interactive Analysis of Observations package \citep[CIAO;][]{ciao}  using sub-pixel event positioning, removing periods of strong background flaring, and merging data from both obsids (with {\tt merge\_obs}) to produce an event file equivalent to a single long exposure.  We describe the use of other data reduction techniques as appropriate in subsequent subsections.   We generated all radial profile maps for X-ray, optical and radio data using the {\it CIAO} tool {\tt dmextract}.

Data processing and production of continuum-subtracted [\ion{O}{3}]\,$\lambda5007\,$\AA\ images for {\it HST} WFC3 data are also described in detail in Paper II.

\subsection{X-ray Imaging and Spectral Features}

\subsubsection{X-ray Continuum Emission}
\label{sec:XConMap}

\begin{figure}[th!]
\vspace{0.1in}
\noindent
\centering
\includegraphics[width=0.47\textwidth]{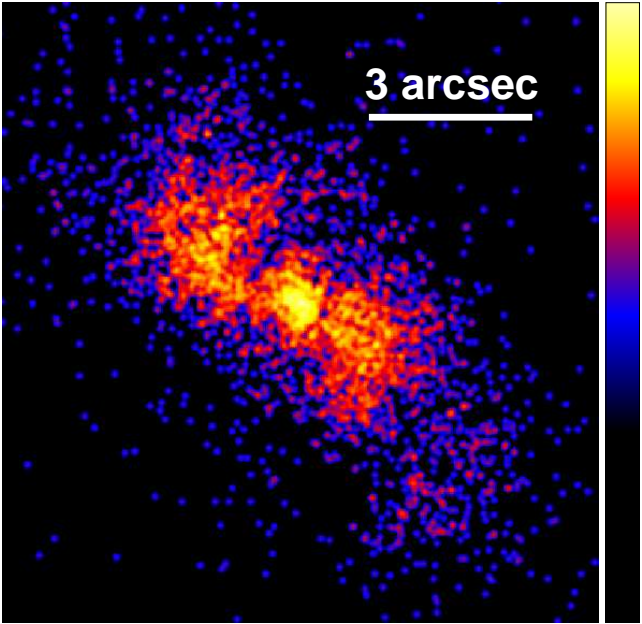}
\caption{{\it Chandra} 0.3-2 keV image of the NGC 3393 nucleus.  Photons are binned to 1/16th ACIS pixel size and smoothed with a gaussian where $r=3\,$pixels.
} 
\label{fig:continuum}
\end{figure}

In Paper II, we investigated the broadband X-ray emission of the NGC 3393 nucleus (Fig. \ref{fig:continuum}).  The brightest part of the soft emission is located in the nucleus (where there is a central radio peak), and wraps around the radio lobes in an S-shape.  We observed that the [\ion{O}{3}] and soft X-rays are effectively co-spatial \citep[as in][]{Levenson06,Koss15}, but find a fuller picture using deconvolved broadband (0.3-8 keV) maps, optical narrow line emission maps, and other multiwavelength data.  Namely, we inferred that although the bulk of the X-ray emission is likely photoionized, shock emission may lead the radio lobes, particularly given that the shocks are likely to be fast enough to be photoionizing.  We also identified extended hard ($2-8$\,keV) emission that traces the brightest parts of the continuum emission.

\begin{figure*} 
\noindent
\centering
\begin{overpic}[scale=0.39]{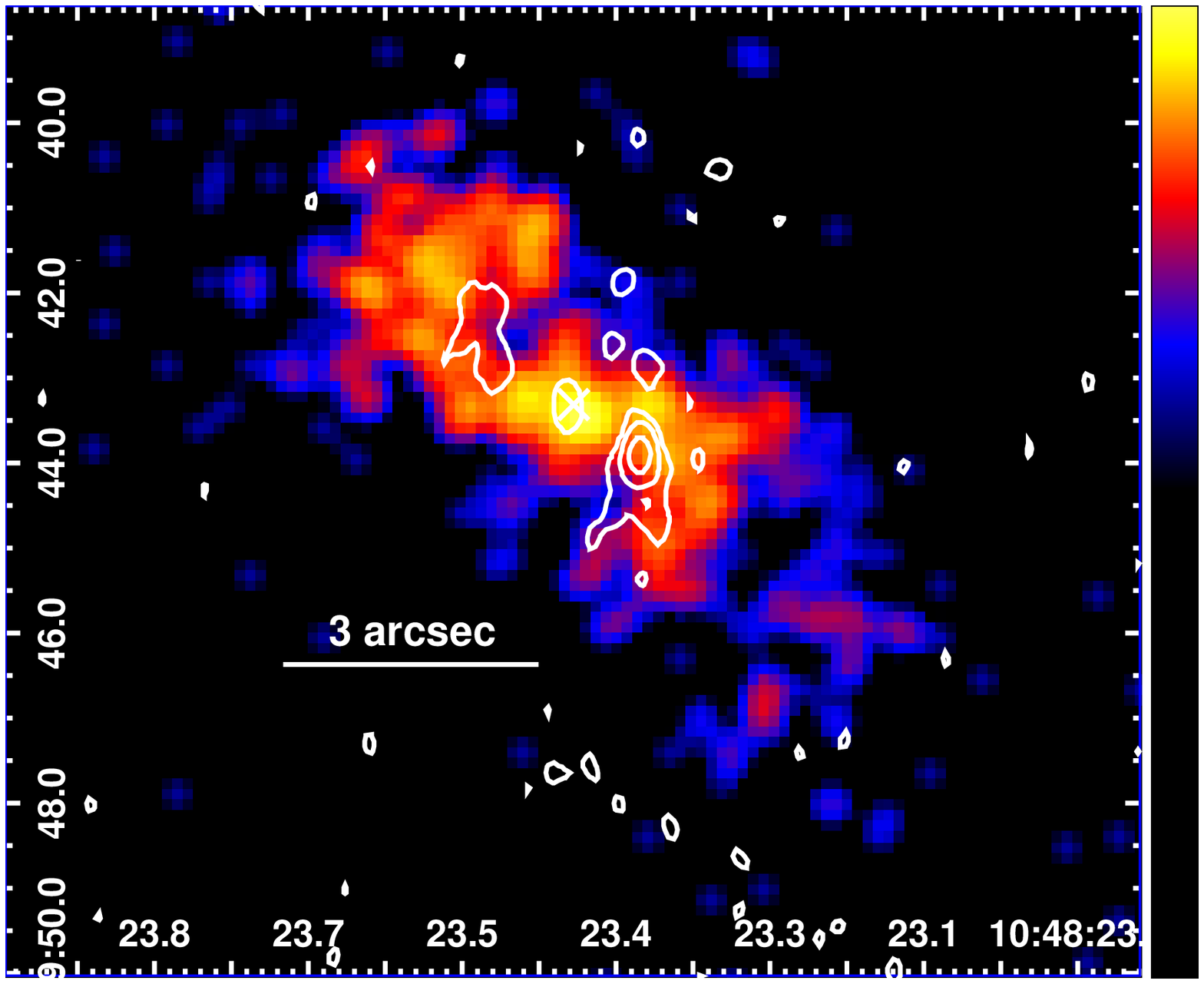}
	\put(60,70){\color{white}\bf (a) \ion{O}{7};}
	\put(70,65){\color{white}\bf radio}
	\put(70,60){\color{white}\bf contours}
\end{overpic}\hspace{0.01\textwidth}%
\begin{overpic}[scale=0.39]{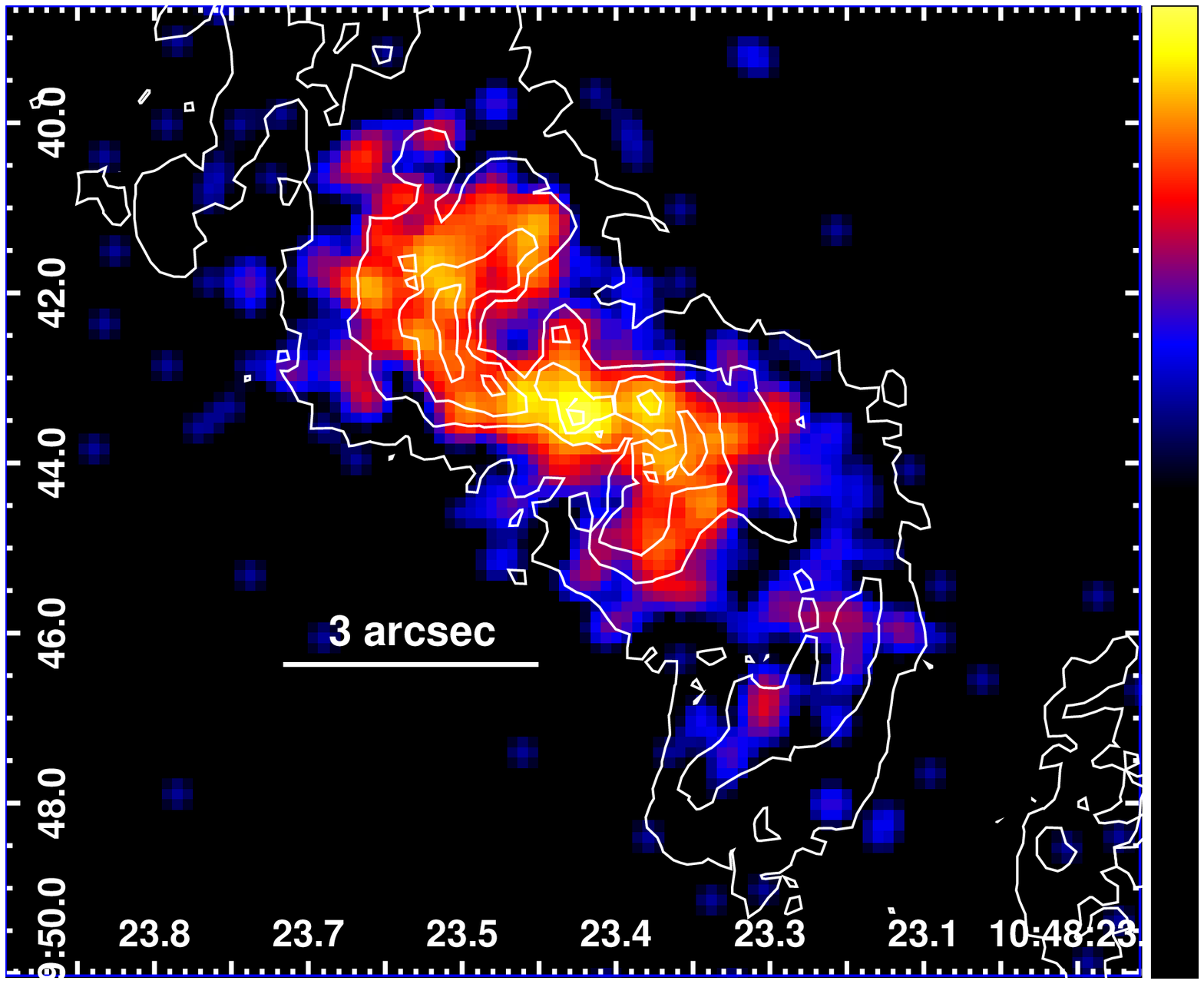}
	\put(60,70){\color{white}\bf (b) \ion{O}{7};}
	\put(68,65){\color{white}\bf [\ion{O}{3}]}
	\put(70,60){\color{white}\bf contours}
\end{overpic}\hspace{0.01\textwidth}\par
\begin{overpic}[scale=0.39]{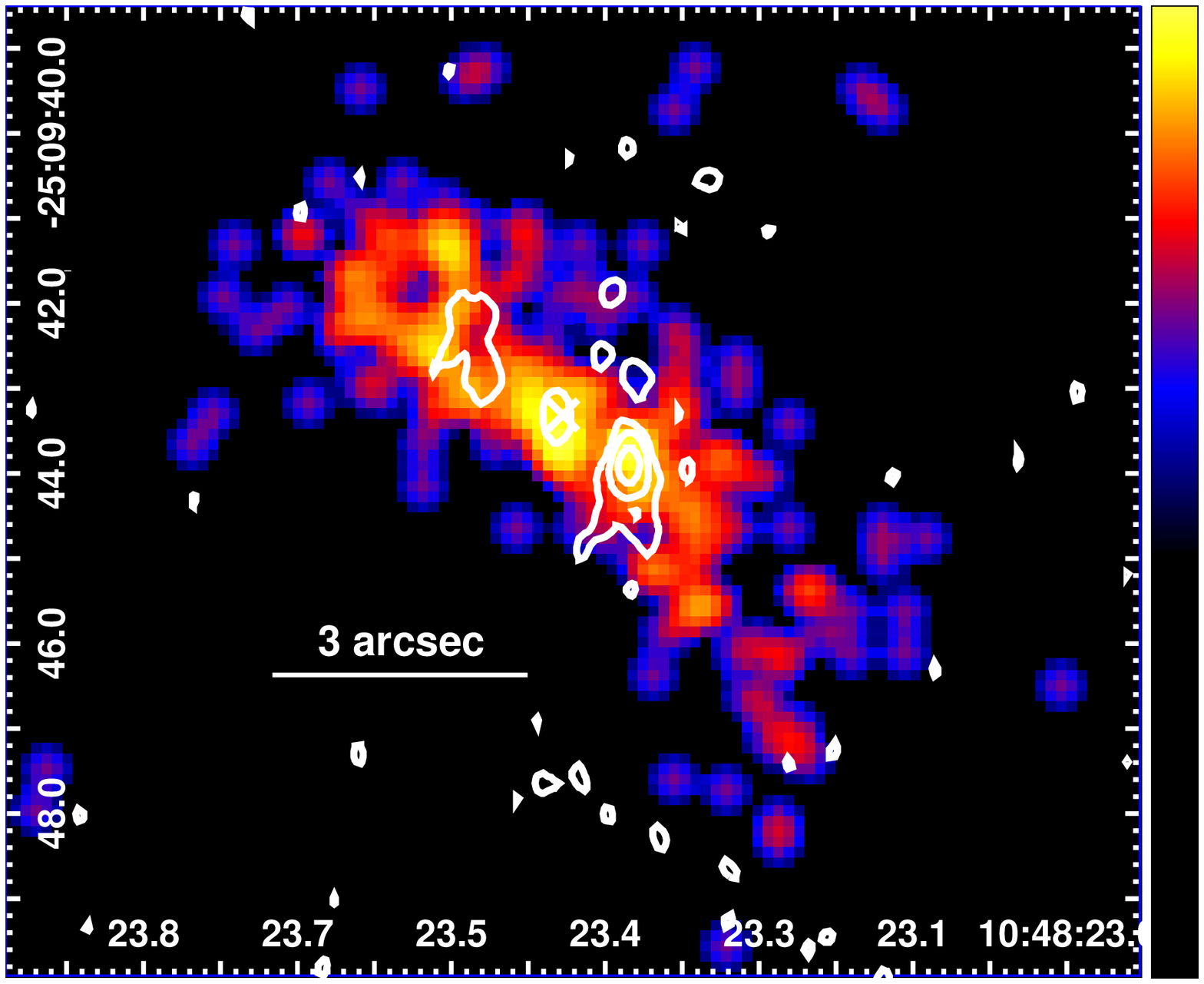}
	\put(60,70){\color{white}\bf (c) \ion{O}{8};}
	\put(70,65){\color{white}\bf radio}
	\put(70,60){\color{white}\bf contours}
\end{overpic}\hspace{0.01\textwidth}%
\begin{overpic}[scale=0.39]{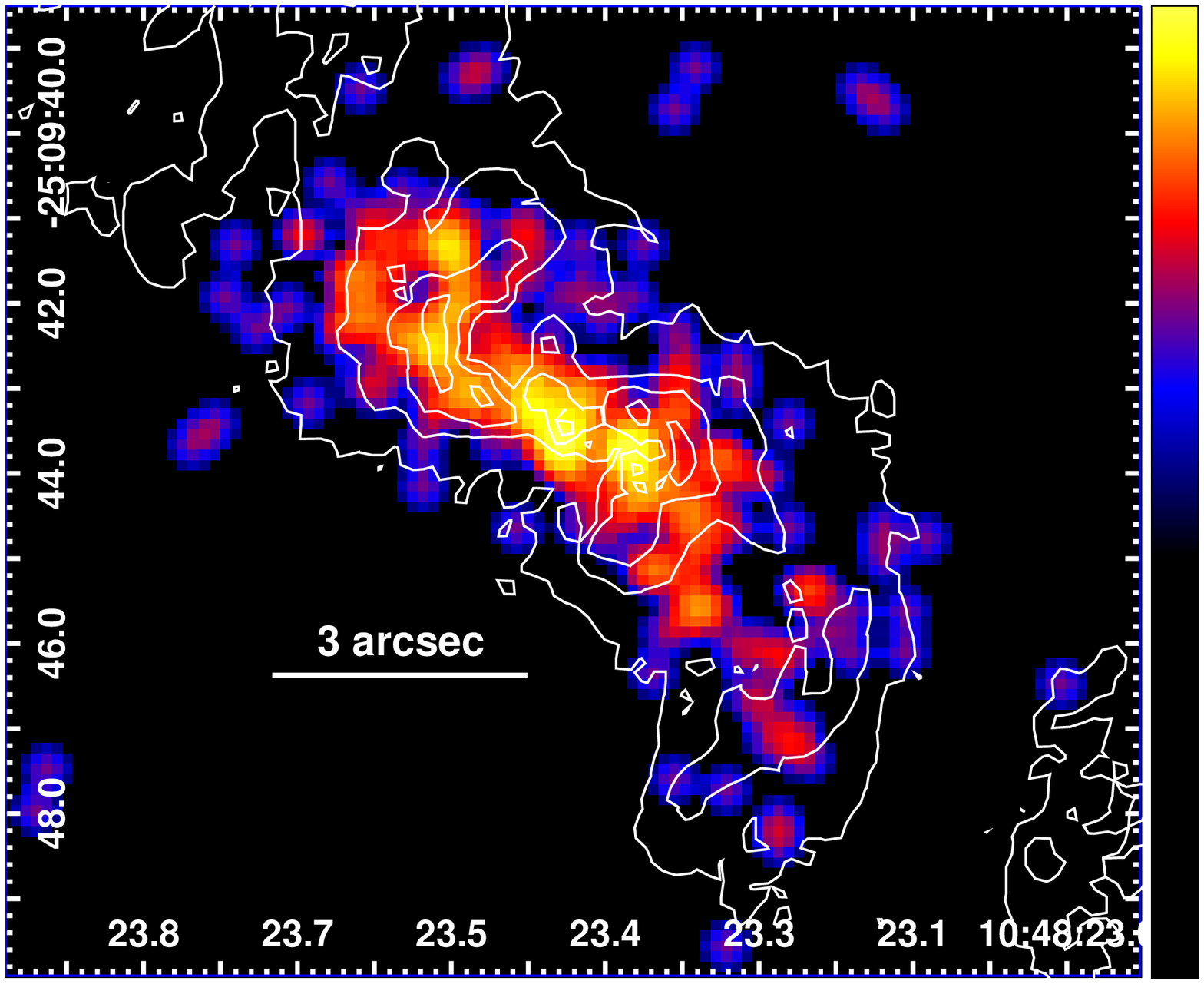}
	\put(60,70){\color{white}\bf (d) \ion{O}{8};}
	\put(68,65){\color{white}\bf [\ion{O}{3}]}
	\put(70,60){\color{white}\bf contours}
\end{overpic}\hspace{0.01\textwidth}\par
\begin{overpic}[scale=0.39]{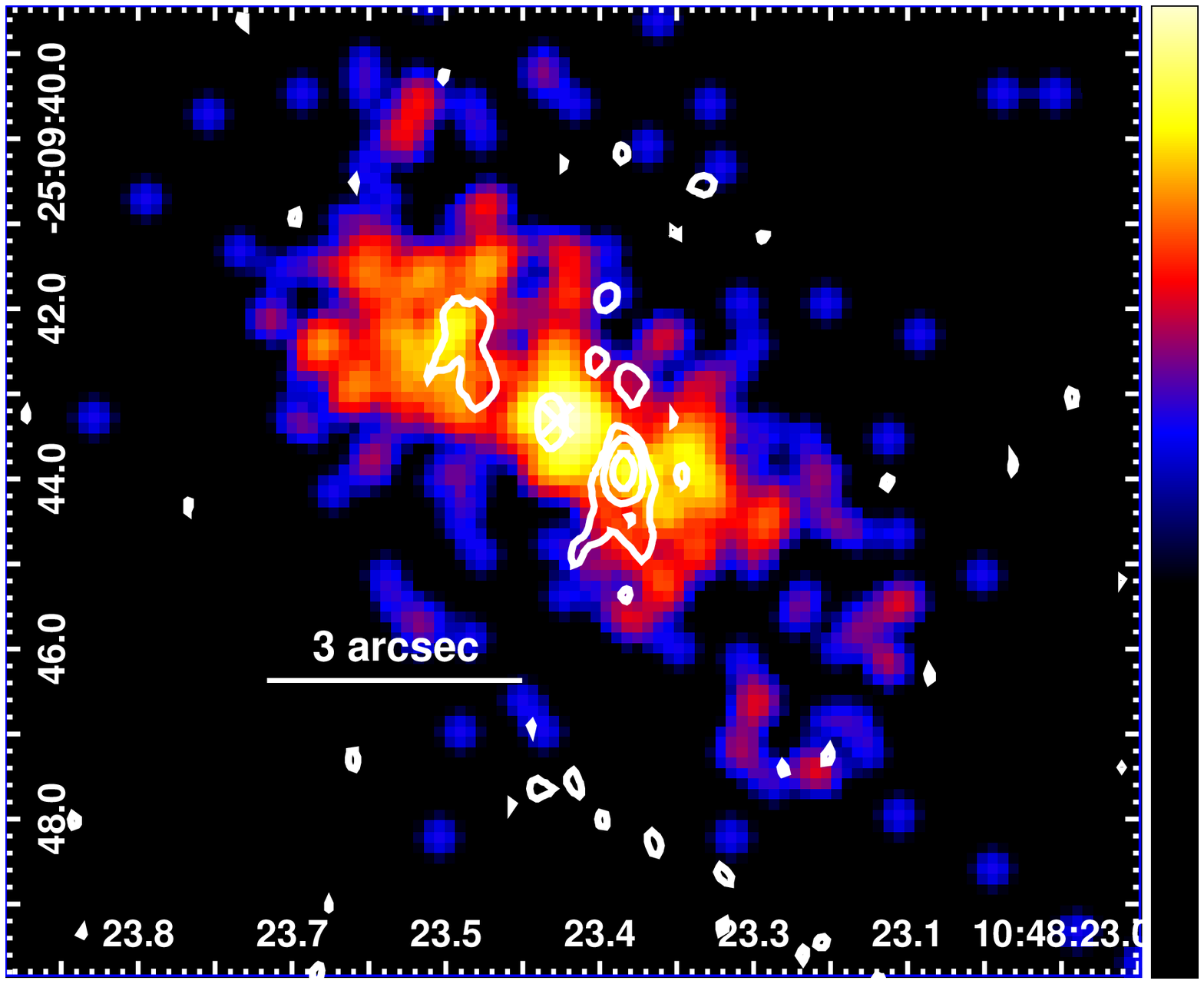}
	\put(60,70){\color{white}\bf (e) \ion{Ne}{9};}
	\put(70,65){\color{white}\bf radio}
	\put(70,60){\color{white}\bf contours}
\end{overpic}\hspace{0.01\textwidth}%
\begin{overpic}[scale=0.39]{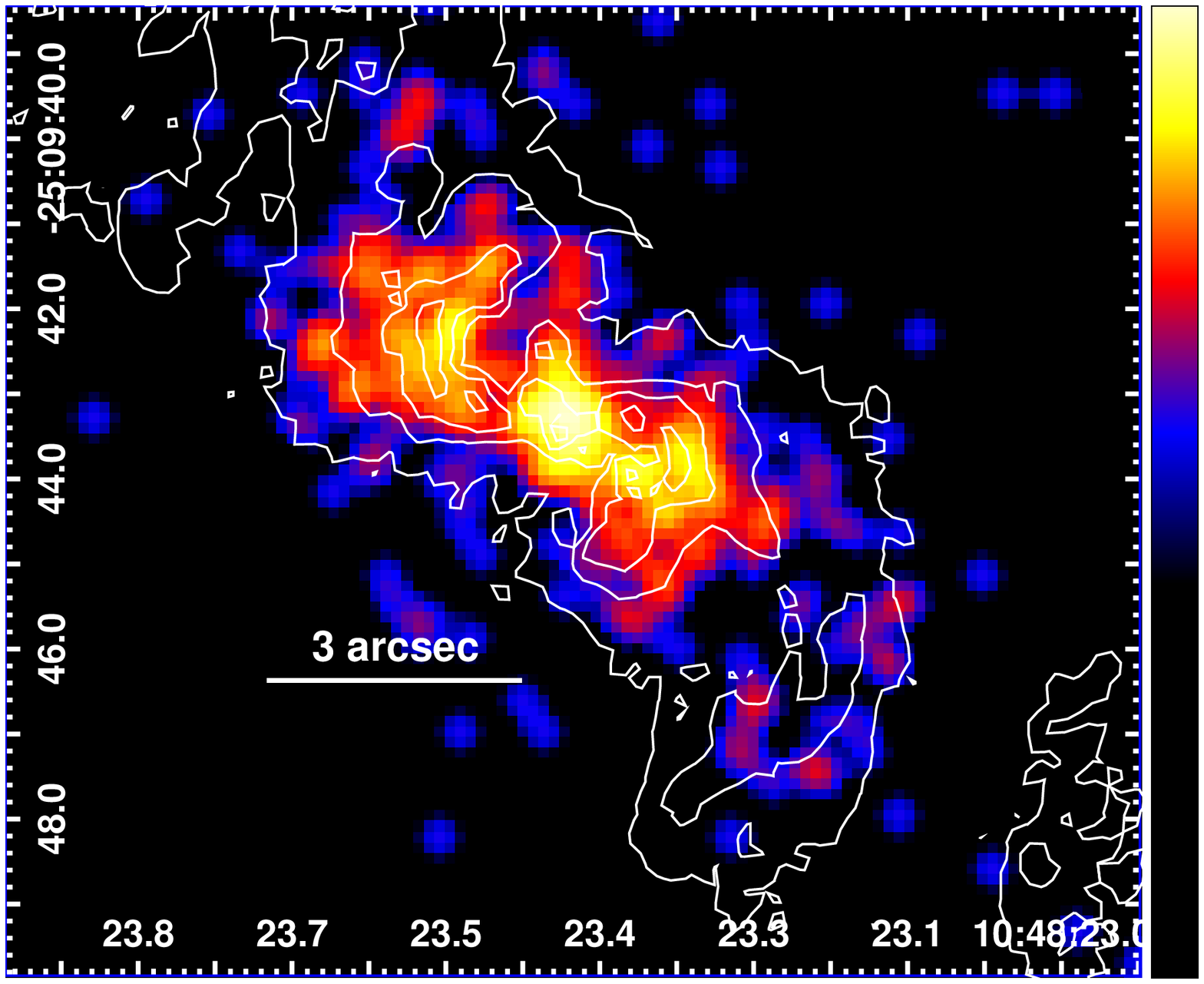}
	\put(60,70){\color{white}\bf (f) \ion{Ne}{9};}
	\put(68,65){\color{white}\bf [\ion{O}{3}]}
	\put(70,60){\color{white}\bf contours}
\end{overpic}\hspace{0.01\textwidth}\par
\caption{X-ray emission line maps from {\it Chandra} observations of NGC 3393.  The {\bf left} column is overlaid with 8.46 GHz contours from FIRST (as in Paper II) and the {\bf right} column has contours from [\ion{O}{3}]\,$\lambda5007\,$\AA, as observed by {\it HST}.  From top to bottom, each row shows {\bf top:} \ion{O}{7} triplet at $0.569\,$keV,  {\bf center:} the \ion{O}{8} Ly$\alpha$ transition at 0.654 keV, and  {\bf bottom:} the \ion{Ne}{9} triplet at $0.915\,$keV.  Photons are binned to 0.123\arcsec\ scale, which is 1/4 of the native ACIS pixel size, and smoothed with a 3-pixel Gaussian filter.
} 
\label{fig:linemaps}
\end{figure*}

\subsubsection{X-ray Emission Line Maps}
\label{sec:XELMap}

We begin our X-ray spectral analysis by investigating the morphological distribution of X-ray emission of known ion species previously studied by \cite{Wang11b} and \cite{Paggi12}.

In each band, we produced merged event files, and images with both 1/8-pixel and 1/4-pixel (0.062\arcsec, 0.123\arcsec) spatial bins.  We generated exposure maps and fluxed images for the combined dataset using the {\it CIAO} script {\tt merge\_obs}.

In Fig. \ref{fig:linemaps}, we map photons in bands of varying width covering the \ion{O}{7} triplet at $\sim0.569\,$keV, the \ion{O}{8} Ly$\alpha$ transition at 0.654 keV, and the \ion{Ne}{9} triplet at $\sim0.915\,$keV.  Given the ACIS-S energy resolution of $\sim100\,$eV\footnote{\raggedright{{\it Chandra} Proposer's Guide; http://cxc.harvard.edu/proposer/POG/html/chap6.html}}, we define these bands such that in the observer frame, $0.526\,\rm{keV}<$\ion{O}{7}$<0.626\,\rm{keV}$,  $0.624\,\rm{keV}<$\ion{O}{8}$<0.674\,\rm{keV}$ and  $0.880\,\rm{keV}<$ \ion{Ne}{9} $<0.954\,\rm{keV}$.

Figure \ref{fig:linemaps} shows two columns with \ion{O}{7} (a,b), \ion{O}{8} (c,d) and \ion{Ne}{9} (e,f) images overlaid with the [\ion{O}{3}] (right) and 8\,GHz radio continuum (left) as white contours. 

According to Figure \ref{fig:linemaps}, the overall distribution of photons in each X-ray band follows similar trends to the deconvolved broadband (0.3-8 keV) maps described in Paper II: the X-rays generally follow the [\ion{O}{3}] maps, and avoid the bubbles corresponding to VLA 8.46 GHz emission.  We identify noticeable differences between the species, however:  
\smallskip
\newline$\bullet$ \ion{O}{7} appears to best trace the low-intensity extended [\ion{O}{3}] contours while \ion{Ne}{9} is associated with narrow, high-intensity [\ion{O}{3}] arcs that wrap around the radio outflows.
\newline$\bullet$ \ion{O}{8} does not have an obvious S-shape.  Rather, \ion{O}{8} appears linear, narrowly confined to a $D\sim1\arcsec$ cylinder oriented NE-SW along the bicone.  There may be a loop to the NE, but there are too few photons to test its putative extent.
\newline$\bullet$ Within $\sim1\arcsec$ of the nucleus, \ion{O}{7} forms a bright E-W bar that is associated with a similar bar in [\ion{O}{3}]; this bar is not seen in  \ion{O}{8} and \ion{Ne}{9}.
\newline$\bullet$ The brightest \ion{Ne}{9}-emitting regions are in closer contact with the radio outflows, whereas \ion{O}{7} is more extended (Fig. \ref{fig:linemaps},f).

The narrow, linear shape of \ion{O}{8} appears simple, so that more rigorous investigation of faint sub-structure is not warranted for this dataset.  But the identification of more complex structural differences between \ion{O}{7} and \ion{Ne}{9} on sub-pixel scales may be unreliable due to the low surface brightness of the S-shaped arms in X-rays (within $r<3\arcsec$ from the nucleus, we find mean surface brightness $\bar{I}\sim0.73\,\rm{count\,arcsec}^{-2}$ in \ion{O}{7}).  We therefore tested for the presence of these line emission features via adaptive smoothing, hardness ratio maps, and radial profile binning.

\subsubsection{Adaptive Smoothing}\label{sec:XELMapSmooth}

\begin{figure}[ht!]
\noindent
\centering
\includegraphics[width=0.49\textwidth]{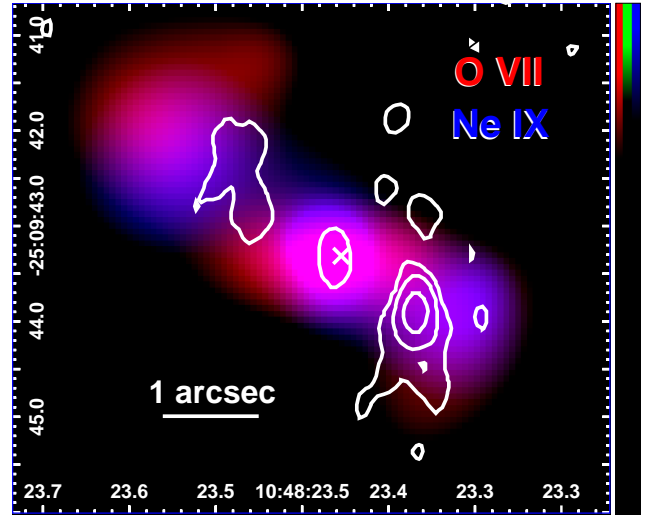}\par
\caption{Adaptively smoothed emission map of \ion{Ne}{9} ({\it blue}) and \ion{O}{7} ({\it red}) with VLA  8.46 GHz contours overlaid.  High-significance emission features of these species are separated by $\sim0.5\arcsec-1\arcsec$ in the S-shaped arms which precede the radio outflows.  The E-W nuclear bar of \ion{O}{7}  identified in Fig. \ref{fig:linemaps} remains prominent here.
} 
\label{fig:ne9o7csm}
\end{figure}

In order to test the veracity of differing line emission morphologies identified in \S\ref{sec:XELMap}, we first created adaptively smoothed images of \ion{Ne}{9} and \ion{O}{7} in the inner $r<10\arcsec$ of NGC 3393.  We used the {\tt csmooth} tool from {\it ciao} to create images at 1/8 native ACIS pixel scale, with minimum feature significance of $2\sigma$ and maximum significance of $5\sigma$.  The 1/8 pixel scale map is displayed in Fig. \ref{fig:ne9o7csm}.  

\begin{figure*}[t]
\noindent
\centering
\includegraphics[width=0.49\textwidth]{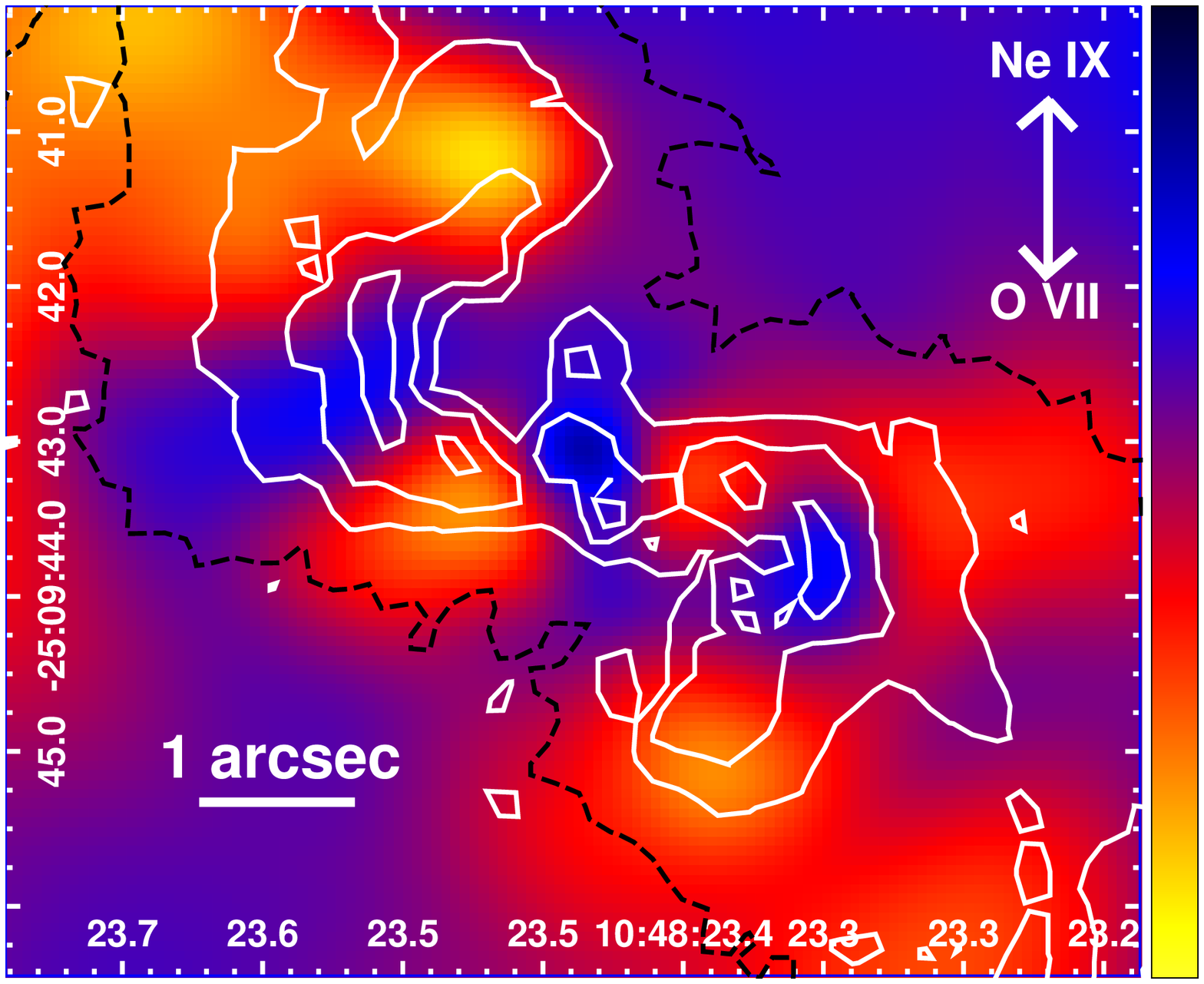}%
\includegraphics[width=0.49\textwidth]{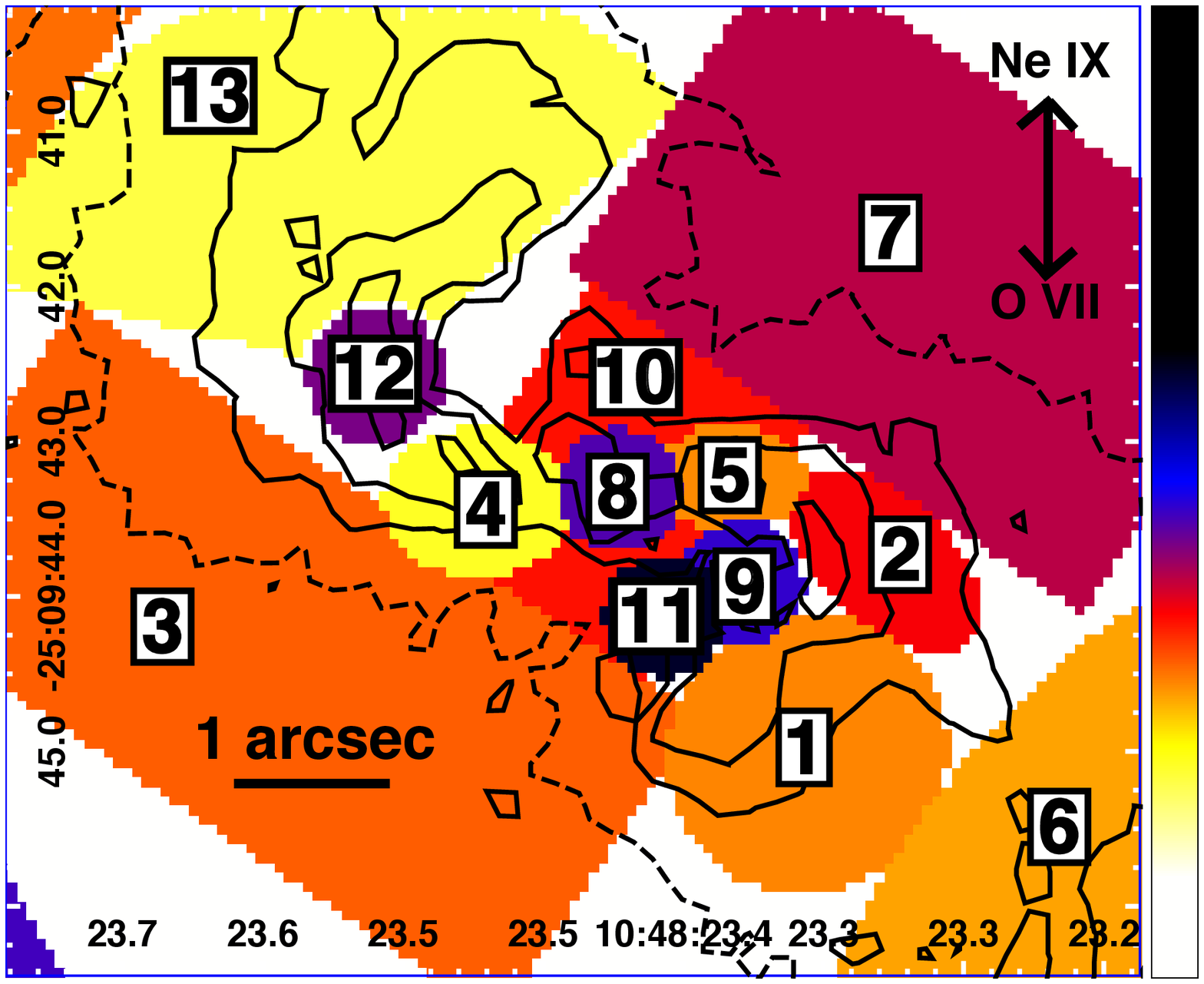}\par
\caption{\newline{\bf Left:} Hardness ratio map of \ion{Ne}{9}/\ion{O}{7} from adaptively smoothed images, with {\it HST} [\ion{O}{3}] contours overlaid ({\bf white}).  {\bf Blue} indicates stronger \ion{Ne}{9}, {\bf red and yellow} indicate stronger  \ion{O}{7}.  X-ray emission outside the  {\bf black dashed contour} represents a low level of [\ion{O}{3}], outside which these X-ray hardness ratios will also have low significance.  Region geometries are simple and may overlap (e.g. reg. 10 is a square and contains all of reg. 8).  Areas of \ion{Ne}{9} excess described in the middle panel are clearly associated with the strongest [\ion{O}{3}] emission in the S-shaped arms.  Red-yellow values associated with the E-W \ion{O}{7} bar overlap a similar E-W [\ion{O}{3}] bar.  Hardness ratios outside the bicone associated with weak [\ion{O}{3}] (NW and SE corners) are also of low significance and can be ignored.
\newline{\bf Right:}  Map of hardness ratios between the \ion{Ne}{9} and \ion{O}{7} bands (as above), but here photons are binned according to defined regions defined by human identification of broadband and single-species features ({\it white} regions are not calculated).  Contours are from smoothed {\it HST} maps of [\ion{O}{3}].  Typical hardness ratio uncertainties range between 0.10 and 0.15.  These binned calculations confirm that \ion{Ne}{9} excesses identified in Fig. \ref{fig:ne9o7csm} are statistically significant.}
\label{fig:ne9o7regs}
\end{figure*}

The enhanced \ion{Ne}{9} is clearly cospatial with the strongest emission in the [\ion{O}{3}] arcs (not shown, but see \S\ref{sec:XELMapHR} and Fig. \ref{fig:ne9o7regs}) leading the radio knots, whereas \ion{O}{7} traces the central [\ion{O}{3}] bar and the weaker [\ion{O}{3}] clouds at larger radii (as shown in Fig. \ref{fig:linemaps}b.  There is some overlap between the strongest \ion{Ne}{9} and \ion{O}{7} in the SW bicone, but in the NE peak emission from the the two lines is physically separate at $d\simgreat0.5\arcsec$.

\begin{table}[h!]
\small
\centering
\caption{Local \ion{Ne}{9}-\ion{O}{7} Hardness Ratios}
\label{table:ne9o7hrs}
\vspace{0.1in}
\begin{tabular}{cc}
\tableline
ID		&	HR			 \\
\tableline
1	&	 $-0.39\,\substack{+0.10 \\ -0.12}$ \\ 
2	&	 $-0.24\,\substack{+0.15 \\ -0.15}$\\ 
3	&	$-0.35\,\substack{+0.11 \\ -0.14}$\\ 
4	&	$-0.56\,\substack{+0.09 \\ -0.14}$\\ 
5	&	 $-0.40\,\substack{+0.14 \\ -0.16}$\\ 
6	&	$-0.43\,\substack{+0.09 \\ -0.11}$\\ 
7	&	$-0.19\,\substack{+0.13 \\ -0.16}$\\ 
8	&	$-0.07\,\substack{+0.12 \\ -0.13}$\\ 
9	&	$-0.03\,\substack{+0.17 \\ -0.19}$\\ 
10	&	$-0.27\,\substack{+0.07 \\ -0.07}$\\ 
11	&	$+0.25\,\substack{+0.29 \\ -0.26}$\\ 
12	&	$-0.11\,\substack{+0.14 \\ -0.15}$\\ 
13	&	$-0.60\,\substack{+0.06 \\ -0.06}$\\ 
\tableline 


\tableline

\end{tabular}
\\
\vspace{0.1in}
\begin{minipage}{0.48\textwidth}
These are the mean {\tt BEHR}-determined Bayesian values for (\ion{Ne}{9}-\ion{O}{7})/(\ion{Ne}{9}+\ion{O}{7}). ID values indicate the corresponding region numbers marked in Figure \ref{fig:ne9o7regs}.
\end{minipage}
\end{table}

\subsubsection{Hardness Ratio Mapping}

\label{sec:XELMapHR}

In order to investigate the relative importance of \ion{Ne}{9} and \ion{O}{7} features identified in \S\ref{sec:XELMapSmooth}, we
 created a hardness ratio map of the adaptively smoothed images in Fig. \ref{fig:ne9o7regs}, such that $HR=(H-S)/(H+S)$, where the hard band $H$ is \ion{Ne}{9} and soft band $S$ is \ion{O}{7}.  [\ion{O}{3}] contours are overlaid for comparison.  This hardness ratio map shows the same patterns described in Fig.  \ref{fig:ne9o7csm} and \S\ref{sec:XELMapSmooth}.

As a second test of the significance of the \ion{Ne}{9}  and \ion{O}{7} features identified in \S\ref{sec:XELMap}, we defined several spatial bins based upon apparent peaks and dips in the \ion{Ne}{9}/\ion{O}{7} map.  We then calculated hardness ratios in each of these bins using Bayesian Estimation of Hardness Ratios \citep[BEHR;][]{behr}, which is well-suited to quantifying hardness ratio uncertainties in the regime of Poisson statistics.  The hardness ratios of each region are depicted in Fig. \ref{fig:ne9o7regs} (right), and listed in Table \ref{table:ne9o7hrs}.  The  \ion{Ne}{9}/\ion{O}{7} structural patterns display typical differences of $\lesssim2\sigma$ in \ion{Ne}{9}-\ion{O}{7} hardness ratio from the region with weakest \ion{Ne}{9}/\ion{O}{7} (region \#13), but regions \#8, \#9, \#10, \#11 and \#12 all differ by $\simgreat3\sigma$ from region \#13.  Region \#13 is within the bicone but not closely adjacent to the brightest radio knots, whereas all high-\ion{Ne}{9}/\ion{O}{7} regions are directly adjacent to bright radio knots.  Regions \#4 and \#5 cover the E-W bar of strong \ion{O}{7}.

\subsubsection{Soft X-ray Emission and [\ion{O}{3}]}

\label{sec:XELO3}

In order to investigate the commonly-observed relationship between soft (0.3-2\,keV) X-ray emission and [\ion{O}{3}], we generated radial surface brightness profiles in each band with $0.25\arcsec$ annular bins (switching to $1\arcsec$ bins at $r\sim5\arcsec$ due to the presence of low surface brightness regions).  We used the {\it CIAO} tool {\tt dmextract} to generate the profiles.  We used the  WFC3 header parameters PHOTFLAM and PHOTBW to calculate [\ion{O}{3}] fluxes\footnote{http://www.stsci.edu/hst/wfc3/documents/handbooks /currentDHB/Chapter2\_data\_structure5.html}, converting RMS bandwidth to FWHM bandwidth.  For the soft X-ray flux we assumed the median in-band (0.3-2\,keV) photon energy at $r<3\arcsec$ ($E=0.911\,\rm{keV}$).  

\begin{figure}[t] 
\noindent
\centering
\includegraphics[width=0.49\textwidth]{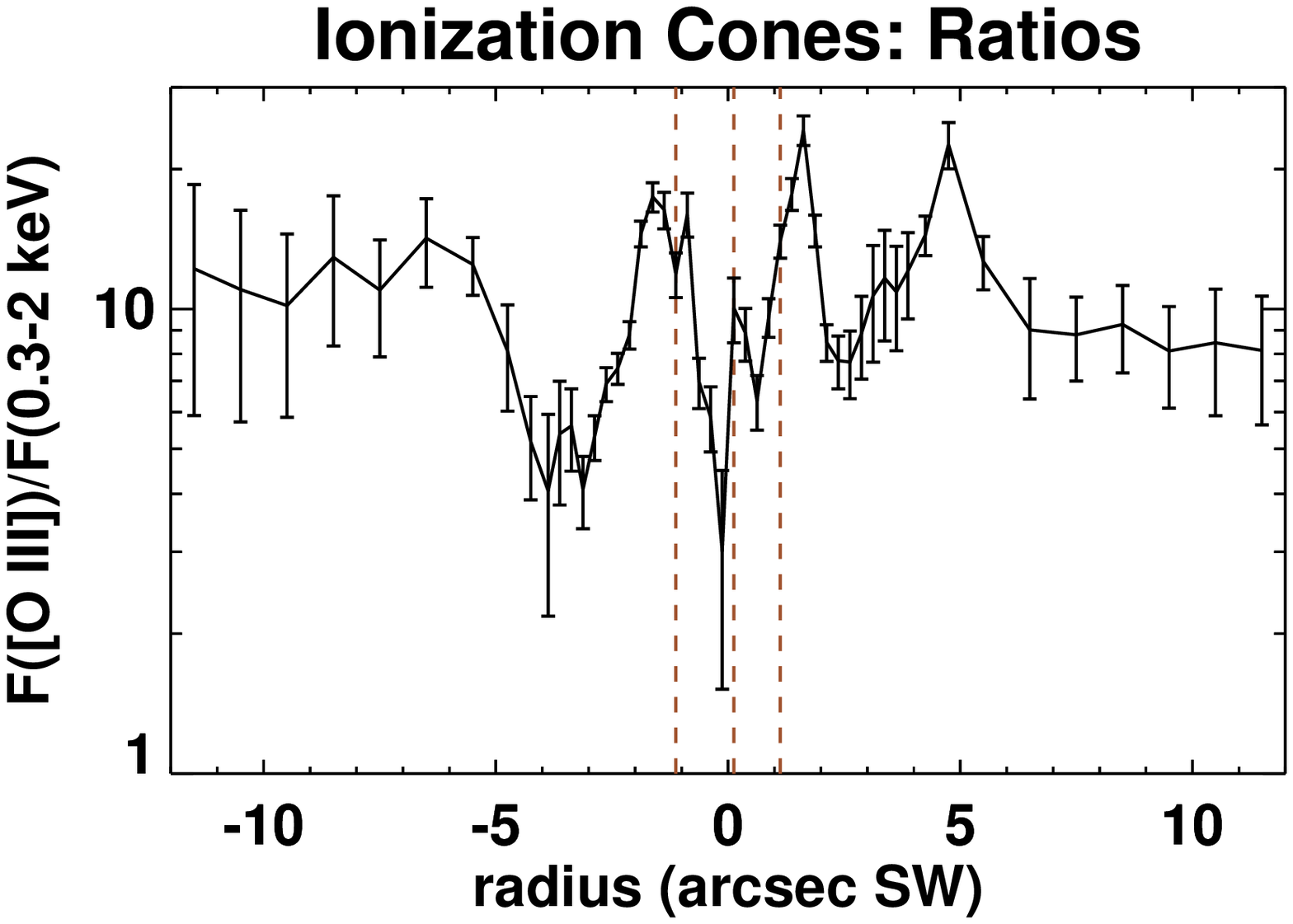}\par
\caption{Radial profile of the ratio between [O\,III]$\lambda5007$\AA\ emission and 0.3-2 keV X-ray flux.  Brown vertical dashed lines indicate peaks of radio emission.  We identify strong peaks immediately exterior to the radio peaks, as well as at $\sim5\arcsec$ SW.  The normalization is comparable values from Mrk 573 \citep{Paggi12} and NGC 4151 \citep{Wang11b} when those are corrected for bandwidth.  Like \cite{Paggi12} and NGC 4151 \cite{Wang11b}, we find spikes associated with high-density features like the leading edges of radio knots.
} 
\label{fig:sxo3rat}
\end{figure}

We show the [\ion{O}{3}]/(0.3-2\,keV) ratio profile in Figure \ref{fig:sxo3rat}, and indicate peaks of the radio emission with brown vertical lines.  We observe that the ratio is flat at $r\simgreat6\arcsec$, but spikes sharply just outside the radio lobes ($r\sim1.5\arcsec$), and at $r\sim5\arcsec$ in the SW cone (but not the NE).  We see [\ion{O}{3}]/(0.3-2\,keV) ratio troughs at the nucleus and immediately outside the  [\ion{O}{3}]/(0.3-2\,keV) spikes which are associated with the radio lobes (at $r\sim3\arcsec-4\arcsec$ (NE) and $r\sim2.5\arcsec$ (SW).

\begin{figure}[th!] 
\noindent
\centering
\begin{minipage}{0.49\textwidth}
\hspace{0.15\textwidth}\includegraphics[width=0.8\textwidth]{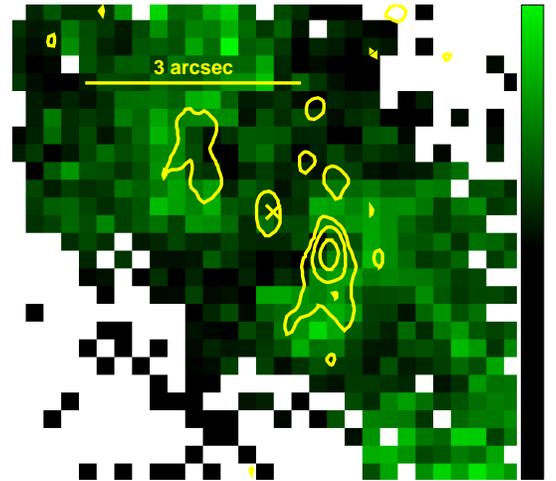}
\end{minipage}\par
\caption{Ratio map of [O\,III]$\lambda5007$\AA\ emission to (0.3-2\,keV) X-ray flux. \hst\ resolution is degraded with a Gaussian kernel to match a \cha\ photon image binned at 0.5-pixel (0.246\arcsec) scale.  Green indicates a high ratio.  White pixels have zero X-ray photons and are hence undefined.   Cyan contours are radio.
} 
\label{fig:sxo3map}
\end{figure}

In order to investigate the [\ion{O}{3}]/(0.3-2\,keV) ratio independently of azimuthal effects, we have also generated a ratio map  shown in Figure \ref{fig:sxo3map}.  We degraded the \hst\ images with a $r=3-\rm{pixel}$ Gaussian kernel.  We then re-sampled the \hst\ image to match 0.5-pixel (0.246\arcsec) scale 0.3-2 keV \cha\ images using {\it CIAO} {\tt reproject\_image}.
Finally, we used {\tt dmimgcalc} to calculate the ratio map.  We find that [\ion{O}{3}]/(0.3-2\,keV) trends are similar to the radial profile, but [\ion{O}{3}]/(0.3-2\,keV) is also elevated counter-clockwise to the radio emission.  For $r\simless3\arcsec$, X-rays are suppressed in the cross-cone and interior to the radio lobes.

\subsubsection{X-ray Emission Line Profiles}

\label{sec:XELProfiles}

\begin{figure}[h!] 
\vspace{0.1in}
\noindent
\centering
\hfill\includegraphics[width=0.4\textwidth]{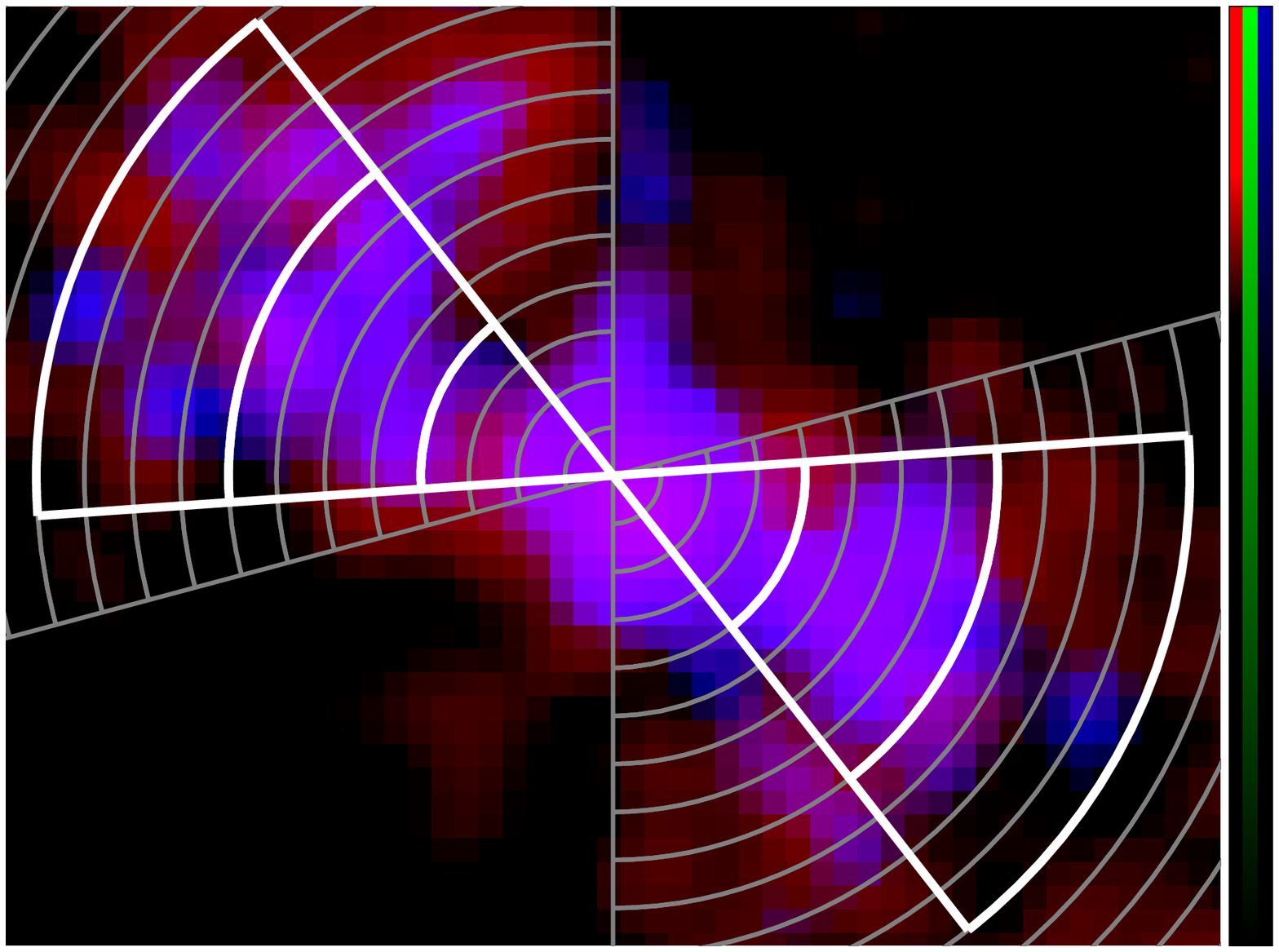}\hfill\\
\includegraphics[width=0.47\textwidth,trim={0.5cm 0 0 0}, clip]{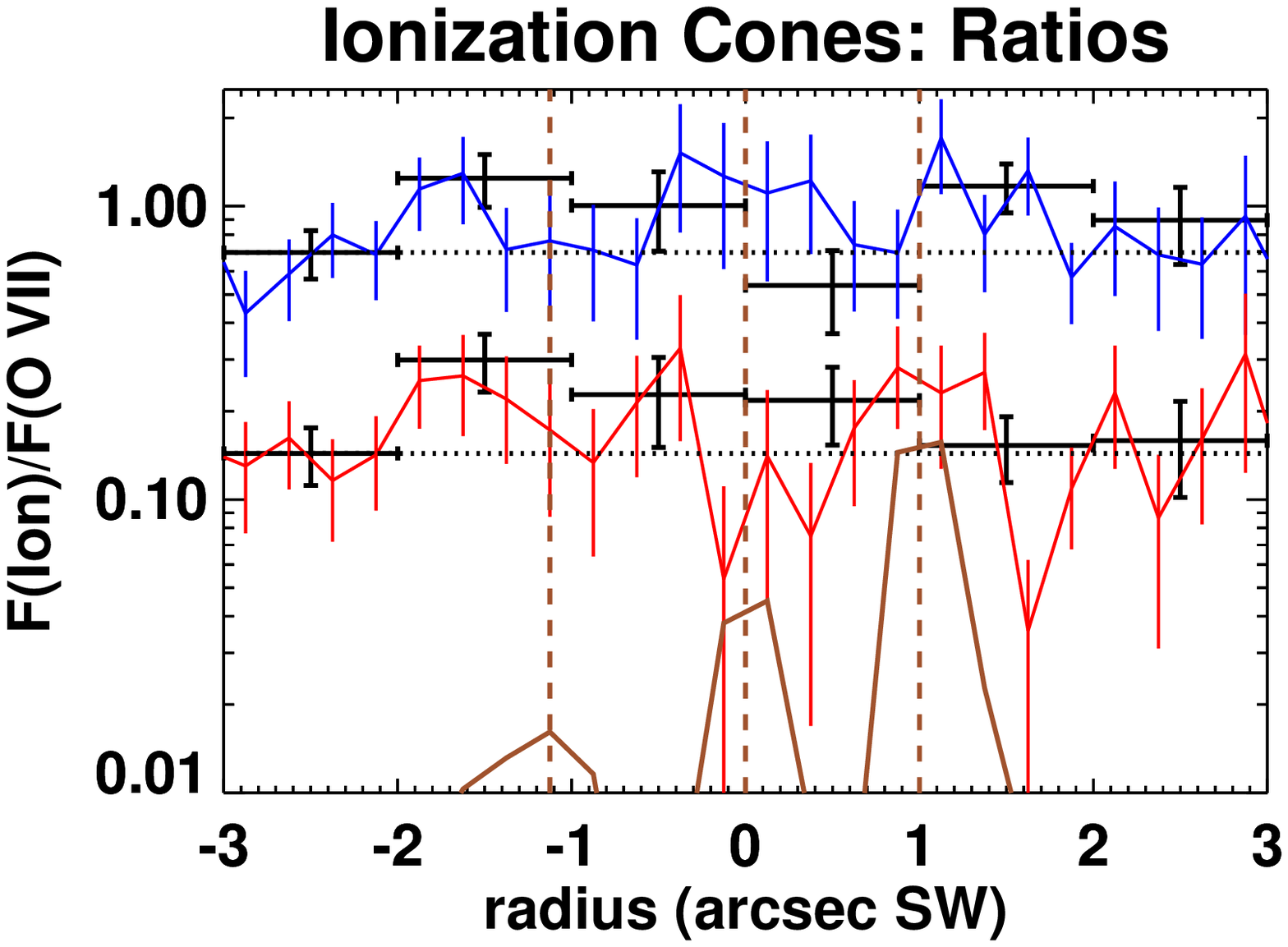}\par
\vspace{0.1in}\caption{\newline\newline{\bf Top:} Emission line image of the inner $r<3\arcsec$ of NGC 3393, binned to 0.25\arcsec\ with $r=2$-pixel gaussian smoothing.  \ion{Ne}{9} is {\bf blue} and \ion{O}{7} is red.  Annular arc contours represent different radial bins, extracted and plotted in the {\bf bottom} panel.  {\bf White} contours have $\Delta r=1\arcsec$ spacing and a narrower angle, {\bf grey} contours are broader, with $\Delta r=0.25\arcsec$. 
\smallskip
\newline {\bf Bottom:} Radial profiles of line ratios for the regions depicted above.  For the $\Delta r=0.25\arcsec$ profile, \ion{Ne}{9}/\ion{O}{7} is {\bf blue} and \ion{O}{8}/\ion{O}{7} is red (with error bars in the same color).  {\bf Black} error bars are overlaid to show uncertainties for the $\Delta r=1\arcsec$ profile.  {\bf Brown} lines show radio emission for the same region, renormalized for ease of comparison, and the peaks of the three main radio blobs are indicated with {\bf vertical dashed lines}.  {\bf Horizontal dashed lines} indicate line ratios at $-3<r<-2$ for ease of comparison.
}
\label{fig:radprof}
\end{figure}

In order to check the observed trends in the emission line ratios  which were described in \S\ref{sec:XELMapSmooth} and  
\S\ref{sec:XELMapHR} (particularly areas of elevated \ion{Ne}{9}/\ion{O}{8}), we also binned emission line maps of \ion{Ne}{9}, \ion{O}{8}, \ion{O}{7} and 8.46 GHz emission radially in conical profiles.  These profiles cover the bicones in both the NE and SW with $105\degr$ azimuthal extent.  We plotted the resulting radial emission profiles in Fig. \ref{fig:radprof}.  We use $\Delta r=0.25\arcsec$ bins in order to approach mirror resolution, which is necessary for direct comparison to \hst\ images.  Consistent with \S\ref{sec:XELMapSmooth}, we observe areas of multiple contiguous bins of increased \ion{Ne}{9}/\ion{O}{7} or  \ion{O}{8}/\ion{O}{7}.  However, the significance is only $\sim1\sigma$ per bin.  
 
\begin{figure}[h!]
\noindent
\centering
\includegraphics[width=0.47\textwidth]{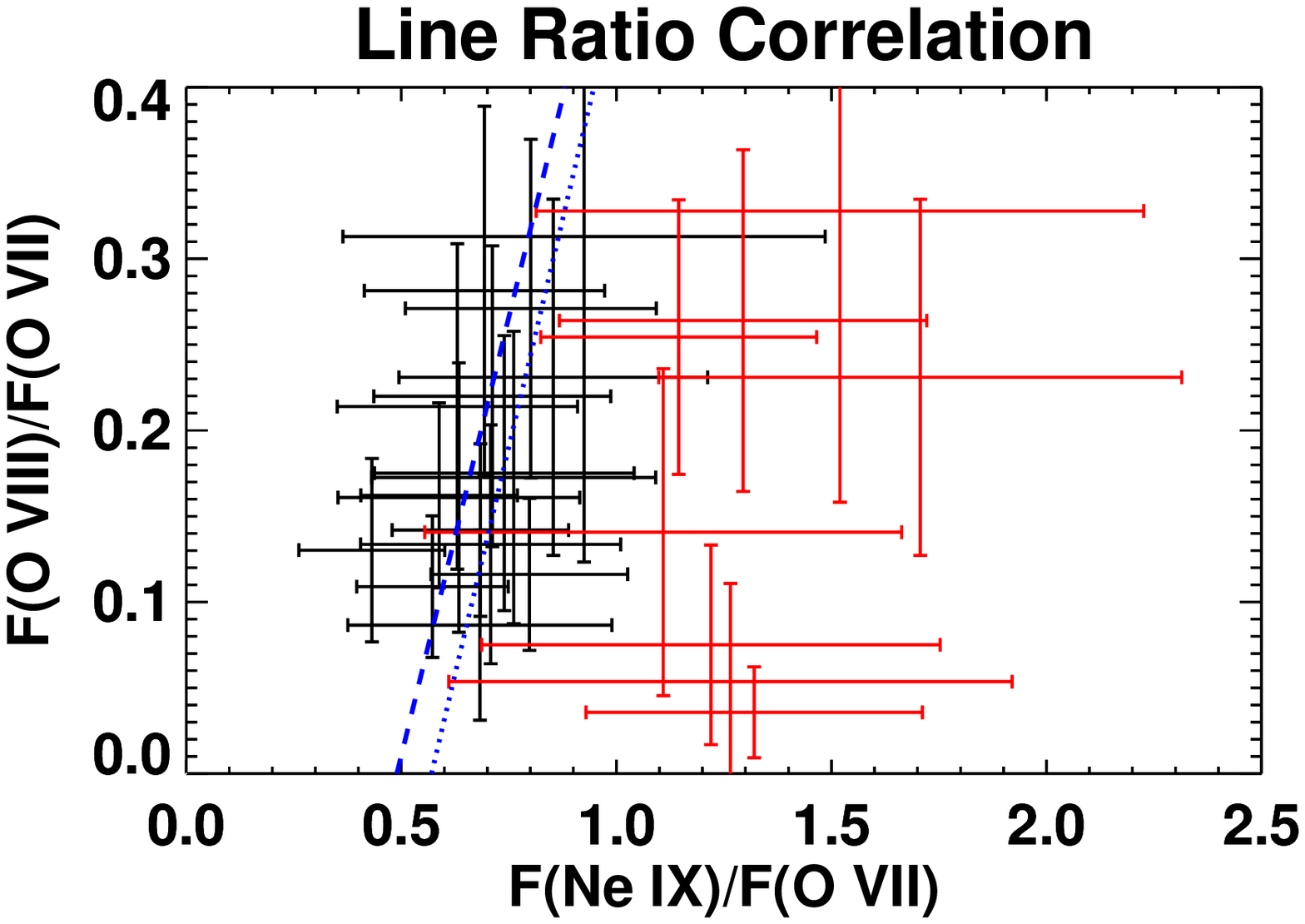}\par
\includegraphics[width=0.47\textwidth]{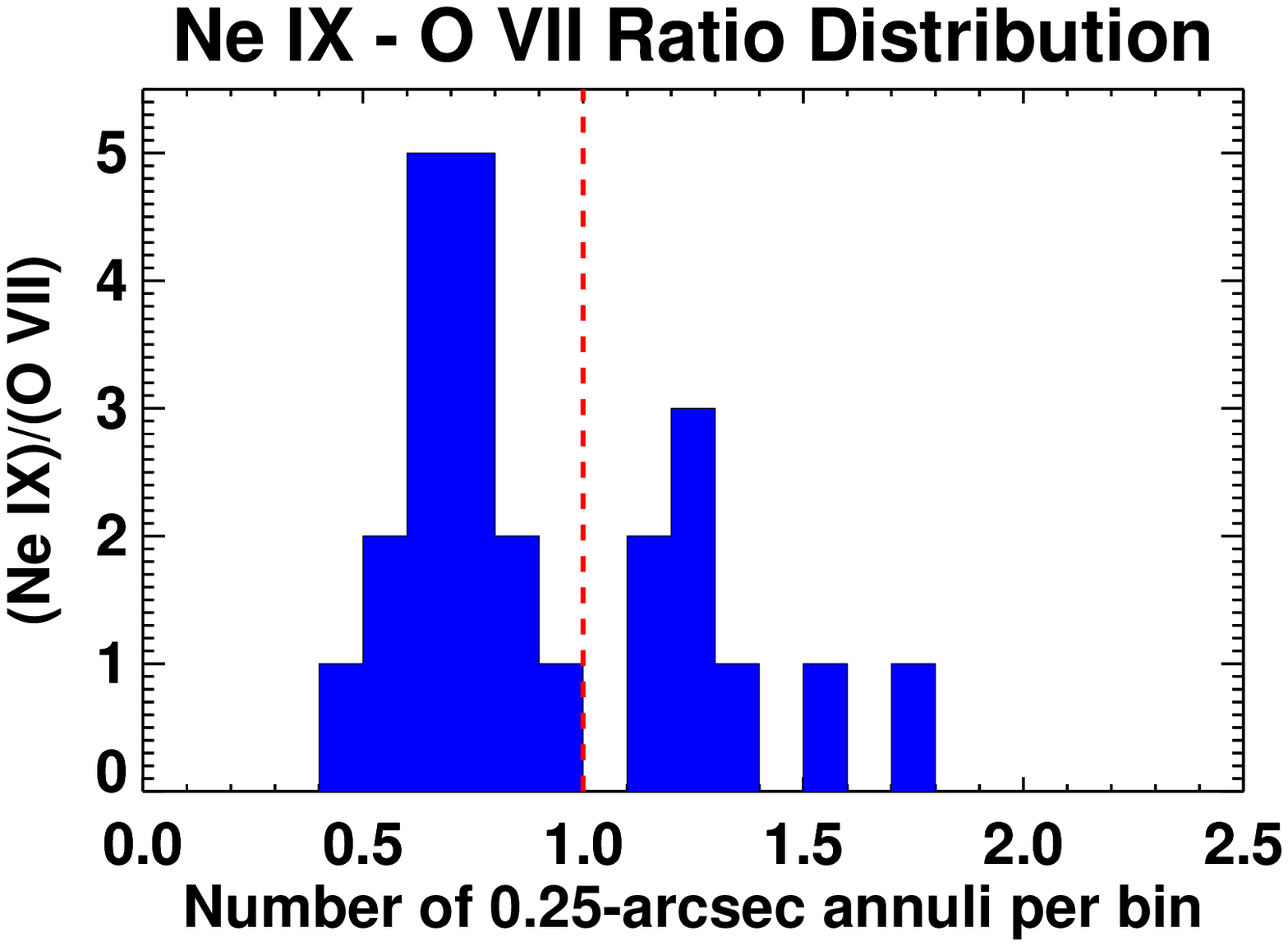}

\caption{
\newline {\bf Top:} \ion{O}{8}/\ion{O}{7} from Fig. \ref{fig:radprof} ($\Delta r=0.25\arcsec$ bins) plotted vs. \ion{Ne}{9}/\ion{O}{7} for correlation analysis.  The best linear fit for the full dataset ({\bf blue dotted line}) is dominated by low ($>1.0$, {\bf black}) \ion{Ne}{9}/\ion{O}{7} values, as demonstrated by the small offset from the best fit  ({\bf blue dashed line}) that excludes high ($\ge1.0$, {\bf red}) outliers. 
\newline {\bf Bottom:}  Histogram of \ion{Ne}{9}/\ion{O}{7} values from above.  The {\bf red dashed vertical line} indicates  \ion{Ne}{9}/\ion{O}{7}=1.0, which marks the separation between  {\bf black and red values in the top plot}.
}
\label{fig:ne9o7-2pop}
\end{figure}

In Fig. \ref{fig:ne9o7-2pop}, we use all three emission lines where both \ion{O}{7} and \ion{O}{8} are related to \ion{Ne}{9}.  Fig. \ref{fig:ne9o7-2pop} (top) shows that there is a range in both \ion{Ne}{9}/\ion{O}{7} and \ion{O}{8}/\ion{O}{7}, but no strong correlation (best-fit linear slope consistent with $\sim1$).  Several bins are outliers at the $1\sigma-2\sigma$ level, but the distribution of outliers is asymmetrical: all outliers have  \ion{Ne}{9}/\ion{O}{7}$\ge 1.0$, compared to typical values of 0.3-1.0 for the bins which are consistent with the best fit.  To illustrate how the outliers affect correlation, we find that the Spearman correlation coefficient is not significant for the full population ($\rho=0.20$ with $\Delta\sigma=-0.98$ from the null hypothesis) but is modestly significant for \ion{Ne}{9}/\ion{O}{7}$< 1.0$ ($\rho=0.56$ with $\Delta\sigma=-2.19$ from the null hypothesis).

We see in Fig. \ref{fig:ne9o7-2pop} (bottom) that a histogram of \ion{Ne}{9}/\ion{O}{7} shows a bi-modal population divided at \ion{Ne}{9}/\ion{O}{7}=1.0.  The \cite{Ashman94} D-statistic for the two populations is $D=4.9$, where $D>2$ is necessary for a clean separation.


To investigate the effects of coarser binning on the systematic trend in line ratios, we generated new line ratio radial profiles $\Delta r=1\arcsec$ bins and a narrower azimuthal extent that primarily covers an opening angle with bright \ion{Ne}{9}.  This better emphasizes regions of elevated \ion{Ne}{9}/\ion{O}{7}.  The SNR is improved (relative to the $r=2\arcsec-3\arcsec$ bins) for multiple regions of enhanced \ion{Ne}{9}/\ion{O}{7} and \ion{O}{8}/\ion{O}{7}, but only at the $\sim2.0-2.5$ level.  

\subsection{Imaging Spectroscopy}

\label{sec:XELImSpec}

\subsubsection{Spectral Extraction and Statistics}

In order to model the hot X-ray emitting plasma in the nucleus of NGC 3393, we extracted spectra from various regions as extended sources using the \ciao\ tool {\tt specextract}.  These spectral regions include the extended emission within $r=5\arcsec$ of the nucleus, as well as various sub-regions defined in \S\ref{sec:XELMapHR} to investigate \ion{Ne}{9}-\ion{O}{7} hardness ratios (see Fig. \ref{fig:ne9o7regs}).  

The regions defined in \S\ref{sec:XELMapHR} typically contain $\sim150-1000$ counts, such that Poisson statistics are relevant, particularly when attempting to take advantage of the ACIS energy resolution.  In this regime, the use of $\chi^2$ for fit minimization can bias model parameters, and the required level of binning for valid $\chi^2$ fitting degrades the energy resolution of the spectrum.  The \cite{cstat} statistic is preferred for Poisson data, but does not permit background subtraction.  Rather, it requires additional modeling of the background such as via a fit to the background region.  To avoid this complexity, we used the L-statistic or ``lstat" in {\tt Xspec} \citep{xspec}, which uses Bayesian analysis of Poisson data with Poisson background, and can be validly used with background subtraction.  We grouped the $r\le5\arcsec$  spectrum at 1 count per energy bin, and all other spectra at 5 counts per energy bin.  In our modeling, we use {\tt Xspec} measurements of $\chi^2$ to illustrate goodness-of-fit.

\subsubsection{Spectral Modeling}

\begin{table}[t]
\vspace{0.1in}
\footnotesize
\centering
\caption{Gaussian Composite Fits for $r<5\arcsec$}
\label{table:spec-gauss}
\tabcolsep=0.02cm
\vspace{0.1in}
\begin{tabular}{lccc}
\tableline
		&				& Fixed $\Gamma$ & Free $\Gamma$ \\
\tableline
Species	&	Rest			& Normalization & Normalization	\\
		&	Energy (keV)				& (1E-07)	& (1E-07)		\\
\tableline

\ion{C}{5} He$\gamma$	 & 0.371 & $540.82^{+141.64}_{-132.36}$ 	& $<63.85$	\\ 
\ion{N}{6} triplet			& 0.426 & ...							& ...	\\ 
\ion{C}{4} Ly$\beta$	 	& 0.436 & $424.92^{+80.36}_{-8.29}$ 		& $135.05^{+114.83}_{-116.56}$	\\ 
\ion{N}{7} Ly$\alpha$	& 0.500 & $290.51^{+48.87}_{-46.82}$		& $78.40^{+53.70}_{-53.50}$	\\ 
\ion{O}{7} triplet			& 0.569 & $732.31^{+53.64}_{-51.90}$		& $519.49^{+63.59}_{-62.75}$ \\ 
\ion{O}{8} Ly$\alpha$	& 0.654 & $216.90^{+25.87}_{-25.23}$		& $115.70^{+29.04}_{-28.54}$ 	\\ 
\ion{Fe}{17} 			& 0.720 & $215.03^{+20.41}_{-19.73}$		& $124.66^{+23.37}_{-22.90}$ \\ 
\ion{Fe}{17} 			& 0.826 & $187.08^{+19.82}_{-19.28}$		& $122.69^{+16.16}_{-15.88}$	\\ 
\ion{Fe}{18} 			& 0.873 & ...							& ...	\\ 
\ion{Ni}{19}			& 0.884 & ...							& ...	\\ 
\ion{Ne}{9} 			& 0.905 & $58.46^{+90.27}_{-58.46}$		& $150.99^{+13.09}_{-12.83}$	\\ 
\ion{Fe}{19}			& 0.917 & ...							& ...  \\ 
\ion{Fe}{19}			& 0.922 & ...							& ...	\\ 
\ion{Ne}{10}			& 1.022 & $95.08^{+7.51}_{-7.31}$			& $57.47^{+8.70}_{-8.50}$	\\ 
\ion{Fe}{24}			& 1.129 & $12.30^{+7.22}_{-7.06}$			& $1.64^{+7.67}_{-1.64}$\\ 
\ion{Fe}{24}			& 1.168 & $45.73^{+7.21}_{-6.99}$			& $26.30^{+6.49}_{-7.45}$	\\ 
\ion{Mg}{11}			& 1.331 & $38.72^{+4.13}_{-4.00}$			& $21.53^{+4.47}_{-4.33}$	\\ 
\ion{Mg}{11}			& 1.352 & ...							& ...	\\ 
\ion{Mg}{12}			& 1.478 & $12.94^{+2.89}_{-2.76}$			& $3.03^{+3.05}_{-2.93}$ \\ 
\ion{Mg}{12}			& 1.745 & $10.80^{+2.83}_{-2.69}$			& $6.28^{+2.85}_{-2.71}$	\\ 
\ion{Si}{13}			& 1.839 & $14.26^{+2.79}_{-9.24}$			& $12.31^{+2.85}_{-4.44}$	\\ 
\ion{Si}{13}			& 1.865 & $<10.81$						& $<4.80$		\\ 
\ion{Si}{14}			& 2.005 & $6.14^{+2.21}_{-2.06}$			& $4.43^{+2.23}_{-2.07}$	\\ 
\ion{Si}{13}			& 2.180 & ...							& ...	\\ 
\ion{S}{15}			& 2.430 & $9.88^{+2.91}_{-2.67}$			& $10.15^{+2.92}_{-2.67}$	\\ 
Fe K$\alpha$	 		& 6.442 & $29.85^{+7.99}_{-7.83}$ & $33.02^{+8.14}_{-7.40}$	\\ 
\tableline
\multicolumn{4}{c}{Basic Parameters}  \\
\tableline
$\Gamma$			& ... & $=1.8^\dagger$ &$=3.11^{+0.13}_{-0.13}$	\\ 
L-statistic				&	& 600.26		&  556.67		\\
$\chi^2$/DOF				& 	&	770.18/545	& 640.24/544		\\
					&	&	=1.41	& =1.18		\\
\tableline

\tableline

\end{tabular}
$\dagger$: For this fit, the power law index of the continuum has been fixed to $\Gamma=1.8$.
\end{table}

\begin{figure} 
\vspace{0.1in}
\noindent
\centering
\includegraphics[width=0.45\textwidth]{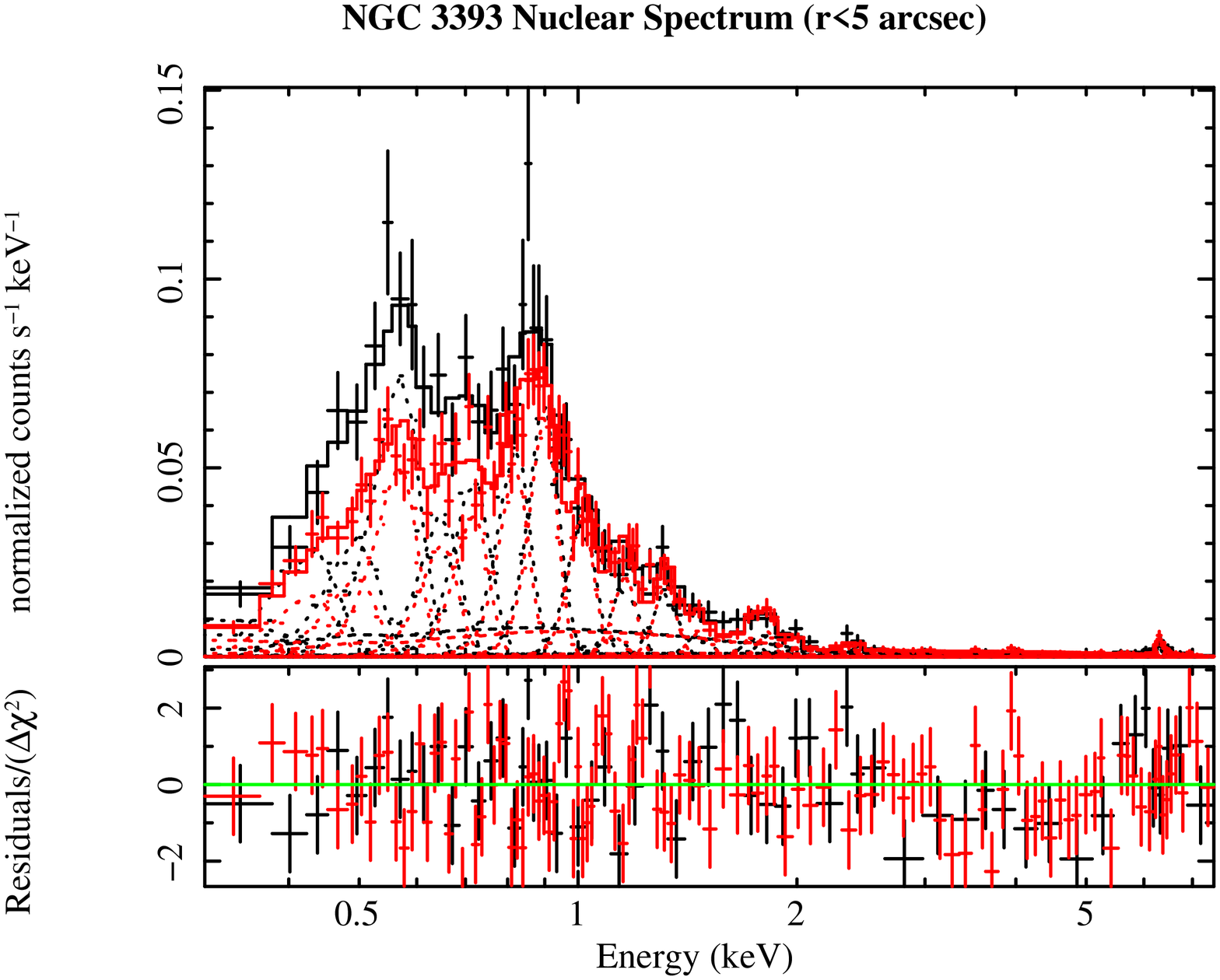}\par
\vspace{0.05in}
\includegraphics[width=0.45\textwidth,trim=0cm 0cm 0cm 1cm,clip]{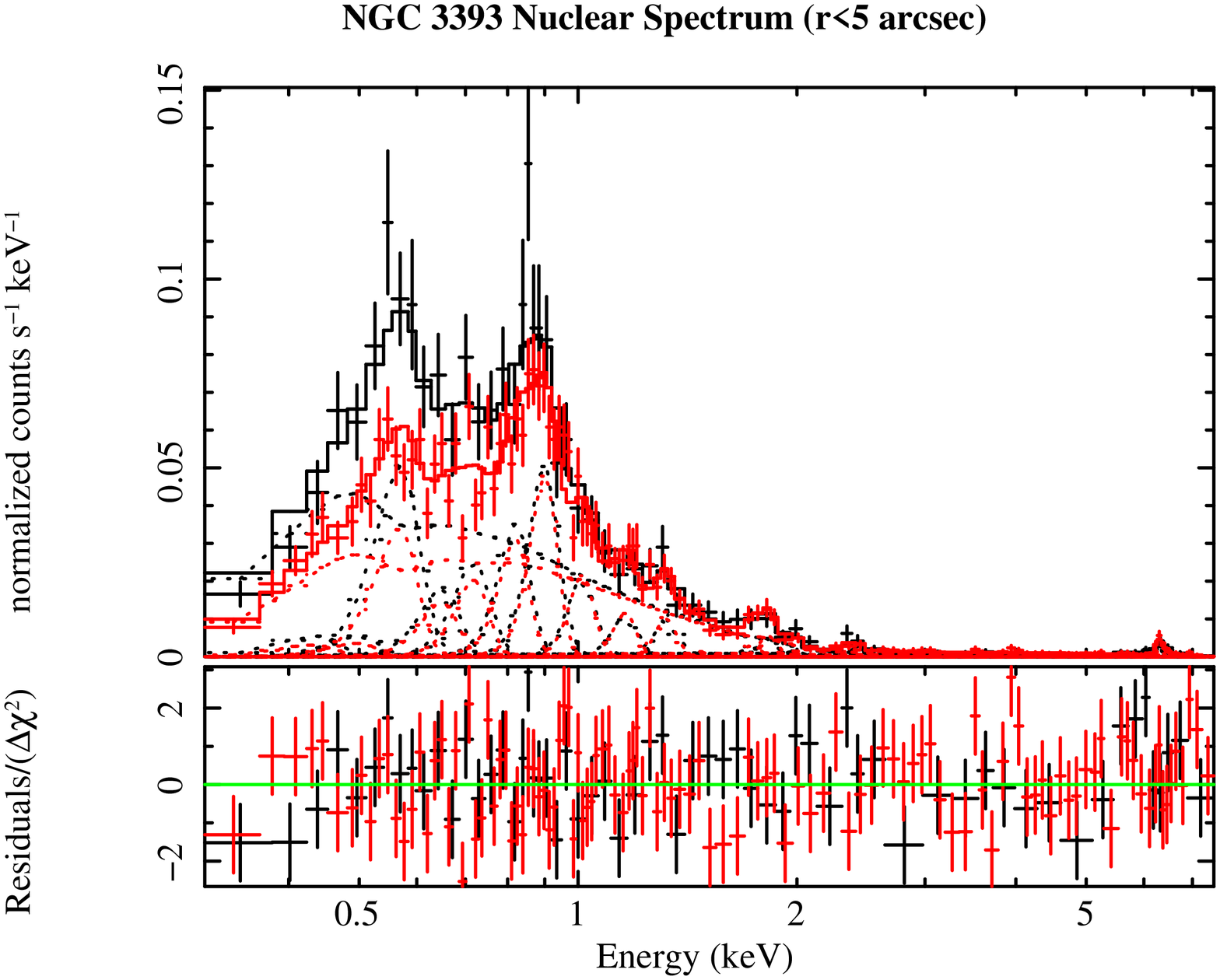}\par
\caption{Spectral fits to X-rays detected by {\it Chandra} within $r=5\arcsec$ from the nucleus of NGC 3393.  In addition to a spatially unresolved nuclear component (fit separately), each model has neutral host absorption component. The extended plasma is modeled a power law continuum plus multiple discrete emission lines described as gaussians with fixed transition energies, as described in Table \ref{table:spec-gauss}.  {\bf Red} and {\bf black} data points are from obsids 12290 and 4868 respectively.  Dashed lines indicate model components.  Residuals from models are depicted in the lower panels.
\newline{\bf Top:} The continuum power law index is fixed to $Gamma=1.8$.
\newline {\bf Bottom:} The power law index is fitted as a free component.
}
\label{fig:linespec}
\end{figure}

We expect harder emission near the nucleus to be at least partially contaminated by the wings of the point spread function from the AGN, so we modeled an AGN component for all spectral fits based on \cite{Koss15} fits of $NuSTAR$ data.  We extracted an AGN-dominated spectrum for $r\le0.4\arcsec$ from the nucleus, using the \ciao\ response correction for extended sources.  Having observed extended X-ray emission up to 4 keV \citep{Maksym17}, we only fit energies above 4 keV.  As per \cite{Koss15}, we included gaussian Fe K$\alpha$ emission ({\tt Xspec} $zgauss$ fixed at 6.4 keV), a warm absorber ($wabs$), absorbed and reflected power laws ($plcabs$ and $pexrav$ with $\Gamma=1.9$, $nH=3.1\times10^{24}\,\rm{cm}^{-2}$, $kT_{cutoff}=200$\,keV, $cos\theta_i=0.45$ and other parameters fixed to values in \citealt{Koss15}).  The final AGN component has only a single free normalization parameter, such that all sub-components are described in terms of that free normalization.  We measured component flux via the convolution model $cflux$ and determined parameter uncertainties via Monte Carlo Markov Chain (MCMC) methods in {\tt Xspec}.

\paragraph{The Composite ENLR}

For the NLR component of the extended $r\le5\arcsec$ region, we begin by assuming a purely phenomenological model of Gaussian emission lines on a power law continuum, comparable to \cite{Paggi12} for Mrk 573 and \cite{Koss15} for {\it Chandra} grating observations of NGC 3393 (which have superior energy resolution but worse counting statistics).  We first assume a typical AGN continuum with fixed $\Gamma=1.8$ for the power law index.  When $\Gamma$ is left free, we find $\Gamma=\vmp{3.1}{0.13}{0.13}$ and an improved fit ($\chi^2$/DOF=1.18 vs 1.41 for fixed $\Gamma$).  The results of these fits are shown in Table \ref{table:spec-gauss} and Figure \ref{fig:linespec}.  

We explore more physically motivated models for the ENLR in NGC 3393 by fitting various combinations of model components used in \cite{Paggi12} for Mrk 573.  These include a collisionally ionized component \citep[APEC][]{apec}, as well as a table model component constructed using CLOUDY \citep{cloudy} by \cite{Paggi12}.  This CLOUDY model describes photoionizing radiation reprocessed by diffuse gas under a grid of ionization parameters ($-2 \le log\ U \le 3$) and column densities ($19.0 \le log\ N_H \le 23.5$).  Single component fits are poor so we explore two-component models, the results of which are indicated in Table \ref{table:spec-mod}.  The best two-component fit includes both thermal APEC and photoionized CLOUDY models in which a collisional APEC component ($kT=\vmp{1.50}{0.11}{0.18}$) that is an order of magnitude weaker than the photoionized component. This collisional component peaks near $\sim1\,keV$ and is most significant relative to the CLOUDY model in a band containing complex emission by \ion{Ne}{9}, \ion{Ne}{10} and \ion{Fe}{24}.  When compared to the Gaussian fit, the slope of the continuum for the APEC+CLOUDY model is better represented by the model where $\Gamma$ is fit as a free parameter.

\begin{table}[t]
\small
\centering
\caption{Plasma Model Fits for $r<5\arcsec$}
\label{table:spec-mod}
\tabcolsep=0.02cm
\vspace{0.1in}
\begin{tabular}{llr}
\tableline
Component	&	Parameter			& Value 	\\
\tableline
\multicolumn{3}{c}{Single APEC and Single CLOUDY}\\
\tableline
zphabs	& $N_H$ ($\times10^{20}\,\rm{cm}^{-2}$)	& $\dagger$  \\
apec		& log $F_X^a$  (\ecmss)					& \vpm{-13.27}{0.09}{0.06}\\
		& $kT$ (keV)						& \vpm{1.50}{0.18}{0.11} \\
cloudy	&  log $F_X^a$  (\ecmss)				& \vpm{-12.42}{0.01}{0.02}\\
		& log U							& \vpm{0.55}{0.03}{0.03}\\
		& log $N_H$ ({cm}$^{-2}$)			& \vpm{21.32}{0.17}{0.15}\\
\tableline
Fit			& L-statistic				& 702.82		\\
			& $\chi^2$/DOF\tablenotemark{a}			& 	780.43/566=1.38		\\
\tableline
\multicolumn{3}{c}{Double APEC}\\
\tableline
zphabs	& $N_H$ ($\times10^{20}\,\rm{cm}^{-2}$)	& $\dagger$  \\
apec		& log $F_X^a$  (\ecmss)					& \vpm{-12.64}{0.03}{0.02}\\
		& $kT$ (keV)						& \vpm{0.13}{0.03}{0.02} \\
apec		& log $F_X^a$  (\ecmss)					& \vpm{-12.77}{0.03}{0.01}\\
		& $kT$ (keV)						& \vpm{0.84}{0.01}{0.11} \\
\tableline
Fit			& L-statistic				& 1065.92		\\
			& $\chi^2$/DOF\tablenotemark{a}			& 	821.03/567=1.45	\\
\tableline
\multicolumn{3}{c}{Double CLOUDY}\\
\tableline
zphabs	& $N_H$ ($\times10^{20}\,\rm{cm}^{-2}$)	& $\dagger$  \\
cloudy	&  log $F_X^a$  (\ecmss)				& \vpm{-12.79}{0.92}{1.01}\\
		& log U							& \vpm{-0.53}{0.03}{0.02}\\
		& log $N_H$ ({cm}$^{-2}$)			& \vpm{23.5}{0.71}{2.00}\\
cloudy	&  log $F_X^a$  (\ecmss)				& \vpm{-12.49}{0.69}{0.76}\\
		& log U							& \vpm{0.80}{0.06}{0.05}\\
		& log $N_H$ ({cm}$^{-2}$)			& \vpm{20.16}{1.69}{0.92}\\
\tableline
Fit			& L-statistic				& 653.95		\\
			& $\chi^2$/DOF\tablenotemark{a}			& 	927.42/574=1.64		\\
\tableline
\tableline

\end{tabular}
\\
$\dagger$: negligible ($N_H<10^{17}$); $a$: 0.3-8.0 keV; $b$: measured for the best lstat under {\tt Xspec}.
\end{table}

\begin{figure} 
\noindent
\centering
\includegraphics[width=0.45\textwidth]{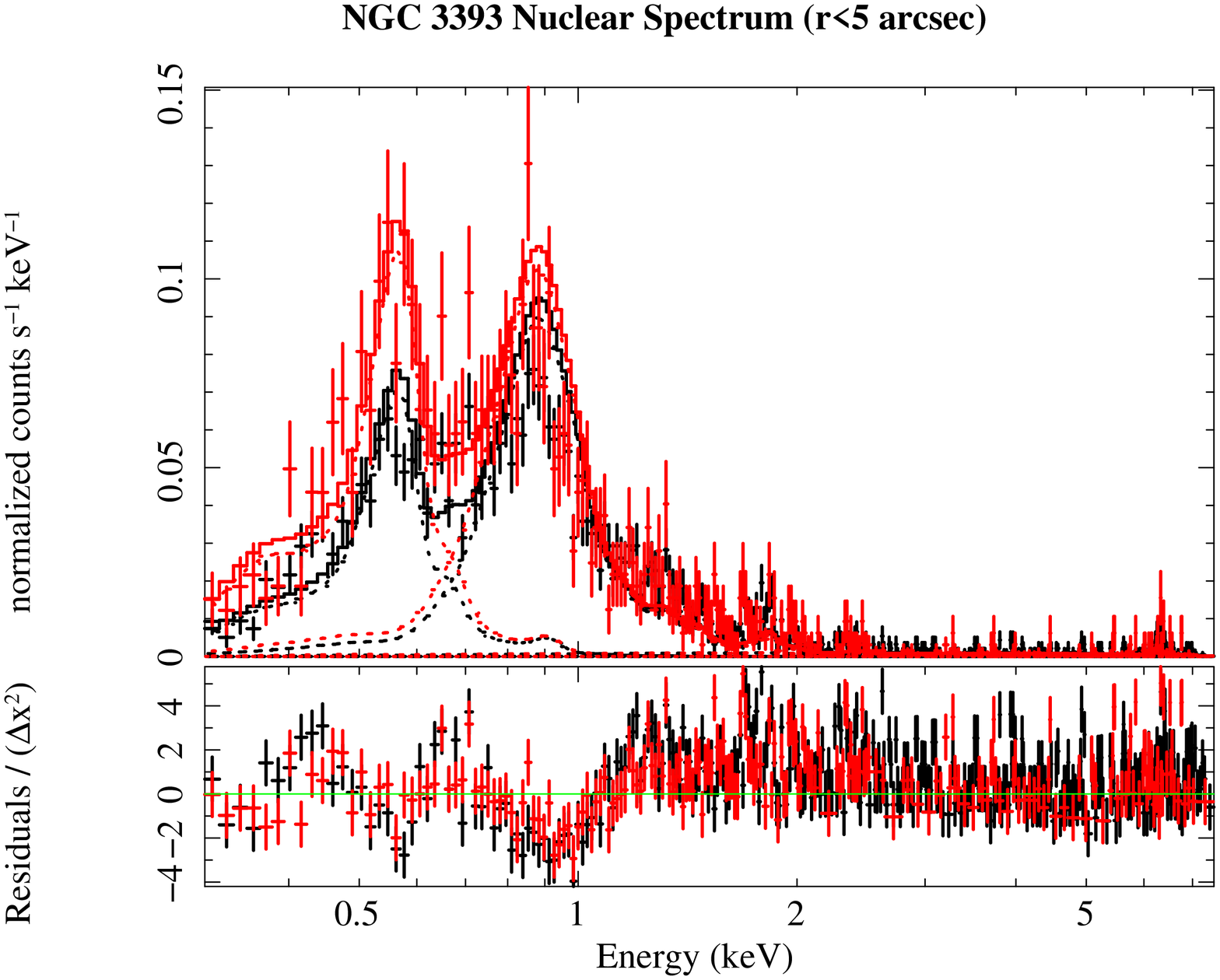}\par
\includegraphics[width=0.45\textwidth,trim=0cm 0cm 0cm 1cm,clip]{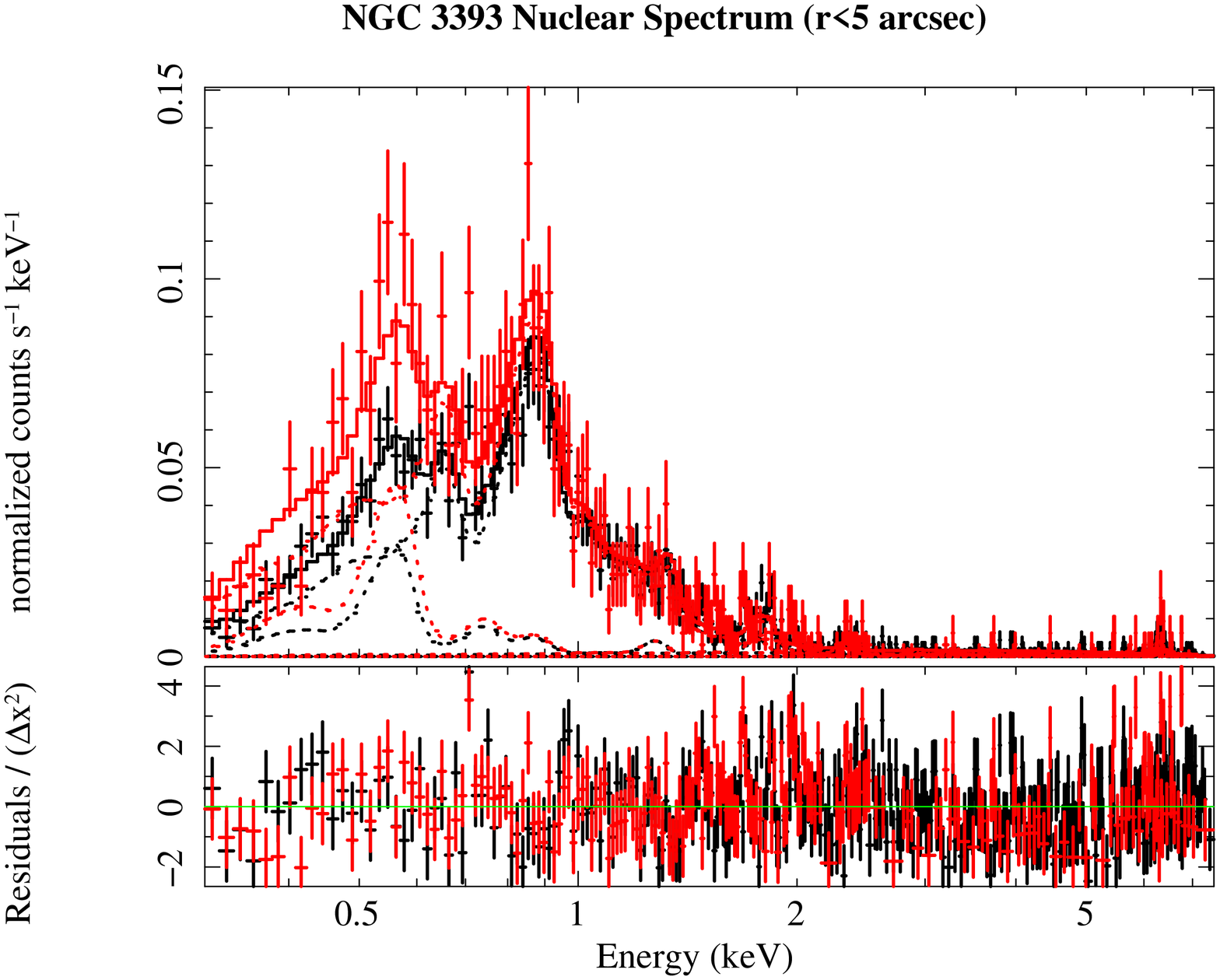}\par
\includegraphics[width=0.45\textwidth,trim=0cm 0cm 0cm 1cm,clip]{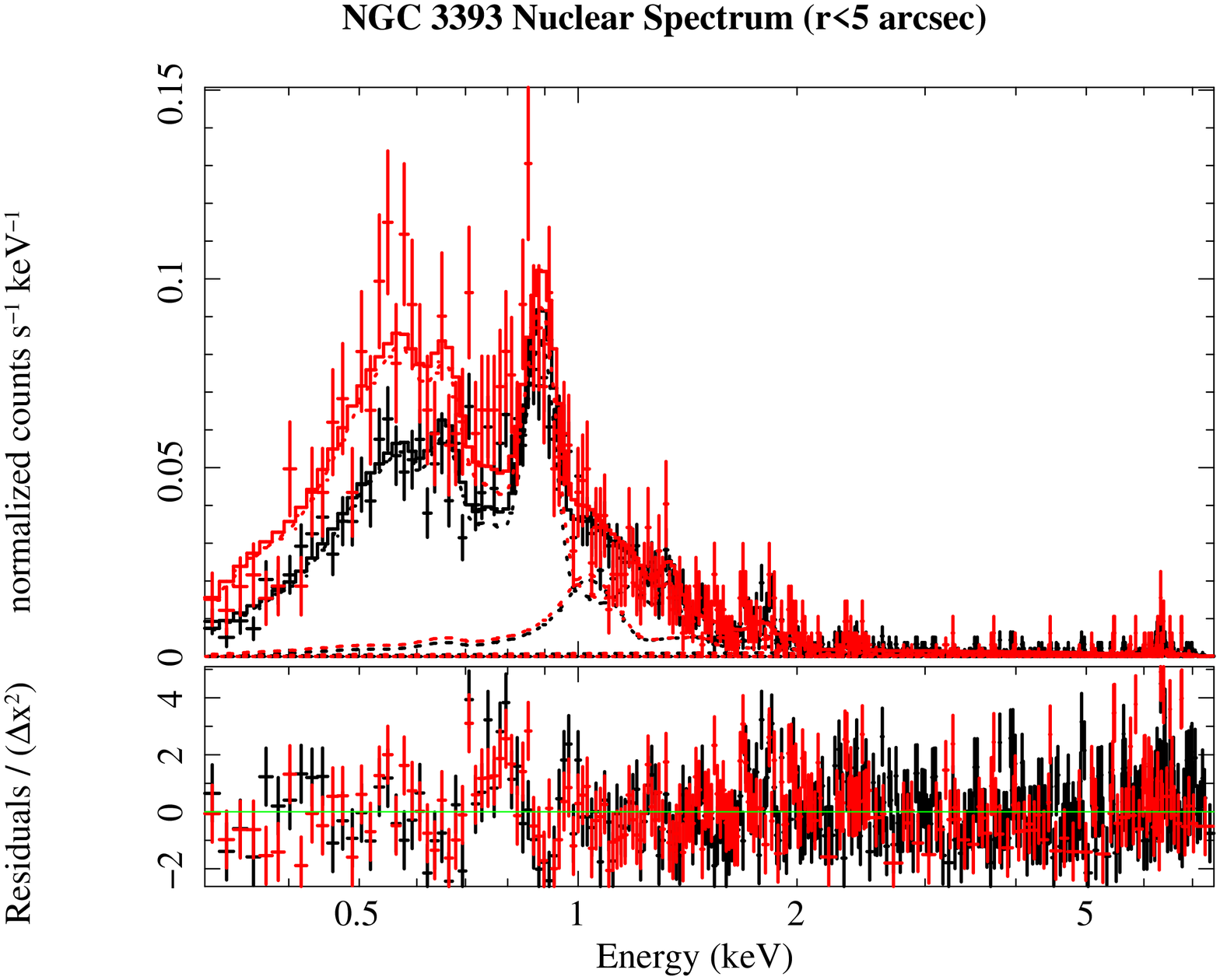}\par
\caption{As in Fig. \ref{fig:linespec}, but for the different combinations of APEC and CLOUDY plasma model components in Table \ref{table:spec-mod}.
\newline{\bf Top:} Two APEC thermal plasma components. \newline {\bf Middle:} Two CLOUDY components.  \newline {\bf Bottom:}  A combination of single APEC and CLOUDY components.  The APEC component peaks near 1\,keV.
}
\label{fig:modspec}
\end{figure}

\paragraph{Spatially Resolving ENLR Spectra}

Our previous analysis of spatial variation in spectral line emission suggests that physical conditions of the NGC 3393 ENLR vary significantly on scales of $\sim10$s of pc, in keeping with analysis by \cite{Maksym17}.  We therefore explore spatial variation of the 0.3-8\,keV spectra X-ray measured by \cha\ by fitting the sub-regions in \S\ref{sec:XELMapHR}.  The low numbers of counts limit the reliability of complex modeling on these spatial scales, so we explore single component APEC and CLOUDY models for the extended emission, as well as a combined APEC+CLOUDY model.  The results of these fits are shown in Table \ref{table:spec-reg} and Figure \ref{fig:regspec}.  

In general, single CLOUDY models are preferred to single APEC models, and the combined APEC+CLOUDY model is preferred in all cases except for region 7 (the NW cross-cone) where the measured APEC contribution is negligible.  An excess of photons near 6.4 keV raises the AGN component in region 11 (a region in the SW cone with low surface brightness in X-rays, optical and UV) such that residuals between 3 and 6 keV become uniformly and significantly negative, so we introduce an additional Gaussian component fixed at 6.4\,keV (host frame) for region 11.  Host column densities tend to be negligible, and fits with non-negligible column densities (such as the single APEC models of regions 3 and 6) tend to imply high component fluxes that are not consistent with the composite $r\le5\arcsec$ ENLR fit.

We caution that there is some degeneracy between different families of model solutions.  For example, in some cases the difference in fit quality may be small between APEC+CLOUDY models with large $log\,U$ and small $kT$, and models with small $log\,U$ and large $kT$.

\section{Results}\label{sec:results}

In this paper, we show that the X-ray line emission morphology reveals structural complexity beyond what is possible with broadband imaging.  This complexity goes beyond Paper II, where we already showed that the ENLR in NGC 3393 that although [\ion{O}{3}]  traces X-ray emission loosely
 (as expected from other studies), high-resolution deconvolved images show that the brightest regions do not perfectly correlate with 
regions of high [\ion{O}{3}]/H$\alpha$ or  [\ion{S}{2}]/H$\alpha$.  The morphologies of these line ratios occupy 
complementary spatial regions, implying multi-phase distribution of the gas.

We now gain a more detailed picture of the physical processes from the X-ray line emission via adaptive smoothing, 
spatially resolved 

\begin{longrotatetable}
\begin{deluxetable*}{llrrrrrr}
\tabletypesize{\scriptsize}
\tablecaption{Region Model Fits \label{table:spec-reg}}
\tablehead{
Component	&				&	&  &	&	&	&	}
\startdata
		&	Parameter		& Reg. 1	 & Reg. 2	& 	Reg. 3 & Reg. 4 & Reg. 5 & Reg. 6 \\
\hline
\multicolumn{8}{c}{\bf Single APEC  component}\\
\hline
zphabs	&  $N_H$ ($\times10^{20}\,\rm{cm}^{-2}$)	 & \vmp{74.44}{8.87}{4.73} & $\dagger$ 
	& $\dagger$ 				& $\dagger$ 				&$\dagger$ 		 		&$\dagger$ \\
apec		&	log $F_X$\tablenotemark{a}  (\ecmss)			&   \vmp{-11.69}{0.35}{0.14} 	& \vmp{-13.73}{0.02}{0.03} 
	&\vmp{-14.00}{0.04}{0.07} 	&\vmp{-14.37}{0.05}{0.27} 	&\vmp{-13.82}{0.04}{0.02} 	&\vmp{-13.16}{0.03}{0.02} \\
		&	$kT$	 (keV)				&  \vmp{0.147}{0.005}{0.027} 	& \vmp{0.975}{0.052}{0.023} 
	&\vmp{0.355}{0.038}{0.061} 	&\vmp{0.972}{0.221}{1.396} 	&\vmp{0.318}{0.022}{0.040}	&\vmp{0.257}{0.008}{0.009} \\
\hline
zgauss	& $\sigma$ (keV)			& N/A					& N/A
	&	N/A					& \vmp{0.098}{0.001}{0.111}	& N/A & N/A  \\
		& norm. ($\times10^{-6}$)		& N/A					& N/A
	&	N/A					& \vmp{1.15}{0.19}{0.68}		& N/A & N/A  \\
\hline
		&	L-Statistic 				&  89.51 					& 191.71
	& 86.75					& 38.60					& 84.21					& 389.08	\\
		&	$\chi^2/\rm{DOF}$\tablenotemark{b}		&  76.19/43=1.77 			& 168.99/94=1.80
	& 114.27/30=3.81			& 27.50/17=1.618			& 105.72/38=2.78			& 424.78/108=3.93	\\
\hline
\multicolumn{8}{c}{\bf Single CLOUDY table component}\\
\hline
zphabs	&  $N_H$ ($\times10^{20}\,\rm{cm}^{-2}$)	 & \vmp{2.72}{0.10}{4.12} & $\dagger$
	&$\dagger$  				&$\dagger$   				&$\dagger$				 &$\dagger$ \\
cloudy	&	log $F_X$\tablenotemark{a} (\ecmss)			&   \vmp{-13.57}{0.02}{0.08} 	& \vmp{-13.40}{0.10}{0.01} 
	&\vmp{-13.78}{0.05}{0.07} 	&\vmp{-13.69}{0.26}{0.18} 	&\vmp{-13.56}{0.03}{0.07} 	&\vmp{-13.02}{0.04}{0.01} \\
		&	log U					&   \vmp{0.86}{0.09}{0.10} 	& \vmp{1.41}{0.38}{0.04} 
	&\vmp{0.78}{0.11}{0.15} 		&\vmp{-1.98}{0.71}{0.20} 		&\vmp{-3.00}{0.03}{0.40} 		&\vmp{0.45}{0.01}{0.13} \\
		&	log $N_H$ ({cm}$^{-2}$)			&   \vmp{21.51}{0.47}{0.58} 	& \vmp{23.5}{1.72}{0.02} 
	&\vmp{20.88}{0.72}{0.86} 		&\vmp{19.02}{0.25}{2.10} 		&\vmp{20.28}{0.99}{0.54} 		&\vmp{20.35}{0.16}{0.23} \\
\hline
		&	L-Statistic 				&  45.29					& 99.52
	& 39.49					& 11.96					 & 68.83					& 164.06	\\
		&	$\chi^2/\rm{DOF}$\tablenotemark{b}		&  38.94/42=0.93			& 104.82/93=1.13
	& 39.79/2=1.37				& 12.48/16=0.78			& 73.57/37=1.99			& 164.72/107=1.54	\\
\hline
\multicolumn{8}{c}{\bf APEC and CLOUDY table components}\\
\hline
zphabs	&  $N_H$ ($\times10^{20}\,\rm{cm}^{-2}$)	 & \vmp{22.29}{5.65}{4.96} & $\dagger$ 
	&$\dagger$  				&$\dagger$  				& $\dagger$ 				&$\dagger$\\
apec		&	log $F_X$\tablenotemark{a} (\ecmss)			&   \vmp{-12.98}{2.98}{0.54} 	& \vmp{-14.19}{2.77}{0.16} 
	&\vmp{-14.23}{0.39}{0.17} 	&\vmp{-14.89}{7.30}{0.41} 	&\vmp{-14.04}{0.11}{0.05} 	&\vmp{-14.01}{0.21}{0.49} \\
		&	$kT$	 (keV)				&  \vmp{0.061}{0.010}{0.001} 	& \vmp{1.018}{0.100}{0.205} 	
	&\vmp{0.261}{0.042}{0.060} 	&\vmp{1.011}{0.316}{0.180} 	&\vmp{0.760}{0.051}{0.113} 	&\vmp{1.491}{0.031}{0.112} \\
cloudy	&	log $F_X$\tablenotemark{a} (\ecmss)			&   \vmp{-13.22}{0.06}{0.04} 	& \vmp{-13.55}{1.30}{0.04} 
	&\vmp{-14.00}{0.24}{0.11} 	&\vmp{-13.69}{0.45}{0.28} 	&\vmp{-13.72}{0.11}{0.08} 	&\vmp{-13.07}{0.03}{0.55} \\
		&	log U					&   \vmp{0.58}{0.04}{0.10} 	& \vmp{1.27}{0.36}{0.07} 
	&\vmp{1.39}{0.11}{0.17} 		&\vmp{-1.78}{0.56}{0.49} 		&\vmp{-0.71}{0.34}{0.07} 		&\vmp{0.30}{0.02}{0.02} \\
		&	log $N_H$ ({cm}$^{-2}$)			&   \vmp{20.29}{0.54}{0.71} 	& \vmp{23.50}{3.25}{1.59} 
	&\vmp{22.65}{2.08}{0.05} 		&\vmp{20.07}{0.84}{1.97} 		&\vmp{23.13}{2.42}{0.14} 		&\vmp{20.79}{0.91}{0.53} \\
\hline
		&	L-Statistic 				&  44.64					& 88.67 
	& 33.43					& 8.14					& 29.93 					& 138.67	\\
		&	$\chi^2/\rm{DOF}$\tablenotemark{b}		&  38.64/39=0.99			& 90.94/90=1.01
	& 30.82/26=1.185			& 7.90/13=0.607			& 28.42/34=0.83  			& 140.86/104=1.35	\\
\hline
\tablebreak
\hline
		&	Parameter		& Reg. 7	 & Reg. 8	& 	Reg. 9 & Reg. 11\tablenotemark{c} & Reg. 12 & Reg. 13 \\
\hline
\multicolumn{8}{c}{\bf Single APEC  component}\\
\hline
zphabs	&  $N_H$ ($\times10^{20}\,\rm{cm}^{-2}$)	 & \vmp{74.44}{8.87}{4.73} & $\dagger$ 
	& $\dagger$ 				& $\dagger$ 				&$\dagger$ 		 		&$\dagger$ \\
apec		&	log $F_X$\tablenotemark{a}  (\ecmss)			&   \vmp{-11.69}{0.35}{0.14} 	& \vmp{-13.73}{0.02}{0.03} 
	&\vmp{-14.00}{0.04}{0.07} 	&\vmp{-14.37}{0.05}{0.27} 	&\vmp{-13.82}{0.04}{0.02} 	&\vmp{-13.16}{0.03}{0.02} \\
		&	$kT$	 (keV)				&  \vmp{0.147}{0.005}{0.027} 	& \vmp{0.975}{0.052}{0.023} 
	&\vmp{0.355}{0.038}{0.061} 	&\vmp{0.972}{0.221}{1.396} 	&\vmp{0.318}{0.022}{0.040}	&\vmp{0.257}{0.008}{0.009} \\
\hline
zgauss	& $\sigma$ (keV)			& N/A					& N/A
	&	N/A					& \vmp{0.098}{0.001}{0.111}	& N/A & N/A  \\
		& norm. ($\times10^{-6}$)		& N/A					& N/A
	&	N/A					& \vmp{1.15}{0.19}{0.68}		& N/A & N/A  \\
\hline
		&	L-Statistic 				&  89.51 					& 191.71
	& 86.75					& 38.60					& 84.21					& 389.08	\\
		&	$\chi^2/\rm{DOF}$\tablenotemark{b}		&  76.19/43=1.77 			& 168.99/94=1.80
	& 114.27/30=3.81			& 27.50/17=1.618			& 105.72/38=2.78			& 424.78/108=3.93	\\
\hline
\multicolumn{8}{c}{\bf Single CLOUDY table component}\\
\hline
zphabs	&  $N_H$ ($\times10^{20}\,\rm{cm}^{-2}$)	 & \vmp{2.72}{0.10}{4.12} & $\dagger$
	&$\dagger$  				&$\dagger$   				&$\dagger$				 &$\dagger$ \\
cloudy	&	log $F_X$\tablenotemark{a} (\ecmss)			&   \vmp{-13.57}{0.02}{0.08} 	& \vmp{-13.40}{0.10}{0.01} 
	&\vmp{-13.78}{0.05}{0.07} 	&\vmp{-13.69}{0.26}{0.18} 	&\vmp{-13.56}{0.03}{0.07} 	&\vmp{-13.02}{0.04}{0.01} \\
		&	log U					&   \vmp{0.86}{0.09}{0.10} 	& \vmp{1.41}{0.38}{0.04} 
	&\vmp{0.78}{0.11}{0.15} 		&\vmp{-1.98}{0.71}{0.20} 		&\vmp{-3.00}{0.03}{0.40} 		&\vmp{0.45}{0.01}{0.13} \\
		&	log $N_H$ ({cm}$^{-2}$)			&   \vmp{21.51}{0.47}{0.58} 	& \vmp{23.5}{1.72}{0.02} 
	&\vmp{20.88}{0.72}{0.86} 		&\vmp{19.02}{0.25}{2.10} 		&\vmp{20.28}{0.99}{0.54} 		&\vmp{20.35}{0.16}{0.23} \\
\hline
		&	L-Statistic 				&  45.29					& 99.52
	& 39.49					& 11.96					 & 68.83					& 164.06	\\
		&	$\chi^2/\rm{DOF}$\tablenotemark{b}		&  38.94/42=0.93			& 104.82/93=1.13
	& 39.79/2=1.37				& 12.48/16=0.78			& 73.57/37=1.99			& 164.72/107=1.54	\\
\hline
\multicolumn{8}{c}{\bf APEC and CLOUDY table components}\\
\hline
zphabs	&  $N_H$ ($\times10^{20}\,\rm{cm}^{-2}$)	 & \vmp{22.29}{5.65}{4.96} & $\dagger$ 
	&$\dagger$  				&$\dagger$  				& $\dagger$ 				&$\dagger$\\
apec		&	log $F_X$\tablenotemark{a} (\ecmss)			&   \vmp{-12.98}{2.98}{0.54} 	& \vmp{-14.19}{2.77}{0.16} 
	&\vmp{-14.23}{0.39}{0.17} 	&\vmp{-14.89}{7.30}{0.41} 	&\vmp{-14.04}{0.11}{0.05} 	&\vmp{-14.01}{0.21}{0.49} \\
		&	$kT$	 (keV)				&  \vmp{0.061}{0.010}{0.001} 	& \vmp{1.018}{0.100}{0.205} 	
	&\vmp{0.261}{0.042}{0.060} 	&\vmp{1.011}{0.316}{0.180} 	&\vmp{0.760}{0.051}{0.113} 	&\vmp{1.491}{0.031}{0.112} \\
cloudy	&	log $F_X$\tablenotemark{a} (\ecmss)			&   \vmp{-13.22}{0.06}{0.04} 	& \vmp{-13.55}{1.30}{0.04} 
	&\vmp{-14.00}{0.24}{0.11} 	&\vmp{-13.69}{0.45}{0.28} 	&\vmp{-13.72}{0.11}{0.08} 	&\vmp{-13.07}{0.03}{0.55} \\
		&	log U					&   \vmp{0.58}{0.04}{0.10} 	& \vmp{1.27}{0.36}{0.07} 
	&\vmp{1.39}{0.11}{0.17} 		&\vmp{-1.78}{0.56}{0.49} 		&\vmp{-0.71}{0.34}{0.07} 		&\vmp{0.30}{0.02}{0.02} \\
		&	log $N_H$ ({cm}$^{-2}$)			&   \vmp{20.29}{0.54}{0.71} 	& \vmp{23.50}{3.25}{1.59} 
	&\vmp{22.65}{2.08}{0.05} 		&\vmp{20.07}{0.84}{1.97} 		&\vmp{23.13}{2.42}{0.14} 		&\vmp{20.79}{0.91}{0.53} \\
\hline
		&	L-Statistic 				&  44.64					& 88.67 
	& 33.43					& 8.14					& 29.93 					& 138.67	\\
		&	$\chi^2/\rm{DOF}$\tablenotemark{b}		&  38.64/39=0.99			& 90.94/90=1.01
	& 30.82/26=1.185			& 7.90/13=0.607			& 28.42/34=0.83  			& 140.86/104=1.35	\\
\hline
\enddata

\tablenotetext{\ensuremath{\dagger}}{negligible ($N_H<10^{17}$)}
\tablenotetext{a}{0.3-8.0 keV}
\tablenotetext{b}{measured for the best lstat under {\tt Xspec}}
\tablenotetext{c}{Region 11 has an excess of Fe\,K$\alpha$, which is modeled in all fits by a single gaussian at 6.4 keV (host frame).}
\tablenotetext{}{Region 10 contains the larger part of several bright spatially resolved regions (4, 5, 9, 10, 11) and is therefore excluded.}
\end{deluxetable*}
\end{longrotatetable}

\begin{figure*}[ht!]
\noindent
\centering
\begin{overpic}[scale=0.23]{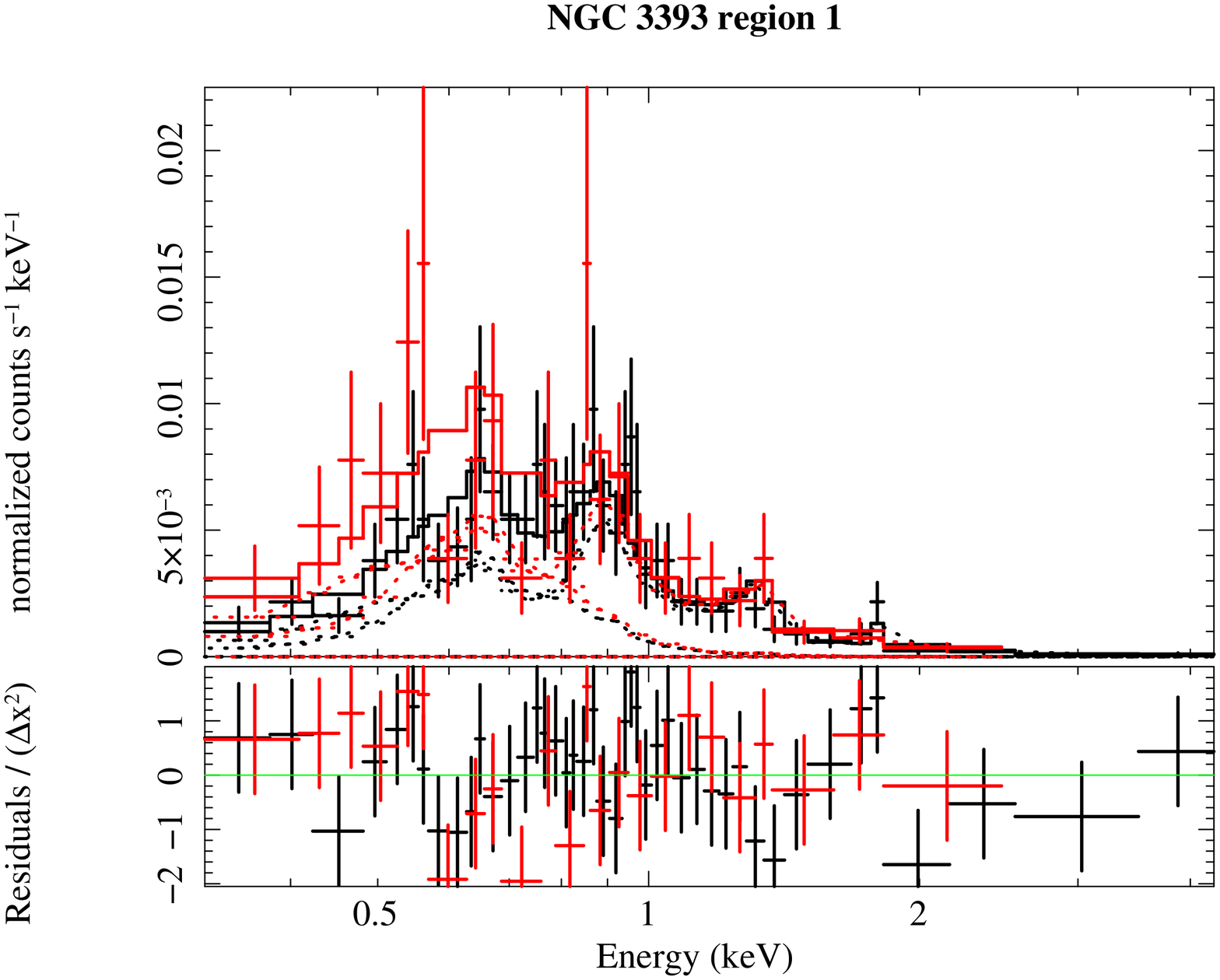}
	\put(58,65){\footnotesize\color{black} $kT=0.23$}
	\put(58,59){\footnotesize\color{black} log $U=1.36$}
	\put(58,53){\footnotesize\color{black} log $N_H=23.5$}
	\put(0,0){\color{white}\rule{0.2cm}{3.5cm}}
	\put(-10,-1){\scriptsize\rotatebox{90}{\parbox{2cm}{%
			\begin{equation}
			\frac{\rm{Residuals}}{\Delta\chi^2} \nonumber
			\end{equation}
			}}}
	\put(-3,32){\scriptsize\rotatebox{90}{\parbox{2cm}{%
			\centering normalized \\ $\rm{counts\,s}^{-1}\,\rm{keV}^{-1}$
			}}}
	\put(45,0){\color{white}\rule{1.3cm}{0.2cm}}
	\put(40,-1){\scriptsize Energy(keV)}
	\put(43,77){\color{white}\rule{1.5cm}{0.2cm}}
	\put(30,75){\bf{\scriptsize NGC 3393 region 1}}
\end{overpic}\hspace{0.01\textwidth}%
\begin{overpic}[scale=0.23]{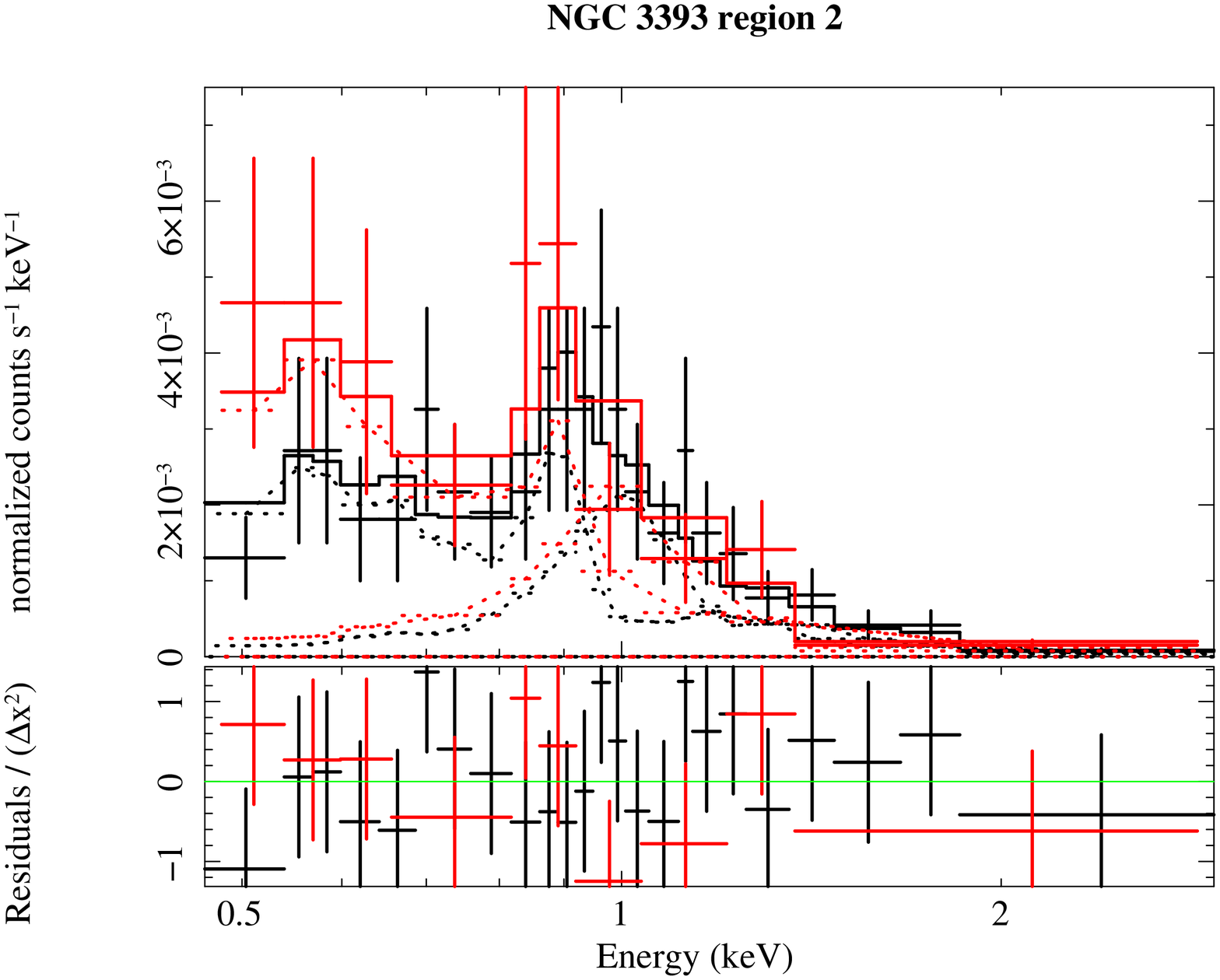}
	\put(58,65){\footnotesize\color{black} $kT=1.26$}
	\put(58,59){\footnotesize\color{black} log $U=0.39$}
	\put(58,53){\footnotesize\color{black} log $N_H=21.1$}
	\put(0,0){\color{white}\rule{0.2cm}{3.5cm}}
	\put(-10,-1){\scriptsize\rotatebox{90}{\parbox{2cm}{%
			\begin{equation}
			\frac{\rm{Residuals}}{\Delta\chi^2} \nonumber
			\end{equation}
			}}}
	\put(-3,32){\scriptsize\rotatebox{90}{\parbox{2cm}{%
			\centering normalized \\ $\rm{counts\,s}^{-1}\,\rm{keV}^{-1}$
			}}}
	\put(45,0){\color{white}\rule{1.3cm}{0.2cm}}
	\put(40,-1){\scriptsize Energy(keV)}
	\put(43,77){\color{white}\rule{1.5cm}{0.2cm}}
	\put(30,75){\bf{\scriptsize NGC 3393 region 2}}
\end{overpic}\hspace{0.01\textwidth}%
\begin{overpic}[scale=0.23]{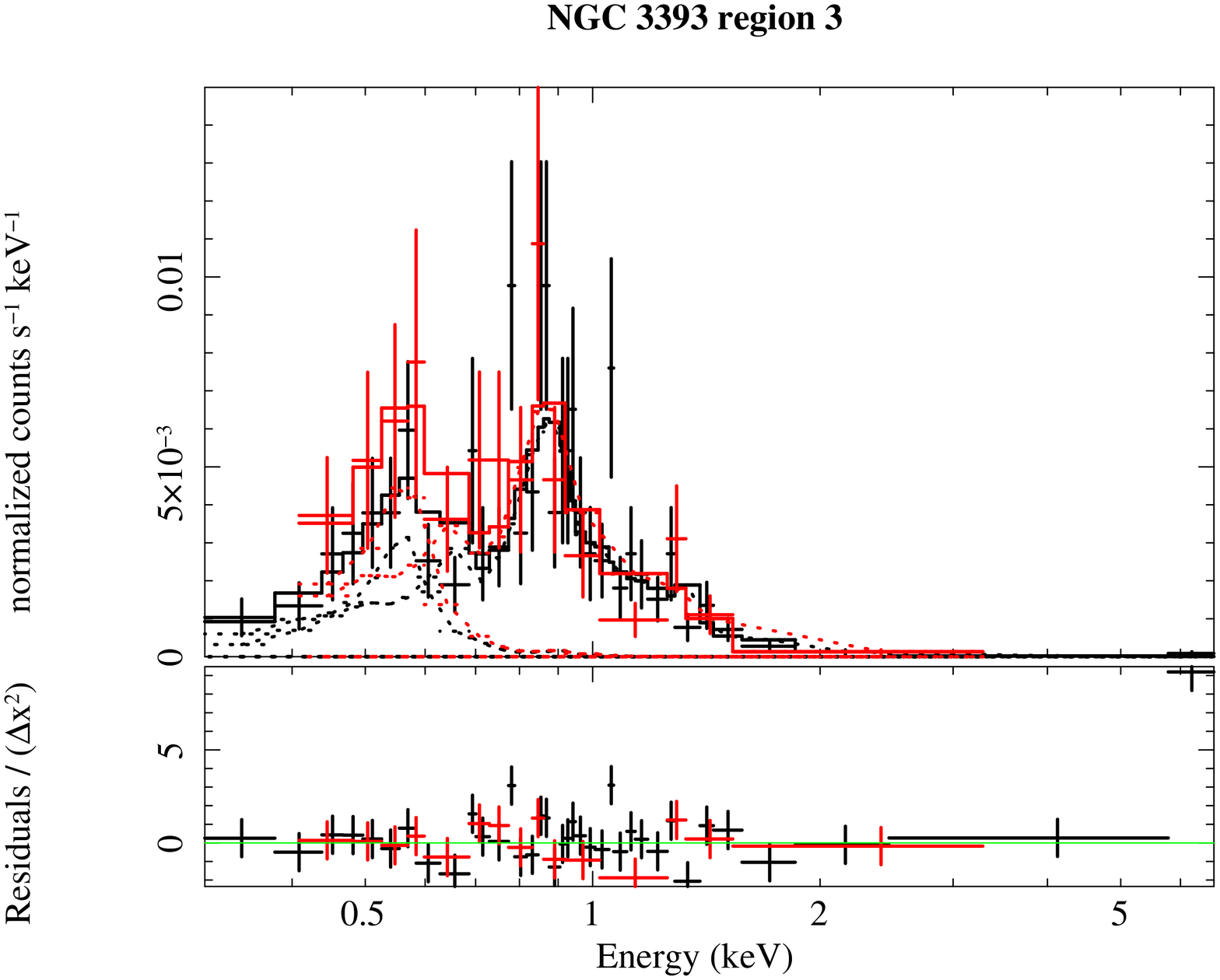}
	\put(58,65){\footnotesize\color{black} $kT=0.10$}
	\put(58,59){\footnotesize\color{black} log $U=0.85$}
	\put(58,53){\footnotesize\color{black} log $N_H=20.1$}
	\put(0,0){\color{white}\rule{0.2cm}{3.5cm}}
	\put(-10,-1){\scriptsize\rotatebox{90}{\parbox{2cm}{%
			\begin{equation}
			\frac{\rm{Residuals}}{\Delta\chi^2} \nonumber
			\end{equation}
			}}}
	\put(-3,32){\scriptsize\rotatebox{90}{\parbox{2cm}{%
			\centering normalized \\ $\rm{counts\,s}^{-1}\,\rm{keV}^{-1}$
			}}}
	\put(45,0){\color{white}\rule{1.3cm}{0.2cm}}
	\put(40,-1){\scriptsize Energy(keV)}
	\put(43,77){\color{white}\rule{1.5cm}{0.2cm}}
	\put(30,75){\bf{\scriptsize NGC 3393 region 3}}
\end{overpic}\par
\vspace{0.1in}
\begin{overpic}[scale=0.23]{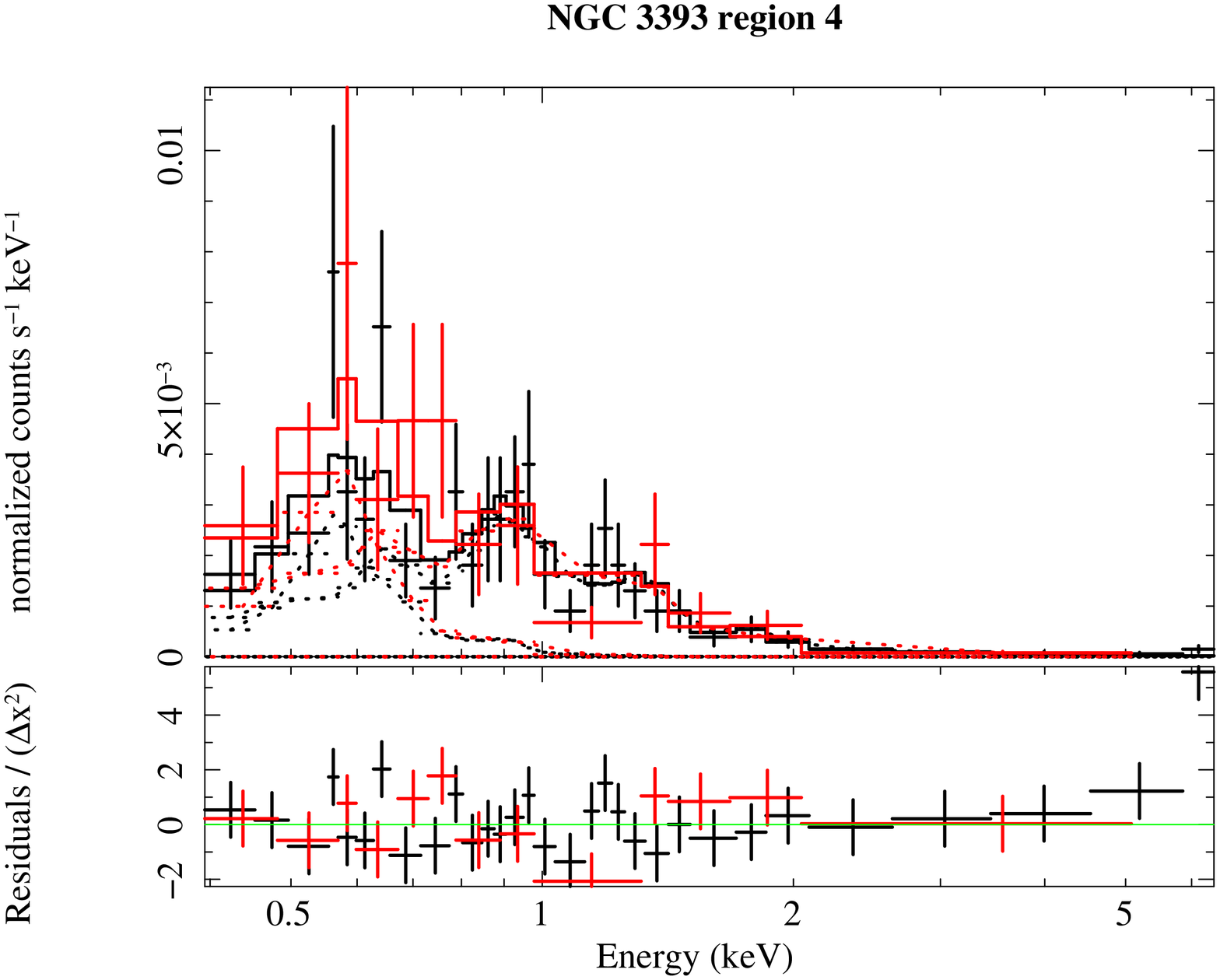}
	\put(58,65){\footnotesize\color{black} $kT=0.16$}
	\put(58,59){\footnotesize\color{black} log $U=1.55$}
	\put(58,53){\footnotesize\color{black} log $N_H=23.5$}
	\put(0,0){\color{white}\rule{0.2cm}{3.5cm}}
	\put(-10,-1){\scriptsize\rotatebox{90}{\parbox{2cm}{%
			\begin{equation}
			\frac{\rm{Residuals}}{\Delta\chi^2} \nonumber
			\end{equation}
			}}}
	\put(-3,32){\scriptsize\rotatebox{90}{\parbox{2cm}{%
			\centering normalized \\ $\rm{counts\,s}^{-1}\,\rm{keV}^{-1}$
			}}}
	\put(45,0){\color{white}\rule{1.3cm}{0.2cm}}
	\put(40,-1){\scriptsize Energy(keV)}
	\put(43,77){\color{white}\rule{1.5cm}{0.2cm}}
	\put(30,75){\bf{\scriptsize NGC 3393 region 4}}
\end{overpic}\hspace{0.01\textwidth}%
\begin{overpic}[scale=0.23]{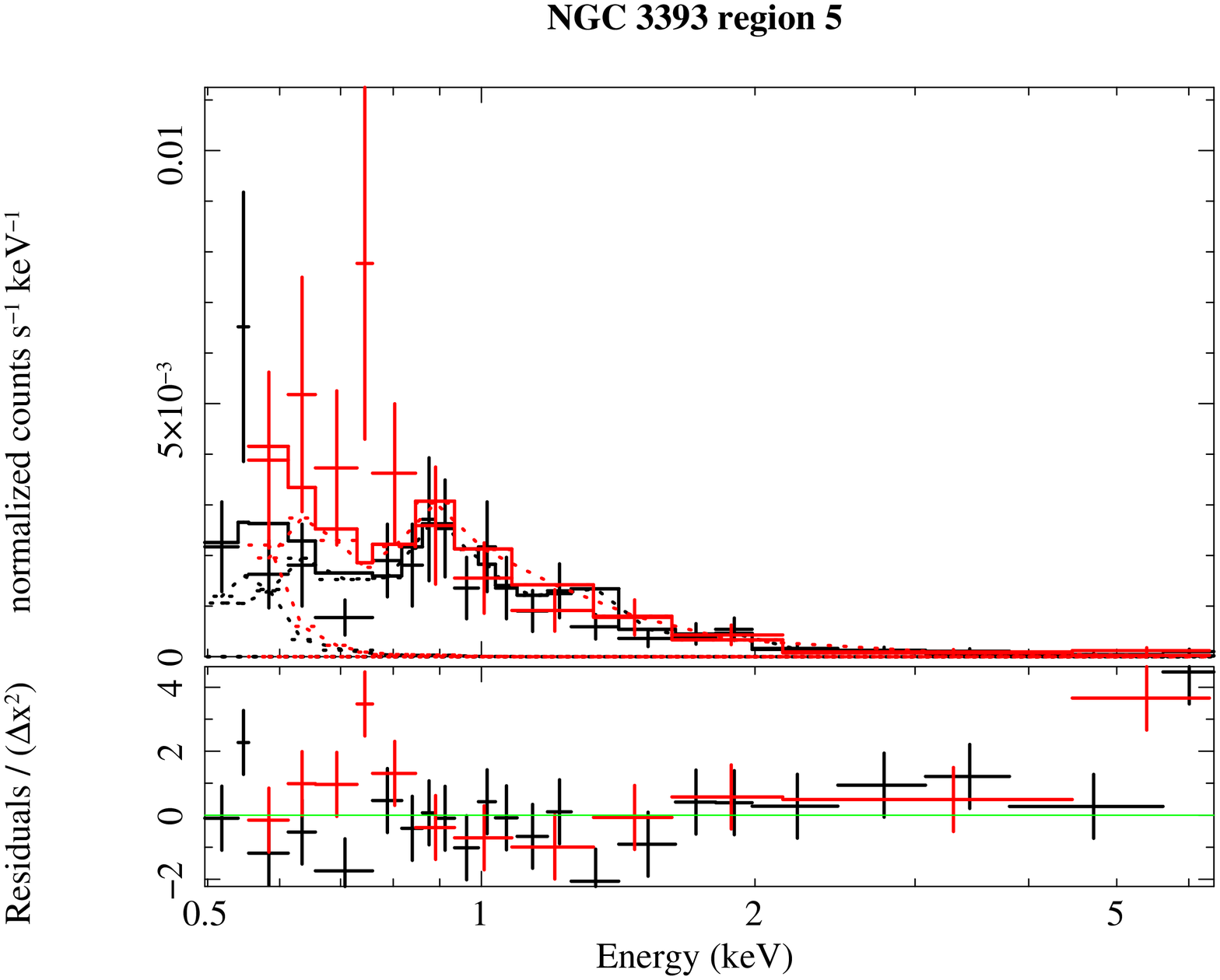}
	\put(58,65){\footnotesize\color{black} $kT=0.08$}
	\put(58,59){\footnotesize\color{black} log $U=1.51$}
	\put(58,53){\footnotesize\color{black} log $N_H=23.5$}
	\put(0,0){\color{white}\rule{0.2cm}{3.5cm}}
	\put(-10,-1){\scriptsize\rotatebox{90}{\parbox{2cm}{%
			\begin{equation}
			\frac{\rm{Residuals}}{\Delta\chi^2} \nonumber
			\end{equation}
			}}}
	\put(-3,32){\scriptsize\rotatebox{90}{\parbox{2cm}{%
			\centering normalized \\ $\rm{counts\,s}^{-1}\,\rm{keV}^{-1}$
			}}}
	\put(45,0){\color{white}\rule{1.3cm}{0.2cm}}
	\put(40,-1){\scriptsize Energy(keV)}
	\put(43,77){\color{white}\rule{1.5cm}{0.2cm}}
	\put(30,75){\bf{\scriptsize NGC 3393 region 5}}
\end{overpic}\hspace{0.01\textwidth}%
\begin{overpic}[scale=0.23]{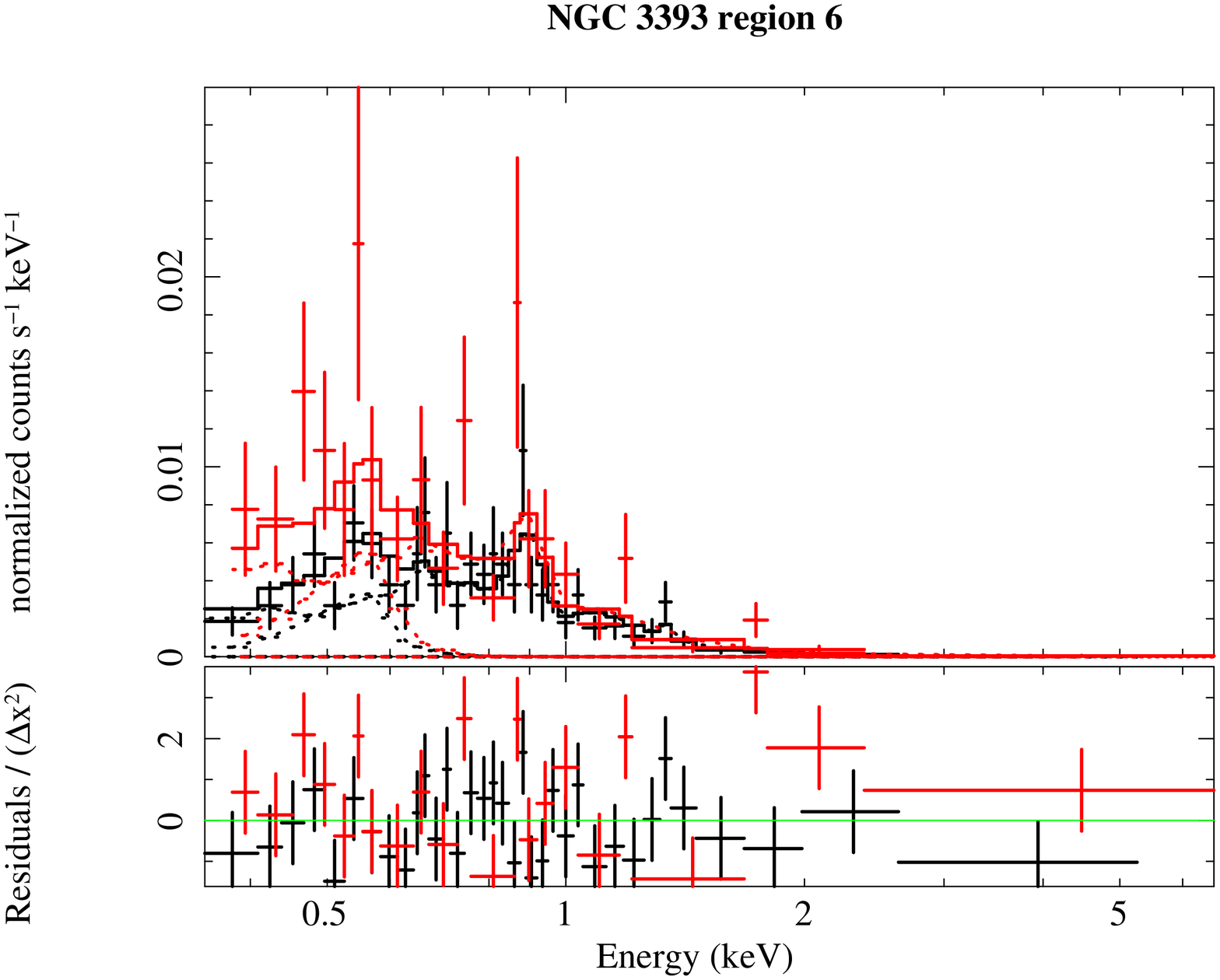}
	\put(58,65){\footnotesize\color{black} $kT=0.05$}
	\put(58,59){\footnotesize\color{black} log $U=0.30$}
	\put(58,53){\footnotesize\color{black} log $N_H=20.2$}
	\put(0,0){\color{white}\rule{0.2cm}{3.5cm}}
	\put(-10,-1){\scriptsize\rotatebox{90}{\parbox{2cm}{%
			\begin{equation}
			\frac{\rm{Residuals}}{\Delta\chi^2} \nonumber
			\end{equation}
			}}}
	\put(-3,32){\scriptsize\rotatebox{90}{\parbox{2cm}{%
			\centering normalized \\ $\rm{counts\,s}^{-1}\,\rm{keV}^{-1}$
			}}}
	\put(45,0){\color{white}\rule{1.3cm}{0.2cm}}
	\put(40,-1){\scriptsize Energy(keV)}
	\put(43,77){\color{white}\rule{1.5cm}{0.2cm}}
	\put(30,75){\bf{\scriptsize NGC 3393 region 6}}
\end{overpic}\par
\vspace{0.1in}
\begin{overpic}[scale=0.23]{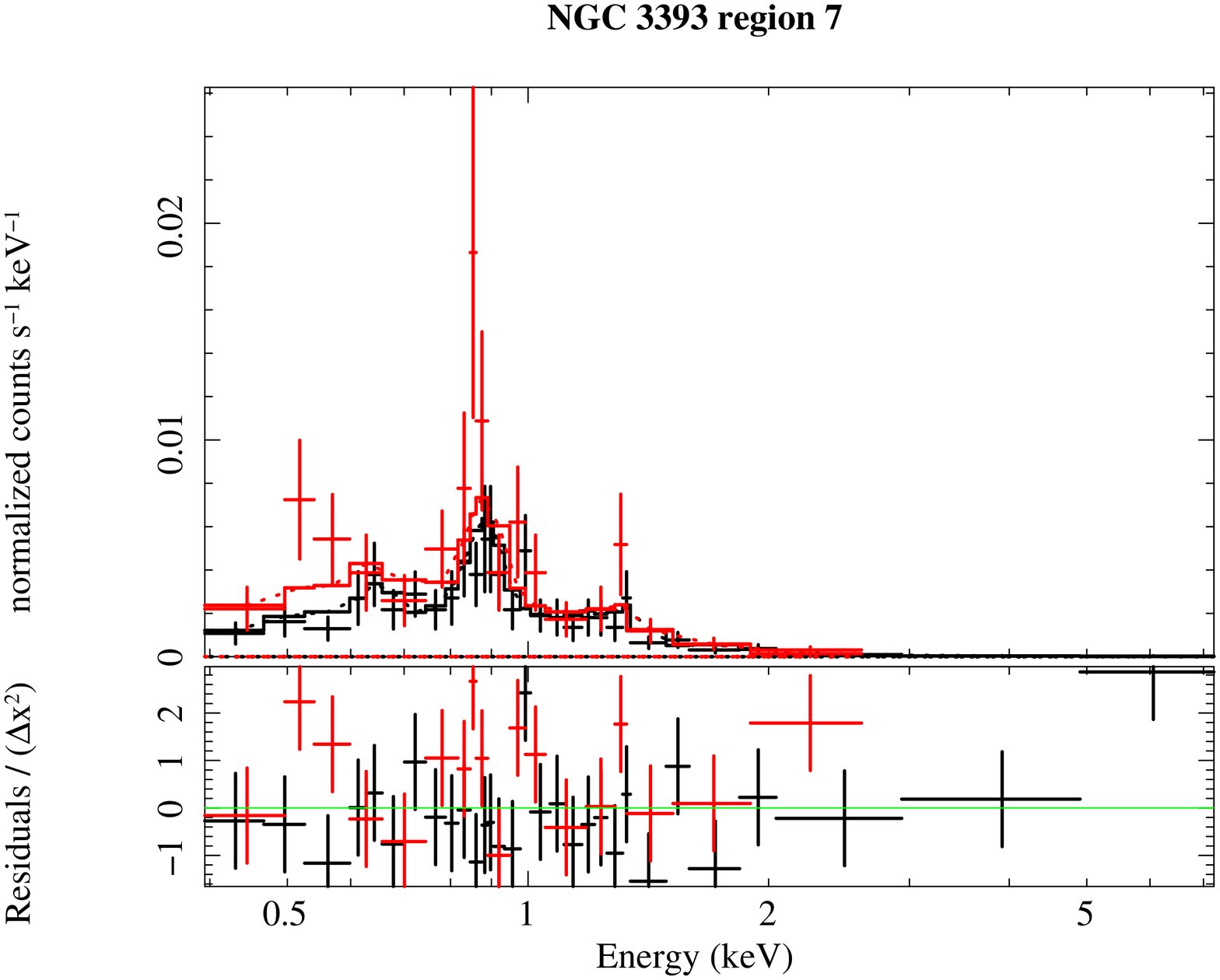}
	\put(58,59){\footnotesize\color{black} log $U=0.86$}
	\put(58,53){\footnotesize\color{black} log $N_H=21.5$}
	\put(0,0){\color{white}\rule{0.2cm}{3.5cm}}
	\put(-10,-1){\scriptsize\rotatebox{90}{\parbox{2cm}{%
			\begin{equation}
			\frac{\rm{Residuals}}{\Delta\chi^2} \nonumber
			\end{equation}
			}}}
	\put(-3,32){\scriptsize\rotatebox{90}{\parbox{2cm}{%
			\centering normalized \\ $\rm{counts\,s}^{-1}\,\rm{keV}^{-1}$
			}}}
	\put(45,0){\color{white}\rule{1.3cm}{0.2cm}}
	\put(40,-1){\scriptsize Energy(keV)}
	\put(43,77){\color{white}\rule{1.5cm}{0.2cm}}
	\put(30,75){\bf{\scriptsize NGC 3393 region 7}}
\end{overpic}\hspace{0.01\textwidth}%
\begin{overpic}[scale=0.23]{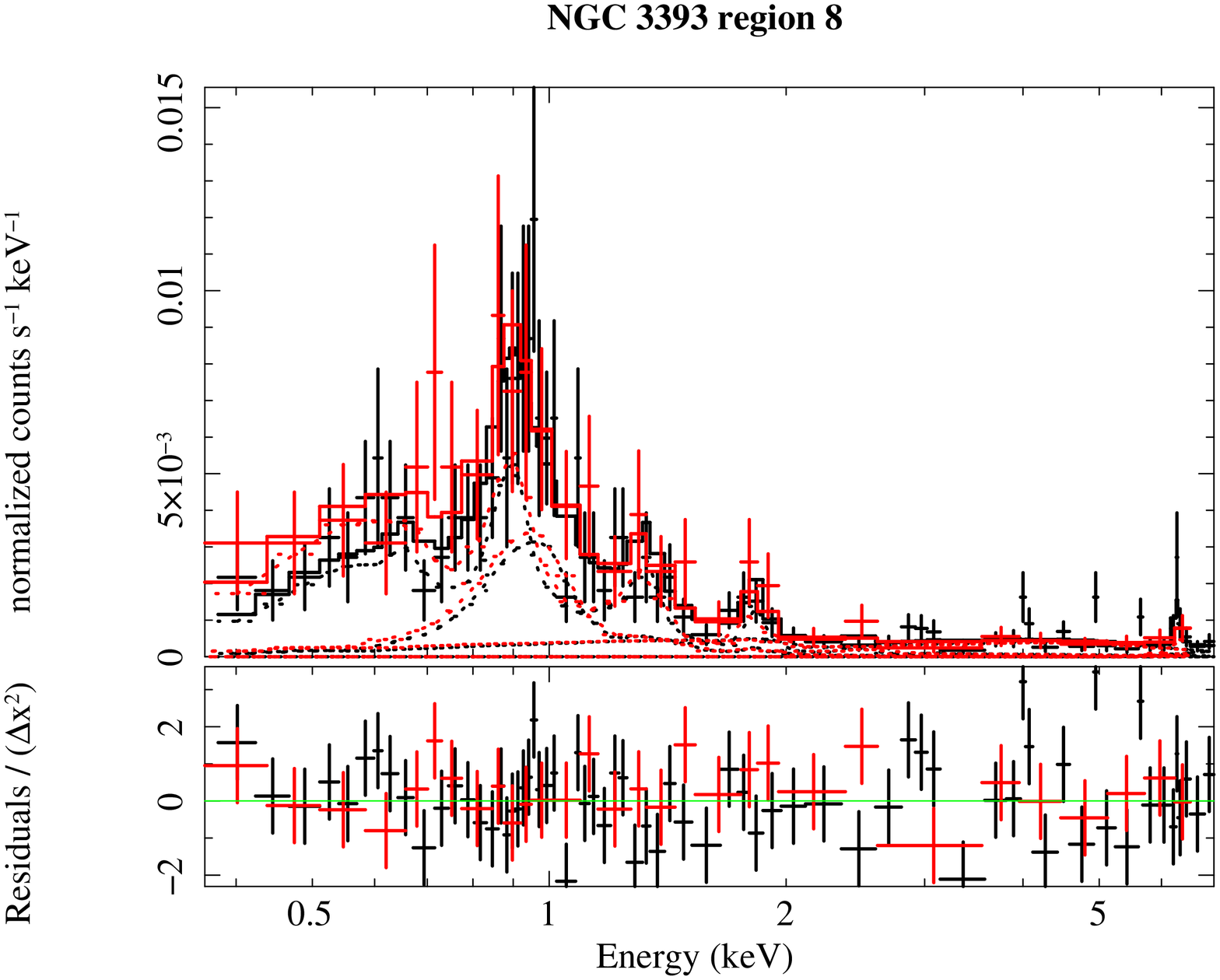}
	\put(58,65){\footnotesize\color{black} $kT=1.02$}
	\put(58,59){\footnotesize\color{black} log $U=1.27$}
	\put(58,53){\footnotesize\color{black} log $N_H=23.5$}
	\put(0,0){\color{white}\rule{0.2cm}{3.5cm}}
	\put(-10,-1){\scriptsize\rotatebox{90}{\parbox{2cm}{%
			\begin{equation}
			\frac{\rm{Residuals}}{\Delta\chi^2} \nonumber
			\end{equation}
			}}}
	\put(-3,32){\scriptsize\rotatebox{90}{\parbox{2cm}{%
			\centering normalized \\ $\rm{counts\,s}^{-1}\,\rm{keV}^{-1}$
			}}}
	\put(45,0){\color{white}\rule{1.3cm}{0.2cm}}
	\put(40,-1){\scriptsize Energy(keV)}
	\put(43,77){\color{white}\rule{1.5cm}{0.2cm}}
	\put(30,75){\bf{\scriptsize NGC 3393 region 8}}
\end{overpic}\hspace{0.01\textwidth}%
\begin{overpic}[scale=0.23]{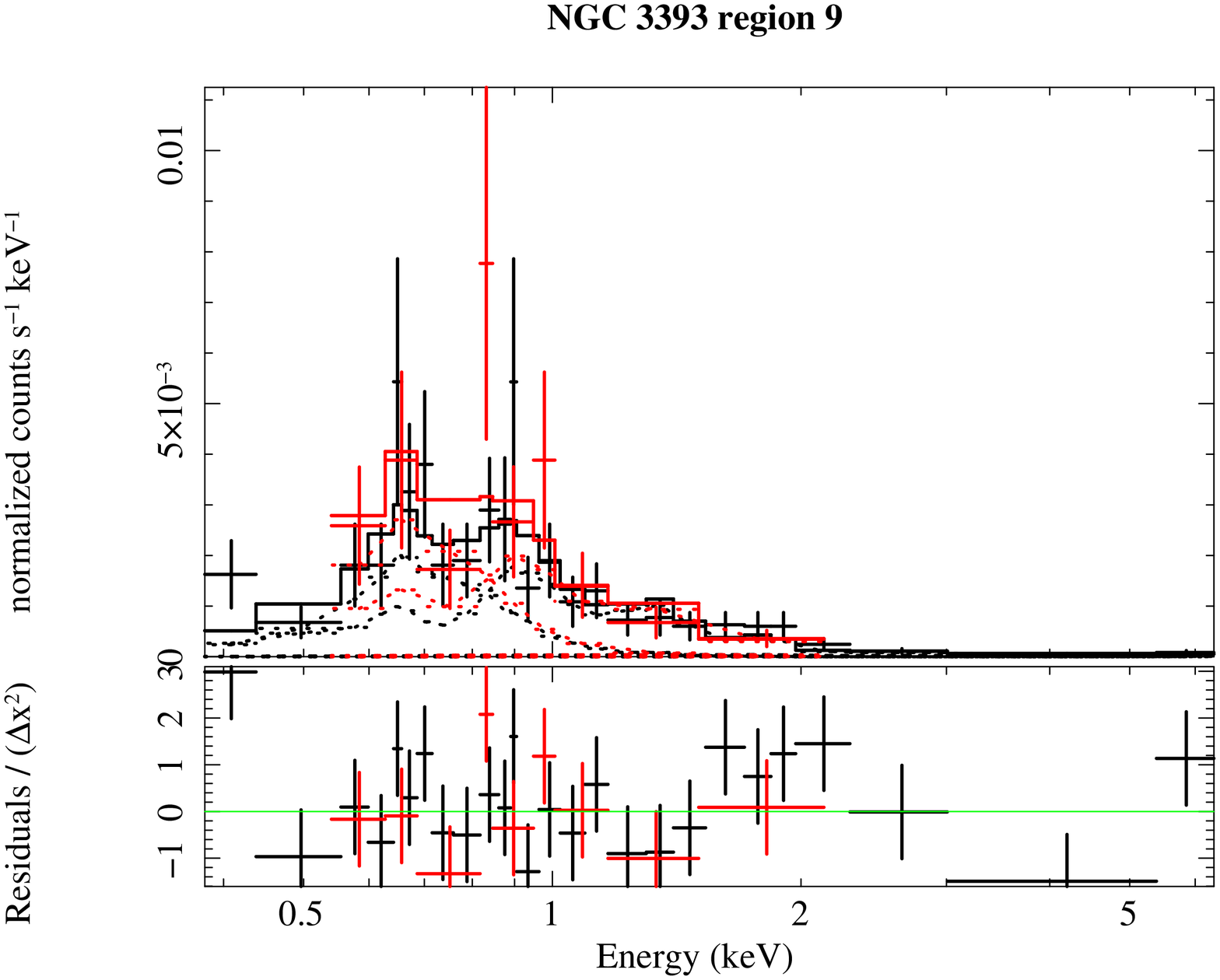}
	\put(58,65){\footnotesize\color{black} $kT=0.26$}
	\put(58,59){\footnotesize\color{black} log $U=1.39$}
	\put(58,53){\footnotesize\color{black} log $N_H=22.7$}
	\put(0,0){\color{white}\rule{0.2cm}{3.5cm}}
	\put(-10,-1){\scriptsize\rotatebox{90}{\parbox{2cm}{%
			\begin{equation}
			\frac{\rm{Residuals}}{\Delta\chi^2} \nonumber
			\end{equation}
			}}}
	\put(-3,32){\scriptsize\rotatebox{90}{\parbox{2cm}{%
			\centering normalized \\ $\rm{counts\,s}^{-1}\,\rm{keV}^{-1}$
			}}}
	\put(45,0){\color{white}\rule{1.3cm}{0.2cm}}
	\put(40,-1){\scriptsize Energy(keV)}
	\put(43,77){\color{white}\rule{1.5cm}{0.2cm}}
	\put(30,75){\bf{\scriptsize NGC 3393 region 9}}
\end{overpic}\par
\vspace{0.1in}
\begin{overpic}[scale=0.23]{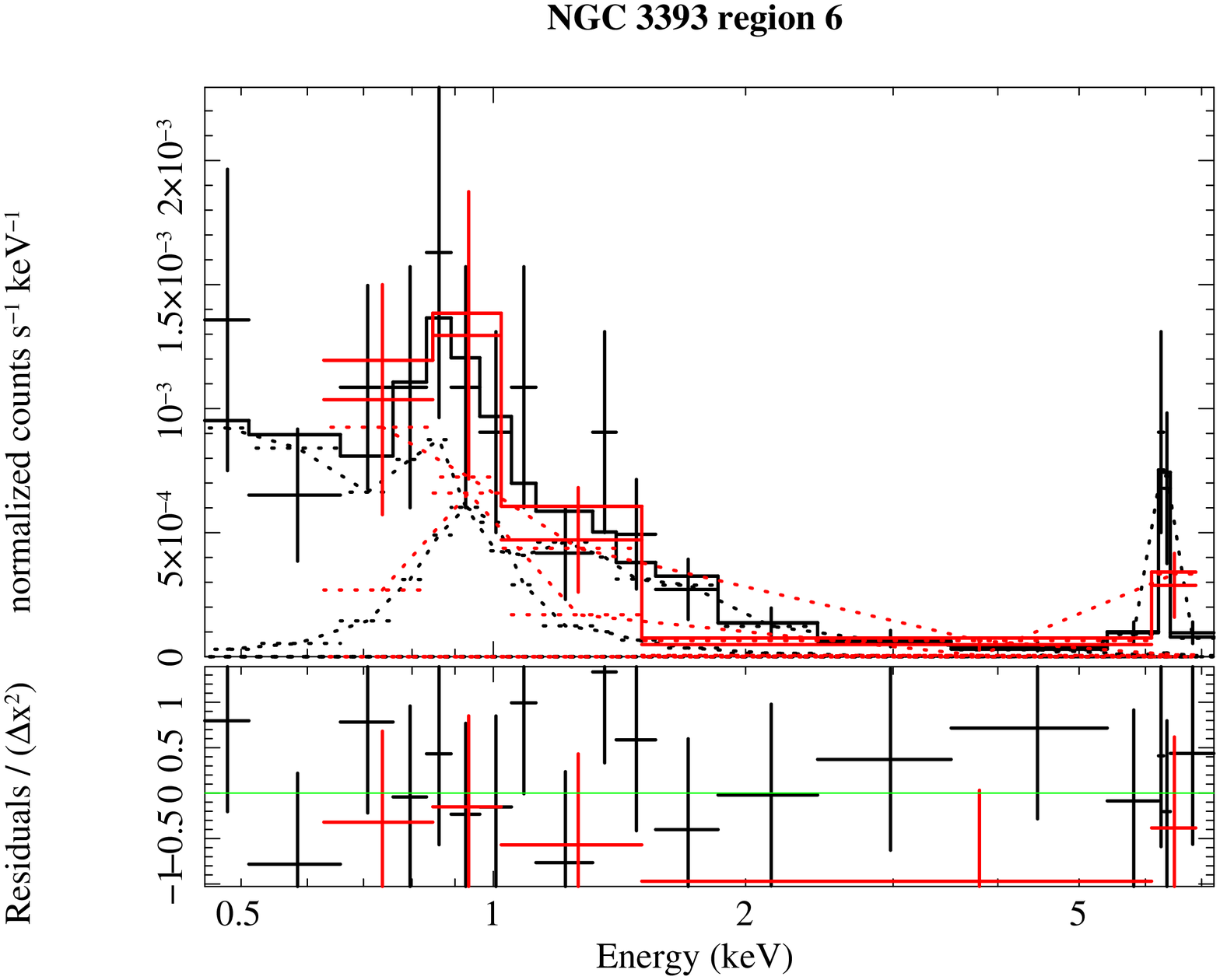}
	\put(58,65){\footnotesize\color{black} $kT=1.011$}
	\put(58,59){\footnotesize\color{black} log $U=-1.78$}
	\put(58,53){\footnotesize\color{black} log $N_H=20.1$}
	\put(58,47){\footnotesize\color{black} Fe K$\alpha$}
	\put(0,0){\color{white}\rule{0.2cm}{3.5cm}}
	\put(-10,-1){\scriptsize\rotatebox{90}{\parbox{2cm}{%
			\begin{equation}
			\frac{\rm{Residuals}}{\Delta\chi^2} \nonumber
			\end{equation}
			}}}
	\put(-3,32){\scriptsize\rotatebox{90}{\parbox{2cm}{%
			\centering normalized \\ $\rm{counts\,s}^{-1}\,\rm{keV}^{-1}$
			}}}
	\put(45,0){\color{white}\rule{1.3cm}{0.2cm}}
	\put(40,-1){\scriptsize Energy(keV)}
	\put(43,77){\color{white}\rule{1.5cm}{0.2cm}}
	\put(30,75){\bf{\scriptsize NGC 3393 region 11}}
\end{overpic}\hspace{0.01\textwidth}%
\begin{overpic}[scale=0.23]{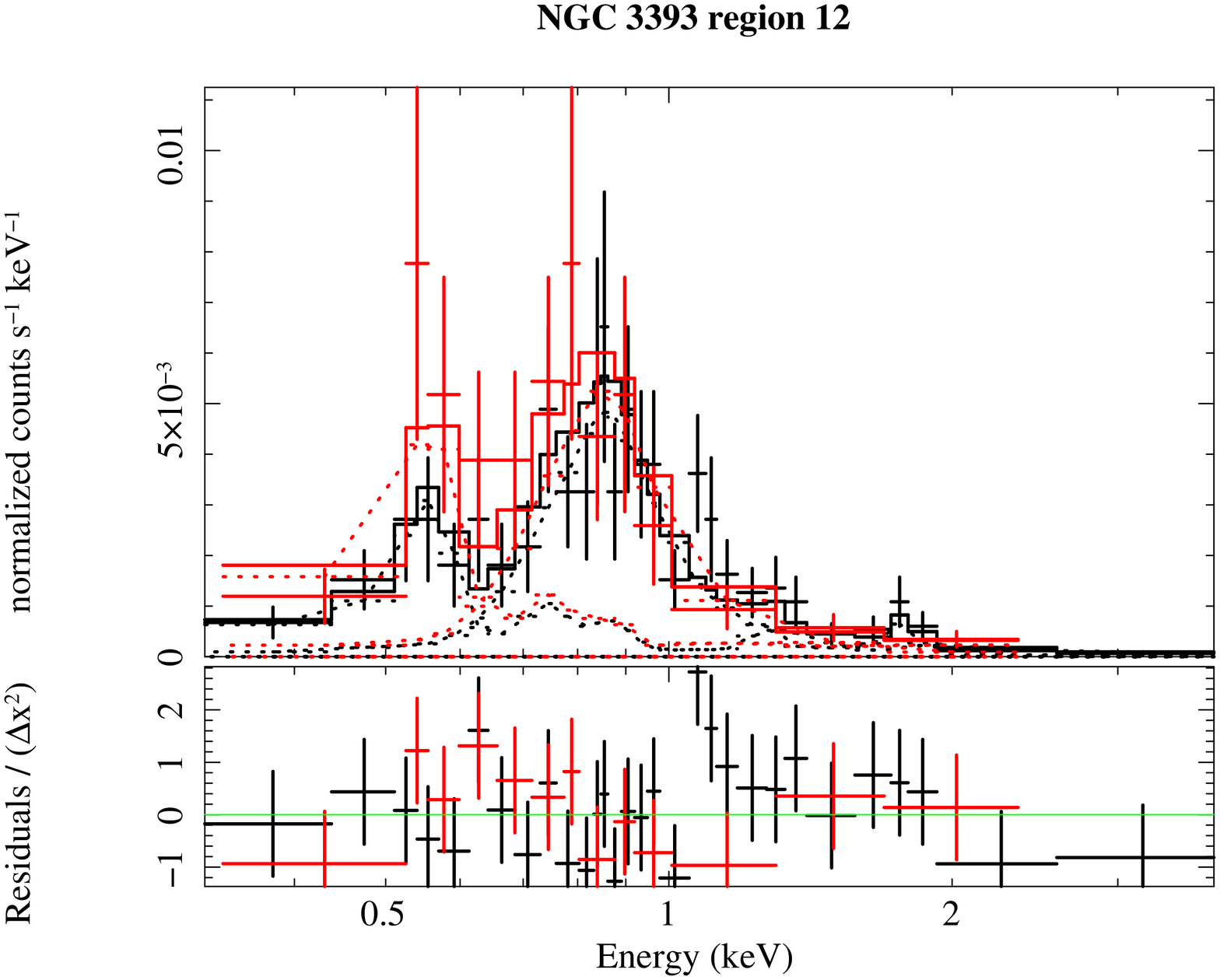}
	\put(58,65){\footnotesize\color{black} $kT=0.76$}
	\put(58,59){\footnotesize\color{black} log $U=-0.71$}
	\put(58,53){\footnotesize\color{black} log $N_H=23.1$}
	\put(0,0){\color{white}\rule{0.2cm}{3.5cm}}
	\put(-10,-1){\scriptsize\rotatebox{90}{\parbox{2cm}{%
			\begin{equation}
			\frac{\rm{Residuals}}{\Delta\chi^2} \nonumber
			\end{equation}
			}}}
	\put(-3,32){\scriptsize\rotatebox{90}{\parbox{2cm}{%
			\centering normalized \\ $\rm{counts\,s}^{-1}\,\rm{keV}^{-1}$
			}}}
	\put(45,0){\color{white}\rule{1.3cm}{0.2cm}}
	\put(40,-1){\scriptsize Energy(keV)}
	\put(43,77){\color{white}\rule{1.5cm}{0.2cm}}
	\put(30,75){\bf{\scriptsize NGC 3393 region 12}}
\end{overpic}\hspace{0.01\textwidth}%
\begin{overpic}[scale=0.23]{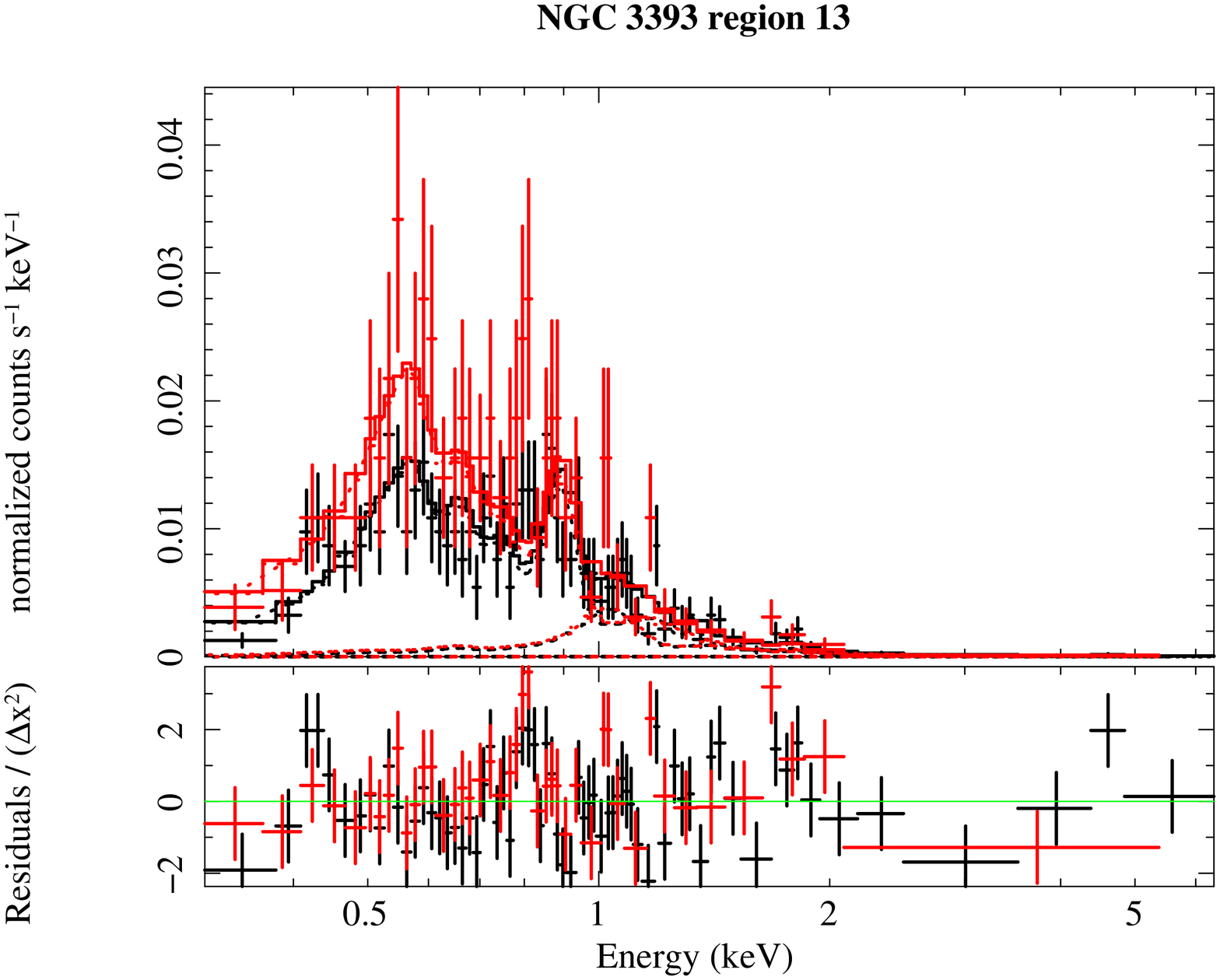}
	\put(58,65){\footnotesize\color{black} $kT=1.49$}
	\put(58,59){\footnotesize\color{black} log $U=0.30$}
	\put(58,53){\footnotesize\color{black} log $N_H=20.8$}
	\put(0,0){\color{white}\rule{0.2cm}{3.5cm}}
	\put(-10,-1){\scriptsize\rotatebox{90}{\parbox{2cm}{%
			\begin{equation}
			\frac{\rm{Residuals}}{\Delta\chi^2} \nonumber
			\end{equation}
			}}}
	\put(-3,32){\scriptsize\rotatebox{90}{\parbox{2cm}{%
			\centering normalized \\ $\rm{counts\,s}^{-1}\,\rm{keV}^{-1}$
			}}}
	\put(45,0){\color{white}\rule{1.3cm}{0.2cm}}
	\put(40,-1){\scriptsize Energy(keV)}
	\put(43,77){\color{white}\rule{1.5cm}{0.2cm}}
	\put(30,75){\bf{\scriptsize NGC 3393 region 13}}
\end{overpic}\par
\caption{Spectra taken from different regions in the nucleus spatially defined in Fig. \ref{fig:ne9o7regs} with best-fit models and residuals as defined in Table \ref{table:spec-reg}.  Column density $N_H$ refers to the reprocessing CLOUDY component, not neutral absorption.  Plots are otherwise similar to Fig. \ref{fig:linespec}.
} 
\label{fig:regspec}
\vspace{0.1in}
\end{figure*}

\noindent
line hardness ratios, and multiple binning approaches.  Such approaches are necessary even for such a bright Sy2 as NGC 3393, due to the limited number of photons per energy band.  The morphology of different X-ray species can then be used to investigate the physical processes behind their excitation.

Inter-species morphological differences are evident to the eye in Fig. \ref{fig:linemaps}, where \cha\ band-specific images including \ion{Ne}{9}, \ion{O}{8} and
\ion{O}{7}, are directly compared with radio emission and [\ion{O}{3}].  A more rigorous approach in \S\S\ref{sec:XELMapSmooth}-\ref{sec:XELImSpec} demonstrates that these differences are significant when comparing \ion{Ne}{9} and \ion{O}{7}.  
\smallskip
\newline$\bullet$ The peaks of \ion{Ne}{9} emission are associated with the leading edges of the radio outflows.  
\newline$\bullet$ Strong \ion{O}{7} coincides with strong \ion{Ne}{9} leading the SW outflow (and stronger but less extended radio emission).
\newline$\bullet$ Strong \ion{O}{7} does not overlap with the \ion{Ne}{9} leading the NE outflow.  
\newline$\bullet$ Radial profiles are less useful for identifying these deviations for X-ray species ratios, which are more significant in e.g. \ion{Ne}{9}-\ion{O}{7} HR maps (Fig. \ref{fig:ne9o7regs}).
\newline$\bullet$ Radial profiles show deviations of  [\ion{O}{3}]/(0.3-2\,keV) ratios in the nucleus from [\ion{O}{3}]/(0.3-2\,keV) at large radii.  
\newline$\bullet$ Elevated  [\ion{O}{3}] is clearly associated with regions at $\simless1\arcsec$ external to the radio peaks.  
\newline$\bullet$ X-ray emission is elevated both inside these radio-dominated regions and immediately outside the [\ion{O}{3}]/(0.3-2\,keV) peaks.  The major exception to this rule is the  [\ion{O}{3}]/(0.3-2\,keV) peak at $r\sim5\arcsec$ SW of the nucleus.

Noting these trends, we examine the spectral fits described in \S\ref{sec:XELImSpec} which are shown in Tables \ref{table:spec-gauss}, \ref{table:spec-mod} \& \ref{table:spec-reg} and Figs. \ref{fig:linespec}, \ref{fig:modspec} \& \ref{fig:regspec}.  
\smallskip
\newline$\bullet$ Gaussian modeling fails to measure \ion{Fe}{19} and other lines around \ion{Ne}{9} which are resolved and detected by \cite{Koss15} using grating spectroscopy.
\newline$\bullet$ Instead, these models of lower-resolution CCD spectra prefer \ion{Ne}{9}, which is inferred to be weak in \cite{Koss15}.
\newline$\bullet$ Several other CCD-derived lines in Table \ref{table:spec-gauss} directly contradict grating measurements by \cite{Koss15} where the separation is too small for CCD energy resolution.
\newline$\bullet$ The simplest models of the ENLR emission suggest domination by photoionizing radiation from the AGN with a possible contribution from a thermal component.
\newline$\bullet$ The nucleus (Reg. 9, containing the unresolved AGN) is well-described by photoionizing radiation from the AGN through a highly obscuring ($log\,N_H\sim23.5\,\rm{cm}^{-2}$) medium.  
\newline$\bullet$ Elevated nuclear emission associated with the \ion{Ne}{9} complex may be due to a collisional plasma, but the flux of this component has a large lower bound.
\newline$\bullet$ The E-W \ion{O}{7} cross-bar (regions 4, 5) is also well-described by photoionizing radiation from the AGN through a highly obscuring ($log\,N_H\sim23.5\,\rm{cm}^{-2}$) medium. 
\newline$\bullet$ In general, spatially resolved spectroscopy of individual regions suggests lower $kT$ values than obtained from the best-fit value of the entire nucleus within $r=5\arcsec$.
\newline$\bullet$ AGN photoionization is most clearly dominant over emission from a collisional plasma in at least two regions in the bicone which are external to the radio emission (2, 13).
\newline$\bullet$ CLOUDY+APEC models of the cross-cone regions (3, 7) is consistent with AGN photoionization through a lower-density ($log\,N_H\sim20.2\,\rm{cm}^{-2}$) medium.  The strength of any thermal plasma is correlated with the assumed obscuration.
\newline$\bullet$ Emission in two regions that have elevated \ion{Ne}{9} and are associated with radio outflows (9, 12) is consistent with having a collisional component with flux comparable to reprocessed photoionizing emission.
\newline$\bullet$ Region 11 in the near-SE bicone has some atypical properties for these regions.  It is bright in the \ion{Ne}{9} complex but not associated with radio emission, and appears to be photoionization-dominated (with an atypically low ionization parameter for this region).  Its models require an additional Fe K$\alpha$ component beyond simple AGN scaling, and at $r\sim1\arcsec$ from the AGN, contamination by the AGN point source should be low.
\newline$\bullet$ Some regions in the bicone (e.g 1, 2, 6) permit a significant collisional component even though they are photoionization-dominated.

\section{Discussion}\label{sec-disc}

\subsection{Emission Line Morphology}

\subsubsection{[\ion{O}{3}] and the AGN Wind}

\cite{Wang11b} investigated the morphology of X-ray emission from the ENLR of NGC 4151 and found spectroscopic evidence for shock excitation on the leading edges of the sub-kpc jets, which coincides with arcs of enhanced \ion{O}{3} emission that wrap around the jet lobes.  \cite{Paggi12} likewise found spectroscopic evidence for shock excitation associated with the arcs of  \ion{O}{3}  emission which lead the jet lobes of Mrk 573.  If the multi-wavelength evidence for shock excitation noted in Papers I and II has its origin in photoionizing shocks leading the bipolar outflows, then spatially resolved spectroscopy of the ENLR  should show similar evidence.  

The  [\ion{O}{3}]/0.3-2\,keV ratio is sensitive to density, and if an AGN wind has a uniform spectrum of densities at a given distance then simple assumptions of an $r^{-2}$ density profile imply that  this ratio should be flat with increasing radius, since the ionization parameter then will not change \citep{Bianchi06,Wang11b}.  But increased  [\ion{O}{3}]/0.3-2\,keV has been observed in association with possible shock locations by \cite{Paggi12}.  We see similar trends here, such that areas of excess  [\ion{O}{3}]/0.3-2\,keV are associated with the leading edges of radio outflows.  The [\ion{O}{3}]/0.3-2\,keV normalization is comparable to values found in previous studies of Mrk 573 and NGC 4151 \citep{Paggi12,Wang11b} when their values are corrected by converting {\it HST} RMS bandwidth to FWHM bandwidth\footnote{RMS bandwidth appears to have been used in \cite{Paggi12} and \cite{Wang11b}, which gives an incorrect low value for [\ion{O}{3}]/(0.3-2\,keV).  We have confirmed this by directly examining their archival data.}.

The [\ion{O}{3}]/0.3-2\,keV spikes which we observe could be consistent with local overdensities, such as clearly associated with a spiral arms intersecting the bicone (at $r\sim5\arcsec$ SW), and possibly associated with shocks or ISM compression at $r\sim1\arcsec$ SW \& NE.  A ``local overdensity" interpretation would be consistent with the spike at $r\sim5\arcsec$ SW, where there is no obvious optical or radio evidence for outflow-ISM interaction.  Rather, the bicone intersects with a spiral dust lane at that point.

\subsubsection{X-ray Line Enhancements as Physical Tracers}

The presence of enhanced emission in the Ne\,IX band is consistent with other findings in Mrk 573 \citep{Paggi12} and NGC 4151 \citep{Wang11b}.  In both cases, although the bulk of the X-ray emission for the ENLR is associated with reflection of an incident AGN spectrum by a photoionized medium, regions immediately adjacent to outflow-ISM interactions (indicated by radio emission and bright optical/X-ray emission) demonstrate spectra that are better represented by emission from a thermal plasma, as might be expected from shock interactions or from overdensities in a hot X-ray emitting outflow.  As noted in Paper II, although the optical line ratios within the bicone are consistent with photoionization, shocks are not excluded since \hst\ STIS kinematics in \cite{Fischer13} are consistent with shocks fast enough to photoionize the precursor ISM.

An E-W \ion{O}{7} cross-bar is evident at sub-arcsecond scales.  It neatly traces a similar structure in [\ion{O}{3}], but is stronger at $r<1\arcsec$ (vs. $r>1\arcsec$ for [\ion{O}{3}]).  This appears to correspond to the nuclear bar \citep[described in greater detail by][]{Alonso-Herrero98, Laesker16, Finlez18}.  The \ion{O}{7} bar therefore likely corresponds to highly ionized gas associated with the black hole's fueling flows, but could trace material of a different density or ionization state than gas in the bar at $r>1\arcsec$.

\subsubsection{CCD Imaging Spectroscopy and Grating Measurements}

Unlike CCD spectroscopy (FWHM $\Delta E\sim100\,$eV), grating spectroscopy with {\it Chandra} can resolve closely spaced diagnostic lines.  We therefore prefer measurements of individual emission lines by \cite{Koss15} in assessing the overall state of the NGC 3393 plasma when those measurements directly contradict our gaussian models of the CCD spectrum (as integrated over the entire nucleus, shown in Table \ref{table:spec-gauss}).

\paragraph{The Nature of the \ion{Ne}{9} Complex}

In \S\ref{sec:XELImSpec}, we show that X-ray spectroscopy suggests that an apparent excess in X-ray emission in the ENLR near \ion{Ne}{9} could be explained by a simple two-component model where the bulk of emission is photoionized, since a single component photoionization model is unable to account for this excess.  

Although we continue to designate this spectral excess as ``\ion{Ne}{9}" due to its proximity relative to the triplet at $E=0.915\,$keV (rest), in practice this narrow band refers to a complex where the low energy resolution of ACIS-S blends emission from multiple lines, including \ion{Fe}{18}, \ion{Fe}{19} and \ion{Ni}{19}.  The complexity of this region is evident in grating spectroscopy from \cite{Koss15}.  They attribute relatively little emission to \ion{Ne}{9}, but two factors complicate this interpretation.  First, their spectrum is spatially unresolved and emission from a collisional plasma (for example) could be diluted by the bulk of the emission which is due to photoionization.  Second, the 0.906 keV 1s$^2$\,$^2\rm{S}_0$\,--\,1s2s\,$^3\rm{S}_1$ \ion{Ne}{9} transition is the forbidden component of a triplet where the two higher-energy lines are difficult to distinguish from \ion{Fe}{19} even with the grating.  As a result of the expected ratios within the \ion{Ne}{9} triplet
and the ratio of the two \ion{Fe}{19} blends, we can expect about half of the emission from the \ion{Ne}{9} complex to arise from \ion{Fe}{19} in a collisional plasma. Relatively little  \ion{Fe}{19} is expected from a photoionization-dominated plasma, in which case  \ion{Ne}{9} would dominate this band.

The \ion{Ne}{9} associated with shocks in the NGC 3393 outflows could originate from a shock-photoionized precursor, rather than from collisional plasma in the shocks themselves (S. Kraemer; {\it personal communication}).  But temperatures of peak ionization fractions for \ion{Ne}{9}  and  \ion{Fe}{19} (at $kT\sim0.1\,$keV and $\sim0.7\,$keV respectively) are consistent with our best-fit models. \ion{O}{8} is also roughly co-spatial to \ion{Ne}{9}, which is expected given those species experience similar excitation trends with $kT$ and $U$.

\paragraph{Emission Line Ratios from a Collisional Plasma}

Other line measurements by \cite{Koss15} also point to collisional excitation.  The ratio of the \ion{Fe}{17} lines at 0.720 and 0.826 keV is a little above 1 in a collisional plasma \citep[comparable to those observed by][]{Koss15}, but far above 1 in a photoionized one \citep{Liedahl90}.  The \cite{Koss15} \ion{Fe}{17} line ratios indicate a collisionally excited plasma.  The helium-like triplets of \ion{Mg}{11} and \ion{Si}{13} are even more clarifying than \ion{Ne}{9}: the middle intercombination lines are unresolved, the longer forbidden line is weak and the shorter resonance line is strong, all of which point to a collisional plasma \citep{PD00}.  For He-like ions, the forbidden $^3\rm{S}_1$-ground triplet line is  weaker than the resonance ground-$^1\rm{P}_1$ line in collisional equilibrium, but it is stronger in a photoionized plasma due to recombination \citep{BK00}.

Due to the complexity of the system and the hard X-ray luminosity of the AGN, we do expect photoionized excitation to be present alongside collisional excitation in the bicone.  The observed Fe\,K$\alpha$ in NGC 3393, for example, should be produced by photoionization of gas with a low ionization state. But the line ratios here are clearly inverted from the case of AGN known to be dominated by photoionization, such as NGC 1068 \citep[as shown by][using {\it XMM-Newton} spectroscopy]{Kinkhabwala02}.  NGC 1068 (unlike NGC 3393) also shows multiple strong radiative recombination continuum (RRC) lines expected of photoionization, whereas RRC lines are weak or absent in NGC 3393 (The \ion{O}{8} RRC may overlap with the \ion{Fe}{18} 0.873 keV line claimed by \citealt{Koss15}, but in the case of a photoionized plasma the \ion{O}{8} RRC should be twice as bright as the observed \ion{Fe}{18}).

We therefore infer that X-ray grating spectroscopy requires a strong collisional component.  Our APEC models derived from spatially resolved CCD spectroscopy are therefore necessary tools to describe kinetic feedback in NGC 3393, and should therefore be investigated in the context of outflows observed at other wavelengths and shock generation.

\subsection{X-ray Spectroscopy Connections to Optical/Millimeter Data}

Spatially resolved spectroscopy (as described in \S\ref{sec:XELImSpec}) may be key to interpreting the X-ray ENLR near the nucleus of AGN, since the X-rays imply a complex picture.  Almost all ENLR regions which we define and investigate are consistent with a model that combines photoionization with a collisional plasma, but the properties of those models vary across the ENLR.  Photoionization tends to dominate emission in all regions, but the flux contribution from collisional plasma can approach comparable levels, including throughout the S-shaped arms where the \ion{O}{7}-bright nuclear bar (forming part of the S-shaped arm) appears dominated by highly ionizing radiation reprocessed by nearly Compton-thick material.   

\subsubsection{Hot Collisional Gas and Optical Outflows}

``\ion{Ne}{9}-bright" material leading the radio lobes (regs. 9, 12) can be addressed in the context of \cite{Finlez18}, who recently used {\it Gemini} GMOS IFU data to identify regions of NE overdensity, NE outflow, SW radial outflow, and SW equatorial outflow which they label ``O1", ``O2", ``O3" and ``O4" respectively (Figs. 6, 7).  In the shock interpretation for the NE cone advanced by \cite{Finlez18}, we would expect a hot collisional component associated with  outflowing O1 and higher-density O2, and we find this with Reg. 12, which is located between them.  In our Reg. 12, the collisional component is comparable to the reprocessed component and is relatively hot ($kT\sim0.76$\,keV), as might be expected in a shock scenario.

In the SW cone, reg. 9 is clearly associated with outflow O3 and has a significant collisional component, but unlike reg. 12 the ``\ion{Ne}{9}" excess is attributable to the best-fit photoionization component rather than collisional plasma since the best-fit collisional temperature is lower than reg. 12 ($kT\sim0.26\,$keV).  The reason for this is not clear.  Reg. 9 is closer to the nucleus and may thus encounter more direct photoionization ($log\,U\sim1.39$ vs $\sim-0.71$), or the lateral outflow may prevent pileup of the outflowing plasma near O3.  Alternately, the detected number of photons may be insufficient to distinguish the best-fit model in reg. 9 from a true solution that is closer to reg. 12.

Reg. 11 is spatially consistent with equatorial outflow O4 identified by \cite{Finlez18}.  The origin of the Fe K$\alpha$ component here is uncertain.  Reg. 11 is close enough to the off-center Fe K$\alpha$ source \citep[proposed by][as a secondary SMBH]{Fabbiano11} that it could be contaminated by photons from that source.  We do not find evidence for obscuration, which might be expected from the disk if the outflow were behind it.  The collisional component appears hot ($kT\sim1\,$keV) but weak, $F_X\simless$ an order of magnitude below the reprocessed photoionization.

\subsubsection{AGN Outflows and the Cross-Cone}

The cross-cone (regs. 3, 7) is consistent with reprocessing by a smaller column ($log\,N_H\sim20.2\,\rm{cm}^{-2}$), as might be expected from a leaky torus.  These regions permit significant cold absorption ($N_H\simgreat\rm{few}\times10^{20}\,\rm{cm}^{-2}$), which is unsurprising given the cospatial ALMA detection of CO by \cite[e.g. Fig. 9]{Finlez18}.  A cooler ($kT\simless0.1\,$keV) collisional component is permitted and could be associated with the tail end of an equatorial outflow, but has very uncertain $F_X$ that is correlated with column density due to the low temperature (and indeed the single CLOUDY model with no APEC is preferred in reg. 7).  

\cite{Wang11a} describe a similar scenario based on observations of elevated obscuration in the cross-cone of NGC 4151 which is associated with strong CO emission.  A cocoon of collisionally-excited gas may therefore be common in the presence of AGN feedback (as expected by e.g. \citealt{Mukherjee16} when outflows are confined by the local ISM).  Equatorial outflows are commonly seen in optical and infrared IFU data, such as in NGC 5929 \citep{Riffel15}, NGC 1386 \citep{Lena15} and 3C 33 \citep{Couto17}, and may therefore produce similar X-ray signatures.


\subsection{Constraints on Feedback from X-ray Spectroscopy}

\begin{table*}
\vspace{0.1in}
\footnotesize
\centering
\caption{Local Physical Parameters of Collisionally Ionized Gas Derived from Best-Fit APEC Models}
\label{table:physics}
\tabcolsep=0.1cm
\vspace{0.1in}
\begin{tabular}{lr@{}lp{4mm}r@{}lr@{}lr@{}lr@{}lr@{}lr@{}lr@{}lr@{}l}
\tableline
Region				&	\mcc{$n_e$} 			& \multicolumn{3}{c}{$p_{th}$}   					& \mcc{$E_{th}$} 		& \mcc{$t_{cool}$}	& \mcc{$c_s$} 		& \mcc{$v_{sh}$} 		& \mcc{$t_{cross}$} 	& \mcc{$E_{th}/t_{cross}$}\\
 (Fig. \ref{fig:ne9o7regs})	&	\mcc{(cm$^{-3}$)}		&	\multicolumn{3}{c}{($10^{-10}$\,dyne\,cm$^{-3}$)}	& \mcc{($10^{53}$\,erg)}	& \mcc{($10^6$\,yr)}	&	\mcc{(km\,s$^{-1}$)}	& \mcc{(km\,s$^{-1}$)}	& \mcc{($10^5$\,yr)}	& \mcc{($10^{41}$\,\es)}\\
\tableline
 1 & \vpmc{0.33}{0.06}{0.04} && \vpmc{ 2.46}{ 0.22}{ 0.25}  & \vpmc{ 14.86}{  1.49}{  1.65} & \vpmc{ 11.06}{  4.38}{ 11.06}& \vpmc{ 245}{  34}{  37} & \vpmc{ 421}{  59}{  64} & \vpmc{ 5.51}{ 1.00}{ 0.95} & \vpmc{0.86}{0.17}{0.18} \\
 2 & \vpmc{0.48}{0.18}{0.16} && \vpmc{19.52}{ 9.22}{ 3.24}  & \vpmc{ 20.61}{  9.98}{  4.06} & \vpmc{ 41.29}{ 26.23}{ 11.66}& \vpmc{ 574}{ 195}{ 102} & \vpmc{ 984}{ 334}{ 175} & \vpmc{ 1.04}{ 0.21}{ 0.37} & \vpmc{6.27}{3.76}{1.78} \\
 3 & \vpmc{0.33}{0.02}{0.06} && \vpmc{ 1.05}{ 0.19}{ 0.17}  & \vpmc{ 54.77}{  9.93}{  8.84} & \vpmc{ 21.39}{ 20.97}{ 12.98}& \vpmc{ 162}{  34}{  32} & \vpmc{ 277}{  59}{  55} & \vpmc{21.19}{ 4.74}{ 4.97} & \vpmc{0.82}{0.24}{0.23} \\
 4 & \vpmc{0.77}{0.09}{0.09} && \vpmc{ 3.94}{ 0.51}{ 0.43}  & \vpmc{  4.70}{  0.77}{  0.70} & \vpmc{  5.17}{  1.51}{  4.75}& \vpmc{ 205}{  29}{  21} & \vpmc{ 351}{  50}{  37} & \vpmc{ 3.48}{ 0.50}{ 0.61} & \vpmc{0.43}{0.10}{0.09} \\
 5 & \vpmc{3.34}{0.90}{0.89} && \vpmc{ 8.54}{ 3.77}{ 1.38}  & \vpmc{  4.13}{  1.94}{  0.93} & \vpmc{  3.22}{  2.59}{ 3.22}& \vpmc{ 145}{  46}{  14} & \vpmc{ 248}{  80}{  24} & \vpmc{ 3.55}{ 0.49}{ 1.19} & \vpmc{0.37}{0.21}{0.10} \\
 6 & \vpmc{5.46}{6.17}{2.04} && \vpmc{ 8.74}{ 0.37}{ 3.15}  & \vpmc{636.06}{ 27.94}{229.27} & \vpmc{ 11.35}{ 10.12}{  7.12}& \vpmc{ 114}{  11}{  34} & \vpmc{ 196}{  20}{  59} & \vpmc{21.54}{ 6.81}{ 3.05} & \vpmc{9.38}{1.39}{4.50} \\
 7 & \vpmc{1.56}{5.30}{0.74} && \vpmc{ 3.05}{ 0.07}{ 0.50}  & \vpmc{156.05}{  4.24}{ 25.80} & \vpmc{ 14.29}{ 14.28}{ 14.29}& \vpmc{ 126}{   8}{  26} & \vpmc{ 217}{  14}{  44} & \vpmc{29.48}{ 6.66}{ 3.50} & \vpmc{1.68}{0.20}{0.47} \\
 8 & \vpmc{0.70}{0.11}{0.21} && \vpmc{22.90}{ 5.74}{ 4.10}  & \vpmc{ 12.27}{  3.58}{  2.86} & \vpmc{ 18.22}{ 18.95}{  9.16}& \vpmc{ 517}{ 116}{  81} & \vpmc{ 886}{ 199}{ 139} & \vpmc{ 1.14}{ 0.21}{ 0.28} & \vpmc{3.43}{1.31}{1.02} \\
 9 & \vpmc{0.75}{0.19}{0.14} && \vpmc{ 6.24}{ 1.72}{ 1.37}  & \vpmc{  3.35}{  1.05}{  0.89} & \vpmc{  5.45}{  3.65}{  2.99}& \vpmc{ 261}{  63}{  52} & \vpmc{ 447}{ 107}{  90} & \vpmc{ 2.25}{ 0.51}{ 0.59} & \vpmc{0.47}{0.19}{0.16} \\
11 & \vpmc{0.36}{0.06}{0.06} && \vpmc{11.73}{ 2.73}{ 4.06}  & \vpmc{  6.29}{  1.74}{  2.37} & \vpmc{ 46.78}{ 48.54}{ 46.78}& \vpmc{ 515}{ 109}{ 144} & \vpmc{ 882}{ 186}{ 247} & \vpmc{ 1.14}{ 0.34}{ 0.27} & \vpmc{1.75}{0.63}{0.84} \\
12 & \vpmc{0.64}{0.12}{0.10} && \vpmc{15.56}{ 3.03}{ 2.21}  & \vpmc{ 11.89}{  2.75}{  2.25} & \vpmc{ 12.49}{  4.02}{  2.82}& \vpmc{ 446}{  86}{  58} & \vpmc{ 765}{ 147}{  99} & \vpmc{ 1.48}{ 0.24}{ 0.32} & \vpmc{2.54}{0.81}{0.64} \\
13 & \vpmc{0.23}{0.08}{0.07} && \vpmc{10.94}{ 0.90}{ 0.42}  & \vpmc{122.95}{ 10.85}{  6.23} & \vpmc{120.62}{ 47.45}{120.62}& \vpmc{ 625}{  86}{  45} & \vpmc{1071}{ 147}{  77} & \vpmc{ 5.02}{ 0.62}{ 0.85} & \vpmc{7.77}{1.49}{1.04} \\
\tableline
\tableline

\end{tabular}
\\
\vspace{0.1in}
\end{table*}

We examine feedback effects of the AGN on the surrounding ISM in a simlar manner to \cite{Paggi12} and \cite{Sartori16}.  The results of these calculations are displayed in Table \ref{table:physics}. The thermal energy injected into a given region can be described as $E_{th}=[\gamma/(\gamma-1)]p_{th}V$, where $\gamma$ is the heat capacity ratio in the gas ($\gamma=4/3$ for a relativistic gas, but here we assume $\gamma=5/3$ for a non-relativistic monoatomic gas), $p_{th}$ is the thermal pressure of the collisional gas, and $V$ is the volume of the region.  For circular regions, we assume a sphere.  For ellipsoids and rectangular solids, we assume a $z$-axis equal to the smaller dimension in the plane of the sky.  For region 13 in the NE cone, we assume a solid ``wedge" of thickness 200\,pc.

\subsubsection{Pressure and Density}

The emission measure ($EM$) of the collisional gas is directly measured by the normalization of the {\tt APEC} model assuming angular distance $D_A$.  Since 

\begin{equation}
EM=\frac{10^{-14}}{4\pi D_A^2(1+z)^2}\int n_e n_H dV
\end{equation}

\noindent we can derive $p_{th}$ from the best-fit $EM$ and $kT$ from the {\tt APEC} component by assuming an ideal proton-electron gas where $p=2n_ekT$.  Typical $n_e$ for the X-ray emitting collisional plasma is $\sim1\,\rm{cm}^{-3}$, with $p_{th}\sim\rm{few}\times10^{-10}\,\rm{dyne\,cm}^{-3}$.  In general, $p_{th}$ is a factor of $\sim$few lower than for Mrk 573 \citep{Paggi12}.  The lowest-pressure regions include the cross-cones and may be associated with high-velocity material escaping ``nozzle" \citep{Finlez18}, whereas the highest-pressure regions are at the ``elbows" of the S-shaped arms, leading the radio blobs.  In general  the X-ray pressures comparable to those inferred from optical lines such as [\ion{O}{3}]$\,\lambda\lambda4363,4959,5007$ and [\ion{S}{2}]$\,\lambda\lambda6716,6731$ ($\sim10^{-9}\,\rm{dyne\,cm}^{-3}$; \citealt{Cooke00}), suggesting the optical and X-ray gas may be coincident and in equilibrium.

\cite{Cooke00} show that the equipartition pressures in the NE and SW lobes are $2.7\times10^{-8}\rm{\,dyne\,cm}^{-3}$ and $4.3\times10^{-7}\rm{\,dyne\,cm}^{-3}$ respectively.  This is much higher than for the thermal X-ray plasma, and implies that the jet pressure is easily sufficient to sustain shock heating of the hot gas, and suggests the lobes may be in the process of burrowing outwards.

The large pressure difference between radio and X-ray/optical deserves some consideration.   If [\ion{O}{3}] arises from a photoionization
precursor with $n\sim500\,\rm{cm}^{-3}$ (based on the pressure), then we expect a density of 2000 behind the shock, as opposed to $n_e\sim1\,\rm{cm}^{-3}$ (shown in Table \ref{table:physics}).  A self-consistent interpretation might
be that the radio lobes with $p\sim10^{-7}\,\rm{dyne\,cm}^{-3}$ drive shocks into a pre-shock gas with $v\sim100\,\kms$.  The pre-shock gas would not itself produce X-rays.  
If some of the gas has $n\sim10\,\rm{cm}^{-3}$, then the shock speed would be $\sim700\,\kms$, giving the inferred X-ray temperatures.  A filling factor of $\sim0.1$ would be necessary in the X-ray emitting regions, which could be a thin layer between the radio lobe and optical region.  Another
possibility would be that most of the volume is filled with much lower density (and much hotter) gas that has such a low emission measure
that it is not seen.  These scenarios could affect subsequent estimates of the shock crossing time $t_{cross}$ and the cooling time $t_{cool}$.

Since the X-rays and [\ion{O}{3}] are co-spatial, we might alternatively suppose that the X-ray gas cools to 10,000 K, below which photoionization dominates the energetics.  In that case, the pressure in the [\ion{O}{3}] gas would be equal to that of the X-ray gas (as we infer), or even smaller,
because magnetic pressure takes over as the density increases.  In this case, one would again expect that the density in the X-ray gas is at least an order of magnitude higher than $n_e\sim1\,\rm{cm}^{-3}$, and the filling factor correspondingly small.

\subsubsection{Feedback and the Energy Budget}

The thermal energy $E_{th}$ for most regions ranges between $\sim10^{53}$\,erg and $\sim10^{55}$\,erg.  The large volumes of the cross-cone (regions 3, 7) and outer bicone (6, 13) imply that these extended regions can dominate energetically if the collisional component is real.  Their inclusion increases the total $E_{th}$ by a factor of $\sim13$, from $7.8\times10^{54}$\,erg to $1.04\times10^{56}$\,erg.  Our assumptions of region thickness also remain major sources of uncertainty in these calculations; if the hot gas occupies a thin ``skin" immediately above the galactic plane (as might be the case for wind-disk interactions beyond the S-shaped arms), then $E_{th}$ may become much smaller.  

Regardless, our best estimates of $t_{cool}$ are at least factor of a few larger than $t_{cross}$.  In some cases (regions 1, 7, 11, 13) the lower bound of $t_{cool}$ is unconstrained due to uncertainty in the flux.  As in \cite{Paggi12}, the local sound speed $c_s$ is comparable to the shock velocity ($v_{sh}\sim100[kT/0.013\,\rm{keV}]^{1/2}$, as per \citealt{Raga02}), which gives us $t_{cross}$. More sensitive observations (currently in progress) are necessary to improve measurements of $E_{th}$ and $t_{cool}$ at \cha\ resolution.  But these results suggest that no additional heating sources are required for most regions, and possibly all regions.  The highest-velocity gas measured by {\it HST} \citep{Fischer13} and {\it Gemini} \citep{Finlez18} exceeds the local sound speed in these regions, so shocks must occur.  These should also be fast enough ($FWHM\simgreat1000\,\kms$) to generate X-rays.

The kinetic power $L_K$ should exceed $E_{th}/t_{cross}$.  The total for all regions is therefore $L_K\simgreat3.6\times10^{42}\,$\es.  The bolometric luminosity $L_{bol}\sim8\times10^{45}$\,\es\ \cite{Koss15}, so $L_K/L_{bol}\sim0.5\%$.  This is comparable to the $\sim0.5$ efficiency suggested by the \cite{HE10} two-stage feedback model, but is close enough that \cite{HE10} may be relevant.  \cite{Finlez18} suggest that the NGC 3393 jet is launched into the galactic disk.  If so, the pressure difference between the extended X-ray gas and radio jets suggests that the jet is relatively young and may continue to burrow through the circumnuclear ISM.  But models of jet-disk interaction \citep[e.g.][]{Mukherjee18} indicate that a jet in such a scenario may be redirected, ablating and expelling the circumnuclear ISM away from the disk and along the path of least resistance.  Deeper radio observations will help determine whether the jets are escaping or remain trapped by the ISM, as with e.g. NGC 404 \citep{Nyland17}, and some AGN with evidence for recent accretion mode switching \citep{Keel15,Keel17,Sartori16}.

\section{Conclusion}\label{sec-con}

Using spatially resolved X-ray spectroscopy of the ENLR of the Sy2 NGC 3393, we see the impact of a young radio outflow in shaping the inner $\sim$\,kpc of the galaxy.  {\it Chandra} provides a critical window to interpreting this picture independently of optical (as explored by Paper I \& II, and recently \citealt{Finlez18}).  Sub-arcsecond resolution is necessary to resolve emission line structure, and suggests that unresolved spectroscopy can mischaracterize the physical properties of the circumnuclear ISM in such a case.

Comparing X-ray wavebands centered on \ion{O}{7}, \ion{O}{8} and \ion{Ne}{9}, we see clear differences in line morphology, with \ion{O}{7} roughly tracing the optical line-emitting gas (particularly [\ion{O}{3}]) but with localized enhancements of \ion{Ne}{9} associated with radio outflows, as has been seen in e.g. NGC 4151 \citep{Wang11b} and Mrk 573 \citep{Paggi12}.  Emission associated with \ion{Ne}{9} (0.905\,keV) likely contains contributions from intermediate-state iron emission associated with thermal shocked or shock-forming plasma, and appears to be associated either with shocks or possibly precursor material photoionized by strong shocks.  We find that emission line measurements made by \cite{Koss15} using {\it Chandra} gratings support a dominant role for collisional plasma in the X-ray emission.  Regardless of the exact nature of the emission origin, the measured sound speed in the collisional plasma suggests (when compared with optical data) that shocks must occur, and the total inferred kinetic power of the outflows is sufficient to evacuate gas from the galaxy via the \cite{HE10} two-stage feedback model.

Although the jet power is sufficient to remove the gas over the long term, the current situation revealed by X-rays is complex.  If the outflows are ablating and deforming the circumnuclear ISM in the disk plane (see also \citealt{Finlez18}), then they may be redirected, escaping perpendicular to the plane as the jets continue to burrow outwards.  In addition to cross-cone photoionization (as from a ``leaky torus" seen in other places e.g. \citealt{Paggi12,Fabbiano18a}), there is evidence for a cocoon of cooler thermal plasma in the cross-cone that may be partially absorbed as seen by \cite{Wang11a} in NGC 4151.  Alternately, confinement of the the equatorial outflow seen in \cite{Finlez18} may contribute to the heating of this component.  Such a cool component is not expected to be bright in X-rays, but may be confirmed via detection in UV via e.g. \ion{O}{6}$\,\lambda1031,1036$ with $HST$ COS in NGC 3393 and possibly other galaxies with equatorial outflows \citep{Riffel15,Lena15,Couto17}.

A more direct comparison between available optical and {\it ALMA} datacubes would prove informative, as would more sensitive {\it Chandra} and broadband radio observations.  Our ongoing {\it Chandra} Large Program will better constrain spectroscopic measurements of the collisional plasma at $\sim0.25\arcsec$ resolution.  Better sensitivity at {\it Chandra} resolution is critical to confidently measuring the energy in the collisional component, particularly at large ($\simgreat1$\,kpc) radii, where a low-density ionized wind may be present above the disk.  Accompanying deep {\it JVLA} observations will be sensitive to lower-density plasma in the outflows which is no longer strongly interacting with the circumnuclear ISM.

Studies with {\it Chandra} make it clear that subarcsecond resolution is necessary to study sub-kpc AGN outflows even at relatively low redshifts (like NGC 3393) {\it without} spatial confusion.  This groundwork with {\it Chandra} is essential to interpreting any observations made with the future ESA X-ray mission {\it Athena}.  {\it Athena} will provide powerful diagnostics of the complex plasma that we investigate here, thanks to a collecting area and energy resolution greatly superior to any observatory to date \citep{Athena}, but its spatial resolution ($\sim$few\arcsec) will confuse multiple physically distinct AGN phenomena even at the very low redshifts.  {\it Lynx}, currently proposed for NASA in the 2020 Decadal Survey, will enable imaging spectroscopy at {\it Chandra}-like resolution but with effective area and energy resolution orders of magnitude better than {\it Chandra} \citep{Lynx}.  {\it Lynx} is necessary to efficiently measure the spatially resolved multicomponent hot outflows from AGN, and will confidently distinguish collisional plasma from reprocessing of photoionizing radiation by the ISM even in faint ENLR systems.




\acknowledgments

WPM thanks Travis Fischer, Steve Kraemer and Vinay Kashyap for helpful and informative discussions.  WPM was supported by {\it Chandra} grants GO5-16101X, GO8-19096X, and GO8-19099X, as well as  {\it HST} grants HST-GO-14271.009-A and HST-GO-15350.001-A.



\facilities{CXO,HST}






\bibliographystyle{apj}  

\bibliography{apj-jour,pete_tidal,biblio_mel_marc,pete_agn}




\end{document}